\documentclass[11pt]{article}
\pdfoutput=1
\usepackage{amssymb,amsmath,mathrsfs,enumerate}
\usepackage{graphicx,rotate,multicol}
\usepackage{caption,verbatim}
\usepackage{cite}
\usepackage[colorlinks=true,
linkcolor=red,
urlcolor=blue,
citecolor=blue]{hyperref}
\usepackage{color}
\usepackage{textcomp} 
\usepackage{setspace}
\usepackage{subfig}
 
\evensidemargin=\oddsidemargin
\def\mET{E_T \hspace{-1.2em}/\;\:}

\allowdisplaybreaks
\newcommand{\gsim}{\lower.7ex\hbox{$\;\stackrel{\textstyle>}{\sim}\;$}}
\newcommand{\lsim}{\lower.7ex\hbox{$\;\stackrel{\textstyle<}{\sim}\;$}}

\definecolor{darkgreen}{cmyk}{1,0,1,0.4}

\begin{document}
%
\begin{flushright}
	{\small CUMQ/HEP-192}  
\end{flushright}

\begin{center}
	{\Large \bf Dark Matter and Collider Studies
	in the Left-Right Symmetric Model with Vector-Like Leptons} \\
		\vspace*{1cm} {\sf Sahar Bahrami$^{a,}$\footnote{sahar.bahrami@mail.mcgill.ca}, 
			~Mariana Frank$^{b,}$\footnote{mariana.frank@concordia.ca},
			~Dilip Kumar Ghosh$^{c,}$\footnote{tpdkg@iacs.res.in}, \\
			~Nivedita Ghosh$^{c,}$\footnote{tpng@iacs.res.in},
			~Ipsita Saha$^{d,}$\footnote{ipsita.saha@roma1.infn.it}} \\
		\vspace{10pt} {\small } {\em $^a$ Department of Physics, McGill University, \\
		3600 Rue University, Montreal, Quebec, Canada H3A 2T8 ​\\	
		$^b$ Department of Physics,  
			Concordia University, 7141 Sherbrooke St. West,\\
			Montreal, Quebec, Canada H4B 1R6. \\
			$^c$Department of Theoretical Physics, 
			Indian Association for the Cultivation of Science.\\
			2A $\&$ 2B, Raja S.C. Mullick Road, Kolkata 700032, India \\
			$^d$ Istituto Nazionale di Fisica Nucleare, Sezione di Roma, \\
			Piazzale Aldo Moro 2, I-00185 Roma, Italy}   
\end{center}
\normalsize
\begin{abstract}
In the context of a left-right symmetric model, we introduce one full generation of vectorlike lepton doublets (both left and right-handed) together with their mirror doublets. We show that the lightest vectorlike neutrino in the model is right-handed, and can serve as the dark matter candidate. We  find that the relic density as well as the direct and indirect DM 
detection bounds are satisfied for a large range of the parameter space of the model. 
In accordance with the parameter space, we then explore the possibility of 
detecting signals of the model at both the LHC and the ILC, 
in the pair production of the associated vectorlike charged leptons
which decay into final states including dark matter. 
A comprehensive analysis of signal and backgrounds shows that the signals at the ILC, 
especially with polarized beams are likely to be visible  for light vectorlike leptons, even with low luminosity, 
rendering our model highly predictable and experimentally testable.
\end{abstract}
\bigskip


\section{Introduction}
 \label{sec:intro}
 In left-right symmetric models (LRSMs), left- and right-handed particles are treated on the same footing. This represents an improvement over the Standard Model (SM) where, instead of providing an explanation for its origin, parity violation is incorporated {\it ad hoc} into the model. Left-right models assume parity is conserved at high energies, and broken spontaneously at lower energies, providing an alternative for why nature would prefer any left-right discrimination. In addition, these models have several attractive features, such as providing an explanation for matter-antimatter symmetry, and a natural framework for neutrino masses \cite{Mohapatra:1974gc,Senjanovic:1975rk,Mohapatra:1980yp}.  It is reasonable to ask whether these models can accommodate and explain dark matter (DM), another ingredient missing in the SM.
 
This topic has been previously explored in the context of LRSMs-for instance, imposing a discrete $Z_2=(-1)^{3(B-L)}$ symmetry, denoted matter parity that survives after the breaking of global $SO(10)$ into $SU(3)_c \otimes SU(2)_L \otimes SU(2)_R \otimes U(1)_{B-L}$. Identifying some representations of $SO(10)$ as $Z_2$-even, while others are odd, a possible DM candidate 
can be accommodated where the scalars belonging to the odd representation would represent the DM candidates \cite{Kadastik:2009cu,Kadastik:2009dj,Arbelaez:2013nga,Heeck:2015qra,Garcia-Cely:2015quu}. Alternatively, a $Z_2$ symmetry can also be imposed such as the triplet scalars ($\Delta_L$ and $\Delta_R$), introduced in the model, which will transform under the symmetry as ($\Delta_L \to -\Delta_L, ~~\Delta_R \to \Delta_R$).
Furthermore, setting the vacuum expectation value (VEV) of the 
left-handed triplet $\Delta_L^0$ to be zero ($v_L=0$), the neutral components would become 
degenerate and stable, and can thus cater possible dark matter candidates.  
Unfortunately this $\Delta_L^0$ candidate cannot provide the correct relic density, due to small annihilation cross sections~\cite{Guo:2008hy}. This problem has been cured by introducing an additional gauge singlet~\cite{Guo:2008si}.  Leptophilic properties of the decaying left-handed triplet  Higgs have been explored to explain enhancements in neutrino-induced muon fluxes~\cite{Guo:2010vy}, and further properties of the  gauge singlet have been explored in~\cite{Guo:2008si}. LRSMs with fermionic dark matter  have also been explored~\cite{Ma:2012gb,Gu:2010yf,Nemevsek:2012cd}.

Here we shall explore an alternative candidate for dark matter, vectorlike neutrinos by introducing
additional vectorlike lepton doublets to LRSM\cite{Kuchimanchi:2012te}. Vectorlike fermions appear 
naturally in composite Higgs models, warped extra dimensions, little Higgs and extended grand unified theories.  
In left-right models, where left and right chiral representations are naturally connected, vectorlike fermions are germane, as they are characterized by having left- and right-handed components
transforming in the same way under the symmetry group of the theory, and by the fact that the
couplings for the right-handed components are the same as for the left-handed ones.  vectorlike fermions have received much attention lately, being put forth as explanations for hints of new physics at the LHC: the ATLAS diboson \cite{Aad:2015owa}, the CMS $eejj$ excess \cite{Khachatryan:2014dka} and the 750 GeV diphoton  signal \cite{Aaboud:2016tru,CMS:2016owr}.    In particular,  vectorlike particles in the context of left-right symmetric models are inherent in warped space extradimensional models.  If the 
additional dimension, extending between two branes, one at the TeV
scale (IR brane) and the other at Planck scale (UV brane), with
gravity propagating in the bulk , is warped, the resulting geometry
generates naturally the  hierarchy between the electroweak scale
($M_{\rm ew} \sim 200 $ GeV) and the Planck scale ($M_{\rm Pl}\sim 10
^{18}$~GeV) \cite{Randall:1999ee,Randall:1999vf}. In addition
 if  one allows the SM fields to propagate in the
bulk of the  fifth dimension, these models  can explain the observed masses of
the fermions, with  lighter fermions  localized near the
Planck brane and the heavier ones localized near the TeV
brane.  Within this new framework, generic models with warped extra dimensions are still 
very constrained by electroweak and flavor precision tests \cite{Agashe:2003zs}. To reduce the pressure from electroweak precision tests, a common cure is to enlarge the gauge symmetry of the SM by introducing a custodial $SU(2)_L \times SU(2)_R$ symmetry that limits the corrections to various precision observables \cite{Agashe:2006at}. This new symmetry provides a natural framework for vectorlike fermions in LRSMs, which appear as KK excitations of SM chiral fermions.
 
  vectorlike fermions appear in LRSMs also from gauged flavor symmetries, where they are needed to cancel new gauge anomalies \cite{Guadagnoli:2011id}.   Additionally, in left-right models the scale where the parity breaks down, $\Lambda_R$ is expected to be high, $10^{14}$ GeV or higher. One could introduce another intermediate mass scale, associated with the $(B-L)$-breaking scale $\Lambda_{B-L}$ \cite{Aulakh:1998nn}, which could emerge as the scale of some new fermions, in our case the vectorlike leptons, sometimes interpreted as the scale of compositeness  \cite{Banerjee:2013kna}. In addition, it is known that in two-Higgs doublet models, vectorlike fermions, in addition to the extra Higgs bosons, alleviate electroweak precision tests \cite{Garg:2013rba}.
  Some collider studies related to such vectorlike leptons augmented in two-Higgs doublets has been explored in 
  \cite{Dermisek:2015vra,Dermisek:2015oja,Dermisek:2015hue,Dermisek:2016via}.

  Effects of vectorlike fermions to the couplings of the SM-like 125 GeV Higgs
boson, constrained from measurements of the Higgs-
production cross-sections and branching ratios at the LHC have been analyzed before \cite{Dolan:2016eki}. Here we examine
the possibility that, in the context of LRSMs, when they can become dark matter candidates, what is their effect on the parameter space, and how this scenario can be discriminated at the present (LHC) and proposed (ILC) colliders. 

Our work is organized as follows. In  Sec.~\ref{sec:LR} we describe the LRSM with the addition of  vectorlike lepton doublets. For simplicity, we concentrate in particular on scenarios where mixing between ordinary leptons and vectorlike leptons is forbidden by a discrete parity symmetry. The model has one left-handed and one right-handed doublet, and their mirror representations, yielding mixing between same-charge components. We discuss the mass eigenstates, the lightest of which would be the dark matter candidate.  In Sec.~ \ref{sec:dm} we calculate the relic density, the spin-independent and spin-dependent cross sections, the annihilation cross section and muon and neutrino fluxes, and we explore the parameter space which is consistent with the experimental results on dark matter detection. In Sec.~\ref{sec:collider} we explore ways in which this scenario can be tested, in particular, looking for signals of the lightest vectorlike charged lepton pair at the LHC (\ref{subsec:LHC}) and ILC (\ref{subsec:ILC}). We summarize our results in Sec.~ \ref{sec:conclusion}. In the Appendix we provide explicit expressions for the vectorlike lepton mass eigenstates.


\section{The left-right symmetric model} \label{sec:LR}

In the left-right symmetric model \cite{Mohapatra:1974gc,Senjanovic:1975rk,Mohapatra:1980yp}, the Standard Model gauge symmetry is extended to include the gauge group $SU(2)_R$ (with gauge coupling $g_R $). All right-handed fermions are  doublets under
this gauge group. 
Below we give their quantum numbers under
 $SU(2)_L \otimes SU(2)_R \otimes U(1)_{B-L}$. The ordinary fermions are: the leptons
  \begin{eqnarray}
 L_{Li} = \begin{pmatrix}\nu_L \\ \ell_L\end{pmatrix}_i \, \sim
(\mathbf{2},\mathbf{1},\mathbf{-1}) \, , ~&
L_{Ri} =
\begin{pmatrix}\nu_R \\ \ell_R\end{pmatrix}_i \sim
(\mathbf{1},\mathbf{2},\mathbf{-1}) \, , 
\end{eqnarray}
and the quarks
\begin{eqnarray} 
 Q_{Li} = \begin{pmatrix}u_L \\ d_L\end{pmatrix}_i \, \sim
(\mathbf{2},\mathbf{1},\mathbf{1/3}) \, , ~&
Q_{Ri} =
\begin{pmatrix}u_R \\ d_R\end{pmatrix}_i \sim
(\mathbf{1},\mathbf{2},\mathbf{1/3}) \, , 
\end{eqnarray}
where $i=1,2,3$ are generation indices. Note that the right-handed neutrino is automatically included. The electroweak symmetry is
broken by the bidoublet Higgs field
\begin{equation}
 \Phi \equiv \begin{pmatrix} \phi_1^0 & \phi_2^+ \\ \phi_1^- & \phi_2^0 \end{pmatrix} \sim (\mathbf{2},\mathbf{2},\mathbf{0})\, .
\end{equation}
In addition, to break the $SU(2)_R \otimes U(1)_{B-L}$ gauge symmetry and to provide Majorana
mass terms for neutrinos we introduce the Higgs triplets 
\begin{equation}
 \Delta_{L} \equiv \begin{pmatrix} \delta_{L}^+/\sqrt{2} & \delta_{L}^{++} \\ \delta_{L}^0 & -\delta_{L}^+/\sqrt{2} \end{pmatrix} \sim (\mathbf{3},\mathbf{1},\mathbf{2}) \, , ~~~ \Delta_{R} \equiv \begin{pmatrix} \delta_{R}^+/\sqrt{2} & \delta_{R}^{++} \\ \delta_{R}^0 & -\delta_{R}^+/\sqrt{2} \end{pmatrix} \sim (\mathbf{1},\mathbf{3},\mathbf{2})\, .
\end{equation}
The electric charge is given by $Q=T_L^3+T_R^3+\frac{B-L}{2}$. 
 The subscripts $L$ and $R$ are associated with the
projection $P_{L,R} = \frac 12 (1 \mp \gamma_5)$.  We add one family of vectorlike leptons\footnote{vectorlike quarks can also appear; for the present work, we assume them to be much heavier than the leptons, based on the mass limits \cite{Olive:2016xmw}  and thus they decouple from the low-energy spectrum.}
\begin{eqnarray}
 L^\prime_{L} = \begin{pmatrix}\nu^\prime_L \\ \ell^\prime_L\end{pmatrix} \, \sim
(\mathbf{2},\mathbf{1},\mathbf{-1}) \, , ~&
L^\prime_{R} =
\begin{pmatrix}\nu^\prime_R \\ \ell^\prime_R\end{pmatrix}\sim
(\mathbf{1},\mathbf{2},\mathbf{-1}) \, , \nonumber  \\
L^{\prime \prime}_{R} = \begin{pmatrix}\nu^{\prime \prime}_R \\ \ell^{\prime \prime}_R\end{pmatrix} \, \sim
(\mathbf{2},\mathbf{1},\mathbf{-1}) \, , ~&
L^{\prime \prime}_{L} =
\begin{pmatrix}\nu^{\prime \prime}_L \\ \ell^{\prime \prime}_L\end{pmatrix}\sim
(\mathbf{1},\mathbf{2},\mathbf{-1}) \, ,
\label{eq:VLleptons}
\end{eqnarray}
where $L_L^\prime$ and $L_R^\prime$ are new fermion doublets and $L_R^{\prime \prime}$ and $L_L^{\prime \prime}$ are the mirror doublets.
Furthermore, using  the gauge symmetry to eliminate complex phases, the most
general vacuum is 
\begin{gather} 
 \langle \Phi\rangle = \begin{pmatrix} \kappa_1/\sqrt{2} & 0 \\ 0 &
\kappa_2e^{i\alpha}/\sqrt{2} \end{pmatrix}, \quad \langle \Delta_{L}
\rangle = \begin{pmatrix} 0 & 0 \\ v_{L}e^{i\theta_L}/\sqrt{2} & 0
\end{pmatrix}, \quad \langle \Delta_{R} \rangle = \begin{pmatrix} 0 &
0 \\ v_{R}/\sqrt{2} & 0 \end{pmatrix}. 
\end{gather}
Note that only $\Delta_R$ is needed for symmetry breaking, and $\Delta_L$ is included to preserve left-right symmetry.  We assume, as is usual for LRM, $v_L \ll v_R$, to obtain light left-handed neutrino masses, and $\kappa_2<\kappa_1$ to avoid potentially large flavor violation coming from the Higgs sector.
The Lagrangian density for this model contains, in addition to the SM terms, kinetic and Yukawa for ordinary leptons, explicit terms for the vectorlike leptons, and potential terms:
\begin{eqnarray}
\mathcal{L}_{\rm{LRM}}= \mathcal{L}_{\rm{kin}}+\mathcal{L}_{Y}+\mathcal{L}_{\rm VL}-V(\Phi,\Delta_L, \Delta_R), 
\end{eqnarray}
where
\begin{eqnarray}
L_{kin}&=&i\sum\bar{\psi}\gamma^\mu D_\mu\psi 
=\bar{L}^\prime_L\gamma^{\mu}\left(i\partial_{\mu}+g_{L}\frac{\vec{\tau}}{2}\cdot\vec{W_{L\mu}}-\frac{g'}{2}B_{\mu}\right)L^\prime_L \nonumber \\
&+&\bar{L}^\prime_R\gamma^{\mu}\left(i\partial_{\mu}+g_{R}\frac{\vec{\tau}}{2}\cdot\vec{W_{R\mu}}-\frac{g'}{2}B_{\mu}\right)L^\prime_R 
+\bar{L}_L^{\prime \prime}\gamma^{\mu}\left(i\partial_{\mu}+g_{R}\frac{\vec{\tau}}{2}\cdot\vec{W_{R\mu}}-\frac{g'}{2}B_{\mu}\right) L_L^{\prime  \prime} \nonumber \\
&+&\bar{L}_R^{\prime \prime}\gamma^{\mu}\left(i\partial_{\mu}+g_{L}\frac{\vec{\tau}}{2}\cdot \vec{W_{L\mu}}-\frac{g'}{2}B_{\mu}\right)L_R^{\prime \prime} \,,
\end{eqnarray}
where we introduce the gauge fields, $\vec{W}_{L,R}$ and $B$ corresponding to $SU(2)_{L,R}$ and $U(1)_{B-L}$. They mix with the following matrices \cite{Zhang:2007da}
\begin{equation}
 \begin{pmatrix} W_L^{\pm} \\ W_R^{\pm} \end{pmatrix} =
\begin{pmatrix} \cos\xi & \sin\xi\, e^{i\alpha} \\ -\sin\xi\,
e^{-i\alpha} & \cos\xi \end{pmatrix} \begin{pmatrix} W_1^{\pm} \\
W_2^{\pm}\end{pmatrix}.
\label{eq:wlwr_mixing}
\end{equation}
The angle $\xi$ characterizes the mixing between left- and right-handed gauge
bosons, with $\tan2\xi = -\frac{2\kappa_1\kappa_2}{v_R^2-v_L^2}$. 
It follows that
\begin{equation}
 \xi \simeq -\kappa_1\kappa_2/v_R^2 \simeq -2\frac{\kappa_2}{\kappa_1}\left(\frac{m_{W_L} }{m_{W_R} }\right)^2,
\label{eq:zeta_def}
\end{equation}
so that the mixing angle $\xi$ is at most\footnote{ Although the experimental limit is $\xi < 10^{-2}$ \cite{Olive:2016xmw}, for $m_{W_R} = {\cal O}({\rm TeV})$ one has $\xi \le 10^{-3}$
\cite{Langacker:1989xa}; supernova bounds for right-handed neutrinos lighter than 1 MeV are even more stringent ($\xi < 3 \times 10^{-5}$) \cite{Langacker:1989xa,Barbieri:1988av}.} the square of the
ratio of the left and right scales $(\Lambda_L/\Lambda_R)^2$. Here  $\Lambda_L \simeq 10^2$ GeV corresponds to the electroweak scale and $\Lambda_R \simeq$~TeV is the scale of parity breaking, $v_R$.

With negligible mixing the gauge boson masses become, for $g_L=g_R$, 
\begin{equation}
m_{W_L} \simeq m_{W_1} \simeq \frac{g}{2}\kappa_+\, , \quad {\rm and} \quad m_{W_R} \simeq m_{W_2} \simeq \frac{g}{\sqrt{2}}v_R\, , \label{eq:mw1_2}
\end{equation}
with $\kappa_+^2=\kappa_1^2+\kappa_2^2$. The model also has an additional neutral gauge boson, $Z_R$, which mixes with the Standard Model $Z$ boson. The mass eigenstates $Z_{1,2}$ acquire masses
\begin{equation}
 m_{Z_1} \simeq \frac{g}{2\cos\theta_W}\kappa_+ \simeq \frac{m_{W_1}}{\cos\theta_W}\, , \quad {\rm and} \quad m_{Z_2} \simeq \frac{g\cos\theta_W}{\sqrt{\cos2\theta_W}}v_R\simeq
\sqrt{\frac{2\cos^2\theta_W}{\cos2\theta_W}}\,m_{W_2}\, ,
\label{eq:mz1_2}
\end{equation}
where $g = e/\sin\theta_W$ and with the $U(1)_{B-L}$ coupling constant $g_{B-L}\equiv e/\sqrt{\cos2\theta_W}$. Again one expects the mixing between the neutral gauge bosons to be of order $(\Lambda_L/\Lambda_R)^2$, i.e.,
\begin{equation}
\sin 2 \phi = -\frac{g^2\kappa_+^2\sqrt{\cos\,2\theta_W}}{2c_W^2(m_{Z_2}^2-m_{Z_1}^2)} \simeq -\frac{2 m_{Z_1}^2\sqrt{\cos\,2\theta_W}}{m_{Z_2}^2-m_{Z_1}^2} \simeq -
2 \sqrt{\cos\,2\theta_W}\left(\frac{m_{Z_1}}{m_{Z_2}}\right)^2.
\end{equation}
Equations~\eqref{eq:mw1_2} and \eqref{eq:mz1_2} imply that $m_{Z_2} \simeq 1.7 m_{W_2}$~\cite{Patra:2015bga,Lindner:2016lpp}.
The appropriate gauge coupling constants are $g_s$, $g_L=g_R$ and $g^\prime=g_{B-L}$, respectively. 
The right handed $SU(2)_R$-breaking scale is restricted from low energy observables, such as 
$K_L-K_S$, $\epsilon_K$, $B^0-\bar{B}^0$ mixings and $b \to s\gamma$ processes where the right handed
charged current contributes significantly
\cite{Beall:1981ze,Branco:1982wp,Ecker:1983uh,Bigi:1983bpa,Babu:1993hx,Ball:1999mb,Zhang:2007da,Maiezza:2010ic,Blanke:2011ry,Bertolini:2014sua,Bernard:2015boz}. Thus, these processes provide a bound to the scale $v_R$
by means of the charged right-handed $W_R$ boson mass as well as the LR Higgs masses.
In particular, the right-handed $W_R$ mass is restricted to be greater than 3~TeV while the heavy bi-doublet
Higgs mass should at least be 10~TeV~\cite{Zhang:2007da}. In our study, we thus fix the scale $v_R$ at 10~TeV. 

The rest of the Lagrangian terms  
\begin{eqnarray}
\mathcal{L}_Y&=&-\Big[Y_L {\bar L}_L \Phi L_R +Y_R {\bar L}_R \Phi L_L+
{\tilde Y}_L {\bar L}_L {\tilde \Phi} L_R
  +{\tilde Y}_R {\bar L}_R {\tilde \Phi} L_L\nonumber \\
  &+&h_{L_{ij}}\overline {L^{i c}_L}i \tau_2 \Delta_L L^j_L+h_{R_{ij}}\overline {L^{i c}_R}i \tau_2 \Delta_R L^j_R +\rm{h.c.} \Big] 
\label{fermion_yukawa}
\end{eqnarray}
are the Yukawa interaction terms for the ordinary leptons, where $Y_{L,R}$, ${\tilde Y}_{L,R}$  
 are 3$\times 3$ complex matrices, and $h_{L_{ij}}$, $h_{R_{ij}}$ are $3\times 3$ complex symmetric Yukawa matrices and $\tilde \Phi=\tau_2 \Phi^{\star} \tau_2$.  
Additionally, with the vectorlike family of leptons as defined above, the Lagrangian describing  Yukawa interaction terms for vectorlike fermions and their interactions with ordinary fermions, and allowing for both Dirac and Majorana mass terms, is
\begin{eqnarray}
\mathcal{L}_{\rm VL}&=&-\Big [M_{L}{\bar L}_L^\prime L_R^{\prime \prime}+M_{R}{\bar L}_R^{\prime} L_L^{\prime \prime} +Y_L^\prime {\bar L}_L^\prime \Phi L_R^\prime +Y_R^{\prime } {\bar L}_R^{\prime \prime} \Phi L_L^{\prime \prime} +{\tilde Y}_L^\prime {\bar L}_L^\prime {\tilde \Phi}L_R^\prime 
  +{\tilde Y}_R^{\prime} {\bar L}_R^{\prime \prime} {\tilde \Phi}L_L^{\prime \prime} \nonumber\\
&+&  h^\prime_{L}\overline{L_L^{\prime\, c}}i\tau_2\Delta_L L_L^{\prime\, }
+h^{\prime \prime}_{R}\overline{L_L^{\prime \prime \, c}}i\tau_2\Delta_R L_L^{\prime\prime } +h^{\prime}_{R}\overline{L_R^{\prime  \, c}}i\tau_2\Delta_R L_R^{\prime } +h^{\prime \prime}_{L}\overline{L_R^{\prime \prime \, c}}i\tau_2\Delta_L L_R^{\prime\prime }+   \lambda_L^i {\bar L}_L^\prime \Phi L_R^i  \\
&+& \lambda_R^{i }{\bar L}_L^i {\tilde \Phi} L_R^\prime + \lambda^{\prime\,i}_{L}\overline{L_L^{ ic}}i\tau_2\Delta_L L_L^{\prime}
+ \lambda^{\prime\,i}_{R}\overline{L_R^{ ic}}i\tau_2\Delta_R L_R^{\prime} +
\lambda^{\prime \prime \,i}_{L}\overline{L_L^{ ic}}i\tau_2\Delta_L L_R^{\prime \prime} + \lambda^{\prime \prime \,i}_{R}\overline{L_R^{ ic}}i\tau_2\Delta_R L_L^{\prime \prime}   +\rm{h.c.} \Big]\nonumber
\label{vl_lgr}
\end{eqnarray}
Here, in addition to the new Yukawa couplings $Y^{ \prime}_{L,R},~Y^{\prime \prime}_{L,R}$ of the vectorlike leptons with the bidoublet, and $h^\prime_{L,R},~h^{\prime \prime}_{L,R}$, the Yukawa couplings  of the vectorlike leptons with triplet $\Delta_{L,R}$, we also introduce explicit mass terms for the vectorlike leptons $M_L$ and $M_R$. The scalar potential for the bidoublet $\Phi$  and triplet $\Delta_{L,R}$ Higgs fields is 
\begin{eqnarray}
V(\phi,\Delta_L,\Delta_R)& = &-\mu_{1}^2\left({\rm Tr}\left[\Phi^\dagger\Phi\right]\right)-\mu_{2}^2\left({\rm Tr}\left[\tilde{\Phi}\Phi^\dagger\right]+\left({\rm Tr}\left[\tilde{\Phi}^\dagger\Phi\right]\right)\right)-\mu_{3}^2\left({\rm Tr}\left[\Delta_L\Delta_L^{\dagger}\right]+{\rm Tr}\left[\Delta_R\Delta_R^{\dagger}\right]\right) \nonumber \\
&+&\lambda_1\left(\left({\rm Tr}\left[\Phi\Phi^\dagger\right]\right)^2\right)
+\lambda_2\left(\left({\rm Tr}\left[\tilde{\Phi}\Phi^\dagger\right]\right)
+\left({\rm Tr}\left[\tilde{\Phi}^\dagger\Phi\right]\right)^2\right)
+\lambda_3\left({\rm Tr}\left[\tilde{\Phi}\Phi^\dagger\right]{\rm Tr}\left[\tilde{\Phi}^\dagger\Phi\right]\right)\nonumber\\
&+&\lambda_4\left({\rm Tr}\left[\Phi\Phi^{\dagger}\right]\left({\rm Tr}\left[\tilde{\Phi}\Phi^\dagger\right]
+{\rm Tr}\left[\tilde{\Phi}^\dagger\Phi\right]\right)\right) 
+\rho_1\left(\left({\rm Tr}\left[\Delta_L\Delta_L^{\dagger}\right]\right)^2
+\left({\rm Tr}\left[\Delta_R\Delta_R^{\dagger}\right]\right)^2\right)\nonumber\\
&+&\rho_2\left({\rm Tr}\left[\Delta_L\Delta_L\right]{\rm Tr}\left[\Delta_L^{\dagger}\Delta_L^{\dagger}\right]
+{\rm Tr}\left[\Delta_R\Delta_R\right]{\rm Tr}\left[\Delta_R^{\dagger}\Delta_R^{\dagger}\right]\right) \nonumber \\
& +&\rho_3\left({\rm Tr}\left[\Delta_L\Delta_L^{\dagger}\right]{\rm Tr}\left[\Delta_R\Delta_R^{\dagger}\right]
\right)
+\rho_4\left({\rm Tr}\left[\Delta_L\Delta_L\right]{\rm Tr}\left[\Delta_R^{\dagger}\Delta_R^{\dagger}\right]
+{\rm Tr}\left[\Delta_L^{\dagger} \Delta_L^{\dagger}\right]{\rm Tr}\left[\Delta_R\Delta_R\right]\right)
\nonumber\\
&+& \alpha_1\left({\rm Tr}\left[\Phi\Phi^{\dagger}\right]\left({\rm Tr}\left[\Delta_L\Delta_L^{\dagger}\right]
+{\rm Tr}\left[\Delta_R\Delta_R^{\dagger}\right]\right)\right) \nonumber \\
&+&\alpha_2\left({\rm Tr}\left[\Phi\tilde{\Phi}^{\dagger}\right]{\rm Tr}\left[\Delta_R\Delta_R^{\dagger}\right]
+{\rm Tr}\left[\Phi^{\dagger}\tilde{\Phi}\right]{\rm Tr}\left[\Delta_L\Delta_L^{\dagger}\right]\right)
\nonumber \\
&+&\alpha_2^{*}\left({\rm Tr}\left[\Phi^{\dagger}\tilde{\Phi}\right]{\rm Tr}\left[\Delta_R\Delta_R^{\dagger}
\right]+{\rm Tr}\left[\tilde{\Phi}^{\dagger}\Phi\right]{\rm Tr}\left[\Delta_L\Delta_L^{\dagger}\right]\right) \nonumber \\
&+& \alpha_3\left({\rm Tr}\left[\Phi\Phi^{\dagger}\Delta_L\Delta_L^{\dagger}\right]
+{\rm Tr}\left[\Phi^{\dagger}\Phi\Delta_R\Delta_R^{\dagger}\right]\right)\nonumber\\
&+&\beta_1\left({\rm Tr}\left[\Phi\Delta_R\Phi^{\dagger}\Delta_L^{\dagger}\right]
+{\rm Tr}\left[\Phi^{\dagger}\Delta_L\Phi\Delta_R^{\dagger}\right]\right)
+\beta_2\left({\rm Tr}\left[\tilde{\Phi}\Delta_R\Phi^{\dagger}\Delta_L^{\dagger}\right]
+{\rm Tr}\left[\tilde{\Phi}^{\dagger}\Delta_L\Phi\Delta_R^{\dagger}\right]\right)\nonumber\\
&+&\beta_3\left({\rm Tr}\left[\Phi\Delta_R\tilde{\Phi}^{\dagger}\Delta_L^{\dagger}\right]
+{\rm Tr}\left[\Phi^{\dagger}\Delta_L\tilde{\Phi}\Delta_R^{\dagger}\right]\right),
~~~~ \label{eq:pot_htm}
\end{eqnarray}
where we follow \cite{Dev:2016dja} and explicitly indicate the complex parameters. The parameters can be further reduced and simplified by making use of the symmetries of the model.  Assuming a discrete left-right symmetry in addition to the left-right gauge symmetry, the $SU(2)$ gauge couplings become equal ($g_L=g_R=g$) and  the Yukawa coupling matrices for the left and right-handed sector in the
model are related. With a discrete parity symmetry ($L^\prime_L \leftrightarrow L^\prime_R$, $L^{\prime \prime}_L \leftrightarrow L^{\prime \prime}_R$, $\Phi \leftrightarrow \Phi^\dagger$, $\Delta_L \leftrightarrow \Delta^*_R$) it follows that ${h}^\prime_{L,R}={ h}_{L,R}^{\prime \prime \,\star}$, ${ Y^\prime_L}={ Y_L}^{ \prime \, \star}$,
$\tilde{Y}_L=\tilde{Y}_L^\dagger$, ${ Y^\prime_R}={ Y_R}^{ \prime \, \star}$,
$\tilde{Y}_R=\tilde{Y}_R^\dagger$. In addition, using the charge conjugation symmetry  we obtain ${ h}_{L,R}^\prime={ h}_{L,R}^{\prime \prime}\equiv {h}$.
New symmetries can be introduced to restrict the interactions of the vector leptons. For instance,
we can impose {\it (i)} a symmetry under which all the new $SU(2)_R$ doublet fields are odd, while the new $SU(2)_L$ doublets are even, which forces all  Yukawa couplings involving new fermions to vanish,  $Y_{L}^\prime={\tilde Y}_L^\prime=Y_{R}^\prime={\tilde Y}_R^\prime=0$, and the vector fermion masses arise only from explicit terms in the Lagrangian \cite{Joglekar:2012vc}
;  and/or {\it (ii)} a new parity symmetry which disallows mixing between the ordinary fermions and the new fermion fields, under which all the new vectorlike fields are odd, while the others are even \cite{Ishiwata:2013gma},  such that  $\lambda_{L}^{\prime  \,i}=\lambda_{R}^{\prime  \, i}=\lambda_{L}^{\prime \prime \,i}=\lambda_{R}^{\prime \prime \, i}=0$. The latter symmetry is important for light vectorlike leptons, as this scenario would satisfy restrictions from lepton-flavor violating decays, which otherwise would either force the new leptons to be very heavy, $\sim 10- 100$ TeV, or else reduce the branching ratio for the Higgs into di-leptons  to 30-40\% of the SM prediction \cite{Ishiwata:2013gma, Dermisek:2013gta}. In addition, if all vectorlike fermions are odd under the new parity symmetry, the lightest particle can become stable and act as all, or part of, the dark matter in the Universe. Thus, in what follows we will perform the analysis under the simplifying  assumption {\it (ii)}.

\subsection{Constraints on model parameters}
\label{subsec:constraints}
Current bounds on  additional gauge bosons are derived from their hadronic and leptonic decay channels, and constraints are obtained from both ATLAS and CMS searches ~\cite{ATLAS:2015nsi,Sirunyan:2016iap}. These are quite restrictive, with the $W_R$ and $Z_R$ masses being constrained to lie above, or about, $2.7$~TeV. In our numerical investigations we choose $v_R=10$ TeV, and thus the $W_R$ and $Z_R$ masses remain high, with $M_{W_R}=4.2$ TeV. Furthermore, we assume $v_L < 5 ~\rm GeV$, which agrees with the limits from the electroweak precision constraints, see Refs.~\cite{Kanemura:2012rs,BLANK1998113} for the Higgs Triplet Model, and Ref.~\cite{doi:10.1142/S0217751X0603388X} for the LRSM.

The left-handed doubly charged Higgs bosons $H_L^{\pm \pm}$ 
can be light in general LRSM scenarios. 
The masses of these doubly charged scalars are strongly restricted by the LHC searches for same-sign dilepton (electron or muon) signatures while the bound is less stringent for di-tau final states~\cite{Chatrchyan:2012ya,CMS:2016cpz,CMS:2017pet} .
However, the experiments tend to assume 100\% branching ratios for each of the leptonic final states. 
Thus, the constraints can be softened by assuming small couplings to leptons. 
In addition, allowing them to decay into vectorlike leptons would also modify the mass limits. 
In our case, $H_L^{\pm\pm}$ decay mostly into $W_L^\pm W_L^\pm$ pairs, 
with a branching ratio of 85\%.  We take $M_{H_L^{\pm\pm}}=300$ GeV,
which obeys all the experimental bounds. Our analysis does not particularly focus on the
the scalar sector of the model and the masses can be independently taken at high value
without affecting our searches for the vectorlike leptons, as is also explained in the following sections.

\subsection{vectorlike leptons}
\label{subsec:VL}

The spectrum from  Eq.(\ref{eq:VLleptons}) now consists of, in the charged sector, a $(2 \times 2)$-dimensional mass matrix ${\cal M}_c$.  Note that here, as in the case of the neutral vectorlike leptons studied below, the matter parity symmetry introduced in this section forbids mixing with the ordinary fermions, and thus the mass matrix in the charged sector is just $2 \times 2$, while in the neutral sector it is $4 \times 4$ .
 \begin{eqnarray}
\left ( {\bar e}_L^\prime~~{\bar e}_L^{\prime \prime}\right ) \left ({\cal M}_c \right ) \left( \begin{array}{c} e_R^\prime\\e_R^{\prime \prime} \end{array} \right)\,, \quad {\rm with} \quad
{\cal M}_c=\left ( \begin{array}{cc}
m_E^\prime&M_L\\
M_R&m_E^{\prime\prime}\end{array} \right),  
\end{eqnarray}
with $\displaystyle m_E^\prime=\frac{Y_L^{\prime \, e} \kappa_2 e^{i \alpha}+{\tilde Y}^{\prime\, e}_L\kappa_1}{\sqrt{2}}$ and $\displaystyle m_E^{\prime \prime}=\frac{Y_R^{\prime  \, e} \kappa_2 e^{i \alpha}+ {\tilde Y}^{\prime \, e}_R\kappa_1}{\sqrt{2}} $, from the Lagrangian Eq. (\ref{vl_lgr}). The mass matrix can be diagonalized by two unitary matrices $U^L$ and $U^R$ as follows:
\begin{equation}
{U^L}^\dagger {\cal M}_c U^R=\left ( \begin{array}{cc}
M_{E_1}&0\\
0&M_{E_2} \end{array} \right).
\end{equation}
The mass eigenvalues are (by convention the order is $M_{E_1}>M_{E_2}$ {\bf \cite{Joglekar:2012vc, Fairbairn:2013xaa}})
\begin{equation}
M^2_{E_1, E_2}= \frac12 \left [ \left (M_L^2+m_E^{\prime\,2}+M_R^2+m_E^{\prime \prime\,2}\right ) \pm \sqrt{\left (M_L^2+m_E^{\prime\,2}-M_R^2-m_E^{\prime \prime\,2}\right )^2+4(m_E^{\prime \prime}M_L+m_E^{\prime}M_R)^2} \right ], 
\label{eq:eigenvalues}
\end{equation}
In the neutral sector the mass matrix is:
 \begin{eqnarray}
&& \frac12 \left (\bar {\nu_L^\prime}~~\bar {\nu_R^{\prime \,c}}~~\bar {\nu_R^{\prime \prime\,c}}~~\bar {\nu_L^{\prime \prime}}\right ) \left ({\cal M}_\nu \right ) \left( \begin{array}{c}  \nu_L^{\prime \,c} \\ \nu_R^{\prime }\\ \nu_R^{\prime \prime} \\ \nu_L^{\prime \prime \,c} \end{array}  \right )\,\nonumber \\
&& {\rm with}~
{\cal M}_\nu=\left ( \begin{array}{cccc}
\sqrt{2}h_L^\prime v_Le^{i \theta}&m_\nu^\prime& M_L &0\\
m_\nu^\prime&M_\nu^{\prime}&0&M_R \\
M_L&0&\sqrt{2}h_L^{\prime \prime} v_Le^{i \theta}&m_\nu^{\prime \prime} \\
0&M_R&m_\nu^{\prime \prime}&M_\nu^{\prime \prime}
\end{array} \right) , 
\label{eq:neutrinomasses}  
\end{eqnarray}
with Dirac masses $\displaystyle m_\nu^\prime=\frac{ Y_L^{\prime \, \nu} \kappa_1+{\tilde Y}^{\prime \, \nu}_L\kappa_2 e^{-i \alpha}}{\sqrt{2}}$,  $\displaystyle m_\nu^{\prime \prime}=\frac{Y_R^{\prime \, \nu} \kappa_1+{\tilde Y}^{\prime \, \nu}_R\kappa_2 e^{-i \alpha}}{\sqrt{2}}$ and with Majorana masses $M_\nu^{\prime }=h_R^\prime v_R/\sqrt{2} $ and $M_\nu^{\prime \prime \,}= h_R^{\prime \prime}v_R/\sqrt{2} $.
This mass matrix can be diagonalized  by a unitary matrix $V$:
\begin{equation}
{V}^\dagger {\cal M}_\nu V=\left ( \begin{array}{cccc}
M_{\nu_1}&0&0&0\\
0&M_{\nu_2}&0&0 \\
0&0&M_{\nu_3}&0\\
0&0&0&M_{\nu_4}   \end{array} \right). 
\end{equation}
Exact analytic expressions are difficult to find\footnote{ In our analyses, we use exact numerical expressions, and show here approximate analytical expressions for clarity.}. To simplify, we work in the limit where $h^{\prime \prime}_R= Y_R^{\prime  \, \nu}= {\tilde Y}_R^{\prime  \, \nu}=0$ (meaning $M_\nu^{\prime \prime}=m_\nu^{\prime \prime}=0$). In the limit where $m_\nu^\prime \ll M_L,\, M_R$, the neutrino mass matrix can be diagonalized yielding four neutrino masses:
\begin{eqnarray}
M_{\nu_{1,2}}&=& \frac{M_\nu^{\prime}}{2} \mp  \sqrt{\frac{M_\nu^{\prime \,2}}{4}+M_R^2+m_\nu^{\prime\,2}} \, ,\\
M_{\nu_{3,4}}&=& \pm \, M_L \, ,
\label{eq:vneutrino-eigenvalues}
\end{eqnarray}
valid to ${\cal O}(m_\nu^{\prime\, 2}/M_L)$. The lightest of these states will be the dark matter candidate, which, as it is odd under the additional parity symmetry {\it (ii)}, is stable. 
The lightest  state will depend on assumptions made on the masses $M_L, \,M_R$ and triplet Yukawa couplings $h^\prime_R$ and $Y^{\prime}_L$. We also must choose the parameters carefully in the charged sector, to insure that the charged vectorlike leptons are heavier than the neutral ones. Taking into account the constraints $M_{E_2} \ge 101.9$ GeV~\cite{Olive:2016xmw} 
by LEP, this requirement is not difficult to satisfy. 
Moreover, we analyzed the constraints on the parameter space and found that for a fixed $Y^{\prime \nu}_L (= 1.5)$ (our benchmark BP1 value from \ref{subsec:bench}),
there is a narrow region of very small $h^\prime_R~ (0<h^\prime_R< 0.04)$ for the range of 
$M_R \in (150~{\rm GeV}-1~{\rm TeV})$, for which the charged VL lepton is lighter than the neutral one.  
We exclude these points\footnote{In practice, the software we use gives a warning at points where the DM candidate is charged.}.  In the limit in which  $M_L \gg M_R$, there are two heavy (approximately) degenerate eigenvalues of mass $M_L$, which do not mix with the lighter states $\nu_1$ and $\nu_2$. In this limit, the lightest state is $\nu_1$, and it is mostly the $\nu_L^{\prime \prime}$ state, with some admixture   of $\nu_R^{\prime \,c}$, and is right-handed.  

Some comments about this analysis:
\begin{itemize}
\item If we require $M_\nu^{\prime}=m_\nu^{\prime}=0$ instead of $M_\nu^{\prime \prime}=m_\nu^{\prime \prime}=0$, there is no difference in the final result, but we have $m_\nu^{\prime \prime}$ in the mass expressions replacing $m_\nu^\prime$.
\item If we set $M_\nu^{\prime }=m_\nu^{\prime}=0$ we flip between the $^\prime$ and the $^{\prime \prime}$ states and the lightest vectorlike neutrino will be $\nu_L^\prime$. 
\item 
In our scenario, the  states with mass $M_L$ are heavier and decay into lighter vectorlike leptons, making them unsuitable to be DM candidates. 
\item  It is advantageous that we get a right-handed neutrino to be the DM candidate, as it is more likely to produce a relic density in the desired range. The left-handed vectorlike neutrino, much like the ordinary one, annihilates too fast through the $s$-channel mediated by the 
$Z_L$ boson,  resulting in a large annihilation cross section. On the contrary, 
the right-handed candidate can annihilate through $Z_R$ which 
is quite heavy  (TeV),  and through the Higgs scalars (doublet and triplet).  In our case, the DM neutrino candidate is mostly right-handed, but contains a small mixture of left-handed components, and thus it has a small (but nonzero) coupling to the $Z_L$ boson.
Because this mixing is small,  the Higgs Yukawa coupling controls the relic density as well as
the direct detection cross section for light right-handed vectorlike neutrinos. Thus, the relic density for such 
right-handed neutrinos can easily be tuned to within the right ball park. 
In addition, the direct detection cross section will stay in the
experimentally allowed region which is otherwise violated by the left-handed neutrinos due to
their large coupling with $Z_L$.  
In our scenario, the dominant annihilation processes for right-handed vectorlike neutrinos through
the Higgs mediation yield the correct relic density (within the $2\sigma$
limit of Planck results) and do not violate the experimental bounds on the direct detection cross section 
in the specified regions of the parameter space, making this neutrino a good DM candidate. We return to this in more detail in Secs. ~\ref{subsubsec:rd} and in ~\ref{subsubsec:annihilation}.
\end{itemize}

We now explore the parametric dependence of heavy vectorlike leptons in this model.  
While varying some of the parameters, we fix other parameters
mostly to values that we choose as benchmark points in our following studies.
We shall discuss the benchmark points and corresponding parameters 
 in a later section. 
 
\begin{figure}[htbp!]
\centering
\begin{tabular}{cc}
    \includegraphics[width=3in,height=2.2in]{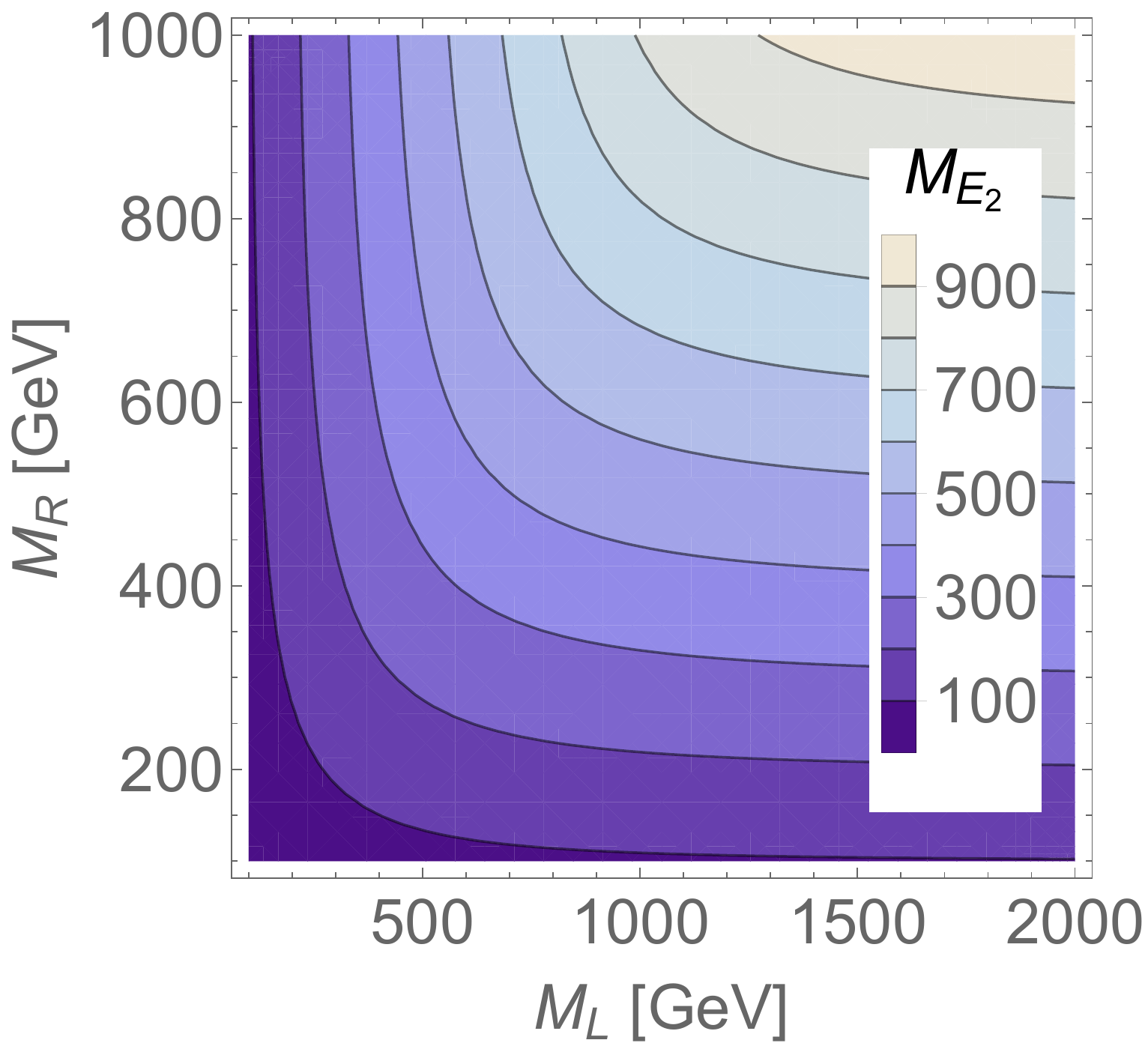} &
    \includegraphics[width=3in,height=2.2in]{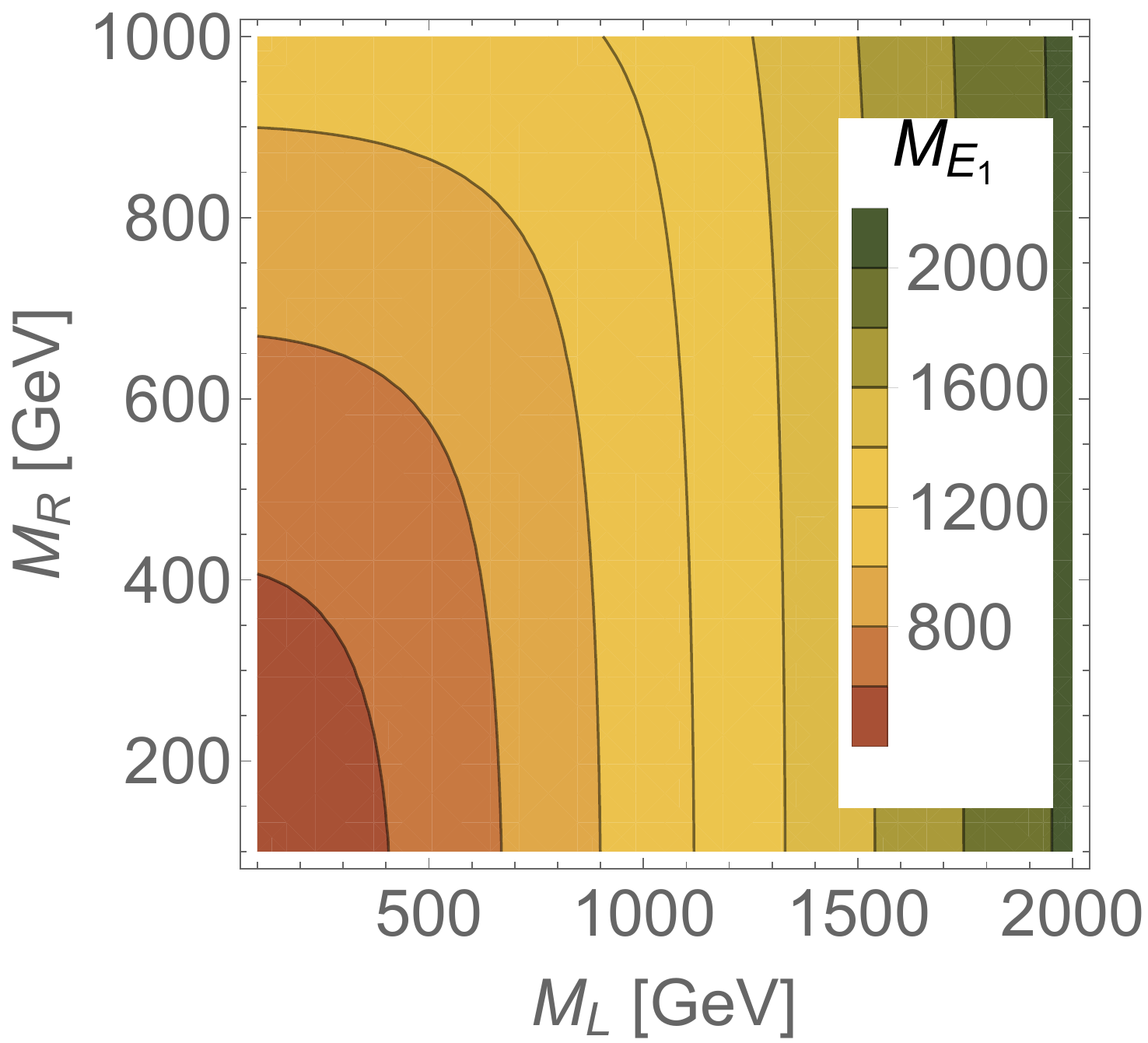} \\
    \includegraphics[width=3in,height=2.2in]{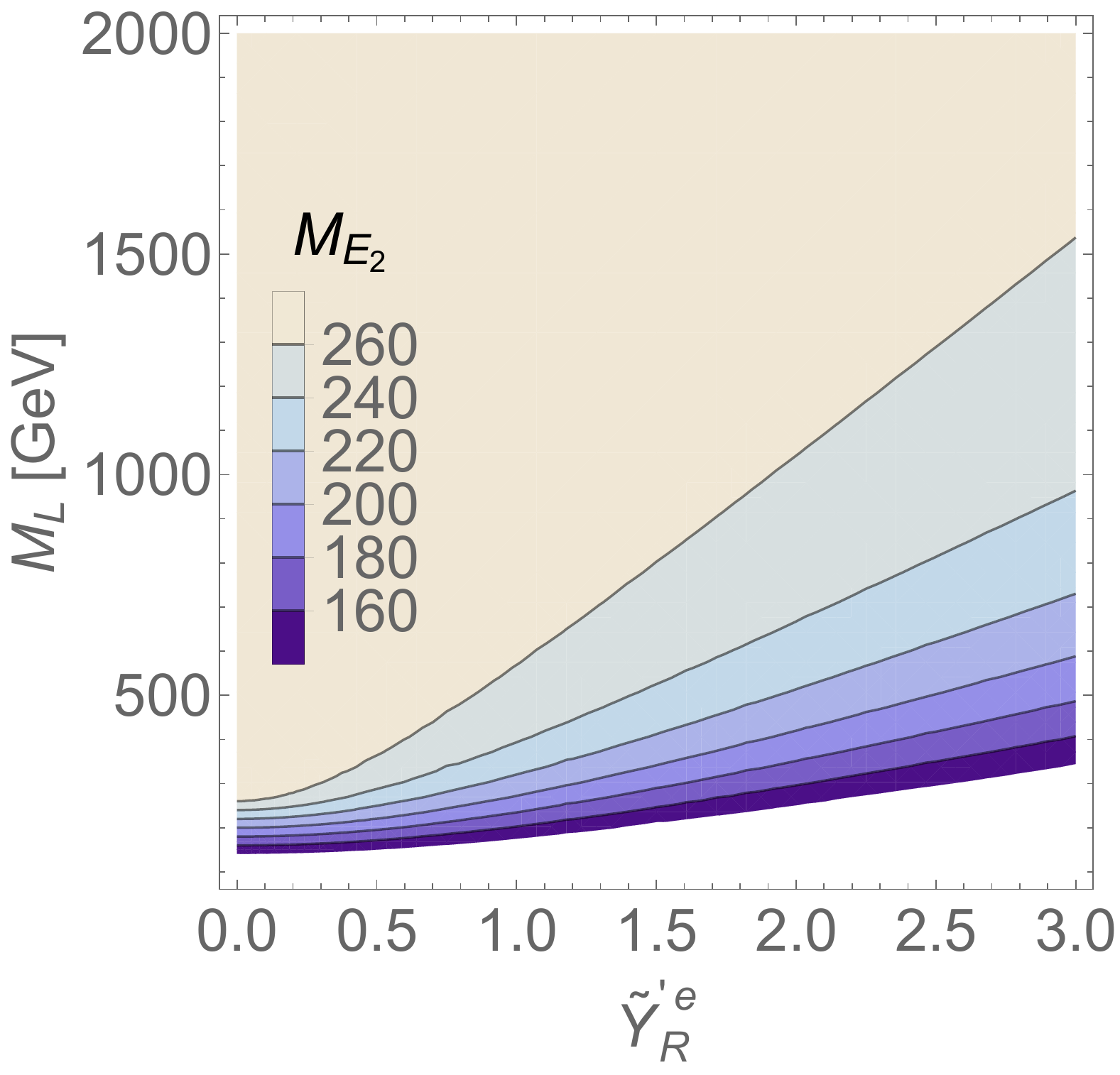} &
        \includegraphics[width=3in,height=2.2in]{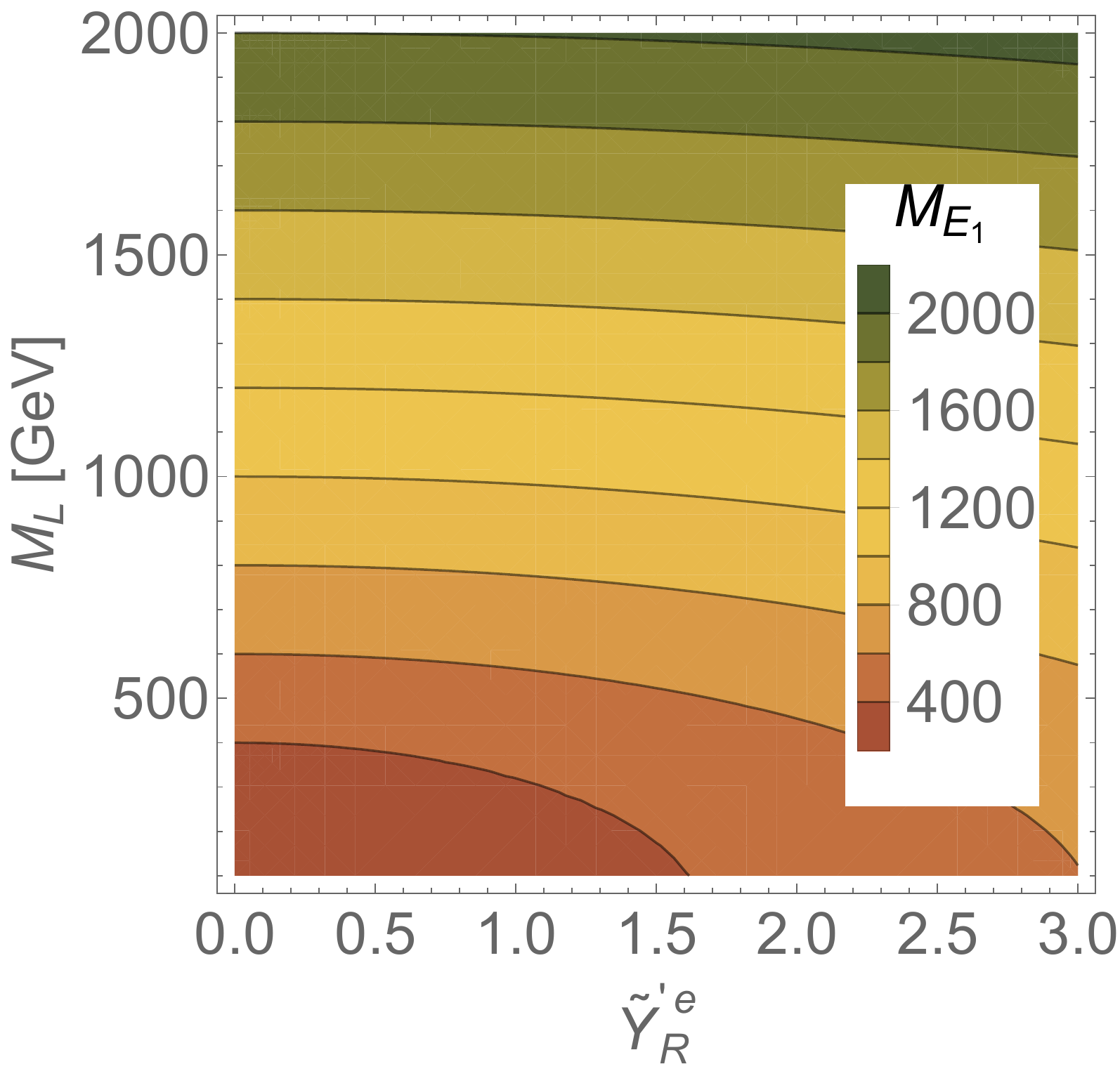} \\
    \includegraphics[width=3in,height=2.2in]{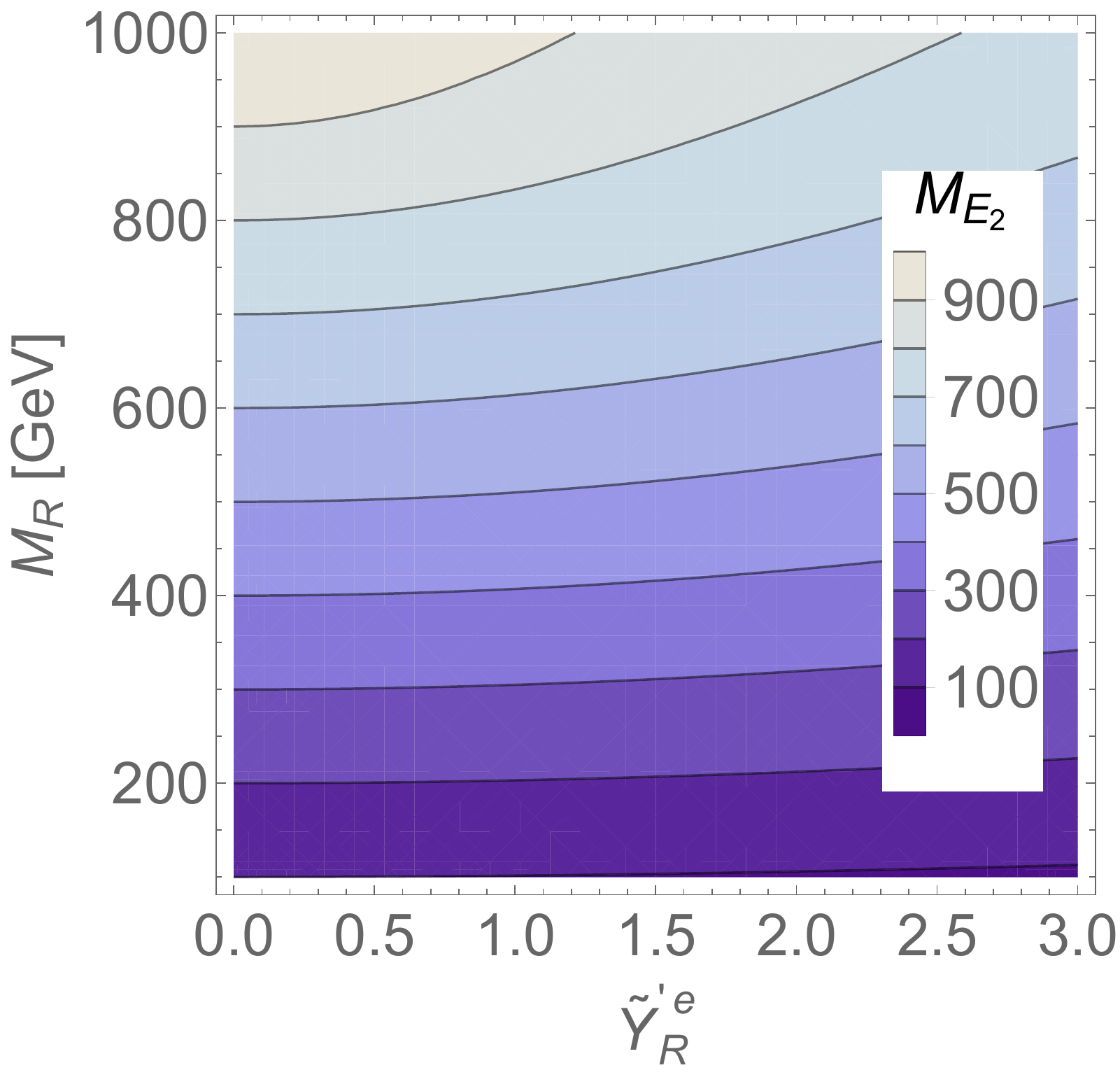} &
        \includegraphics[width=3in,height=2.2in]{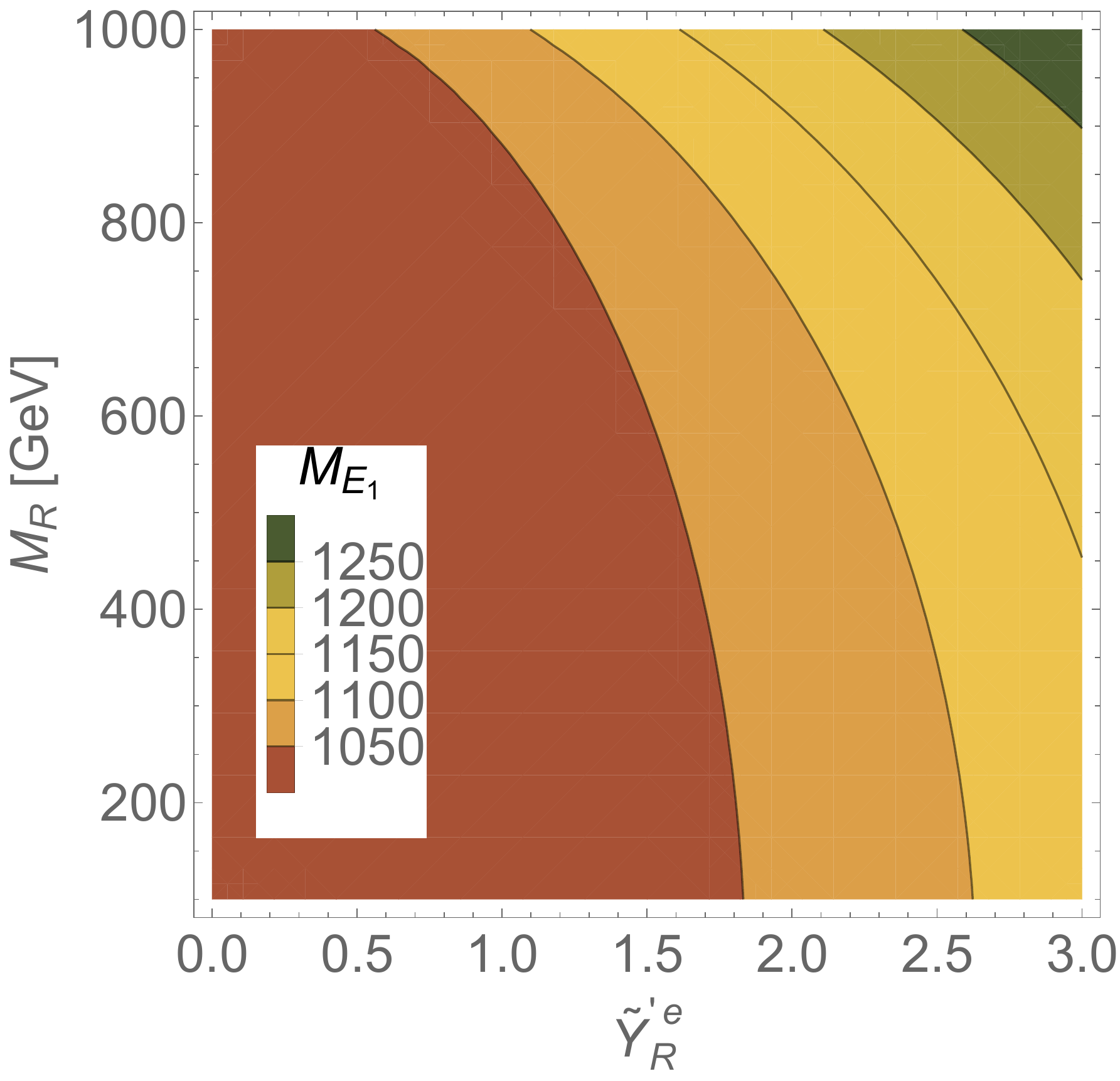} 
        \end{tabular}
     \caption{ \it  
  (Left panel) Contour plots showing the dependence of the  lightest vectorlike charged lepton masses ($M_{E_2}$) on
$M_L, M_R$ for $\tilde{Y}_R^{\prime\,e}=2.5$ (top); on $M_L, \tilde{Y}^{\prime\, e}_R$ for $M_R=275$ GeV (middle); and on $M_R, \tilde{Y}^{\prime\, e}_R$ for $M_L=1000$ GeV (bottom). (Right panel) Same contour graphs showing the dependence of the heavier vectorlike charged lepton mass ($M_{E_1}$). We take Yukawa couplings $Y^{\prime\,e}_L=0.1, Y^{\prime\,e}_R=2.5$. The panels on the right indicate the color-coded mass values for the contours.}
\label{fig:masses_E2}
\end{figure}

\begin{figure}[htbp!]
\centering
\begin{tabular}{cc}
    \includegraphics[width=3in,height=2.2in]{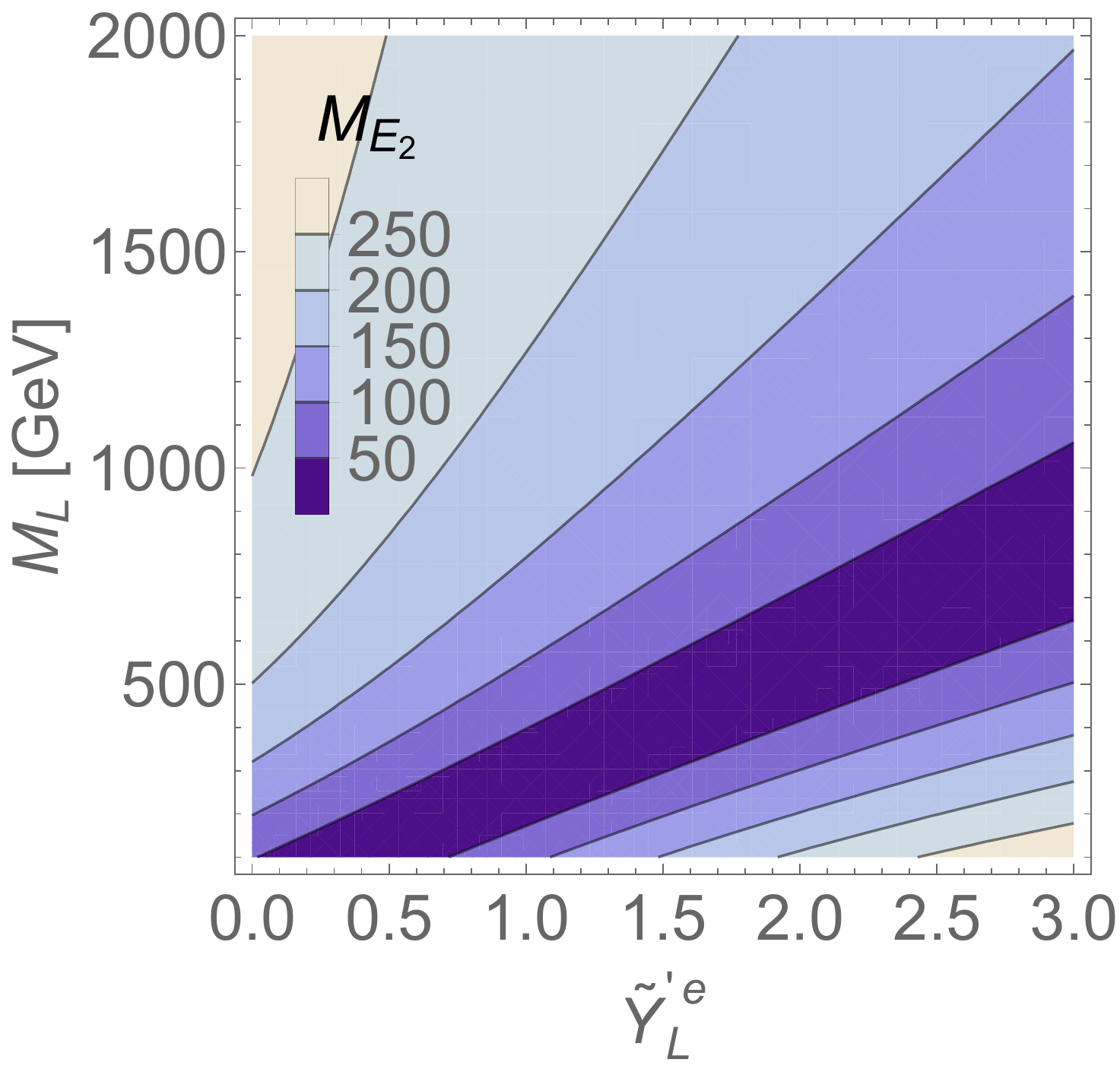} &
        \includegraphics[width=3in,height=2.2in]{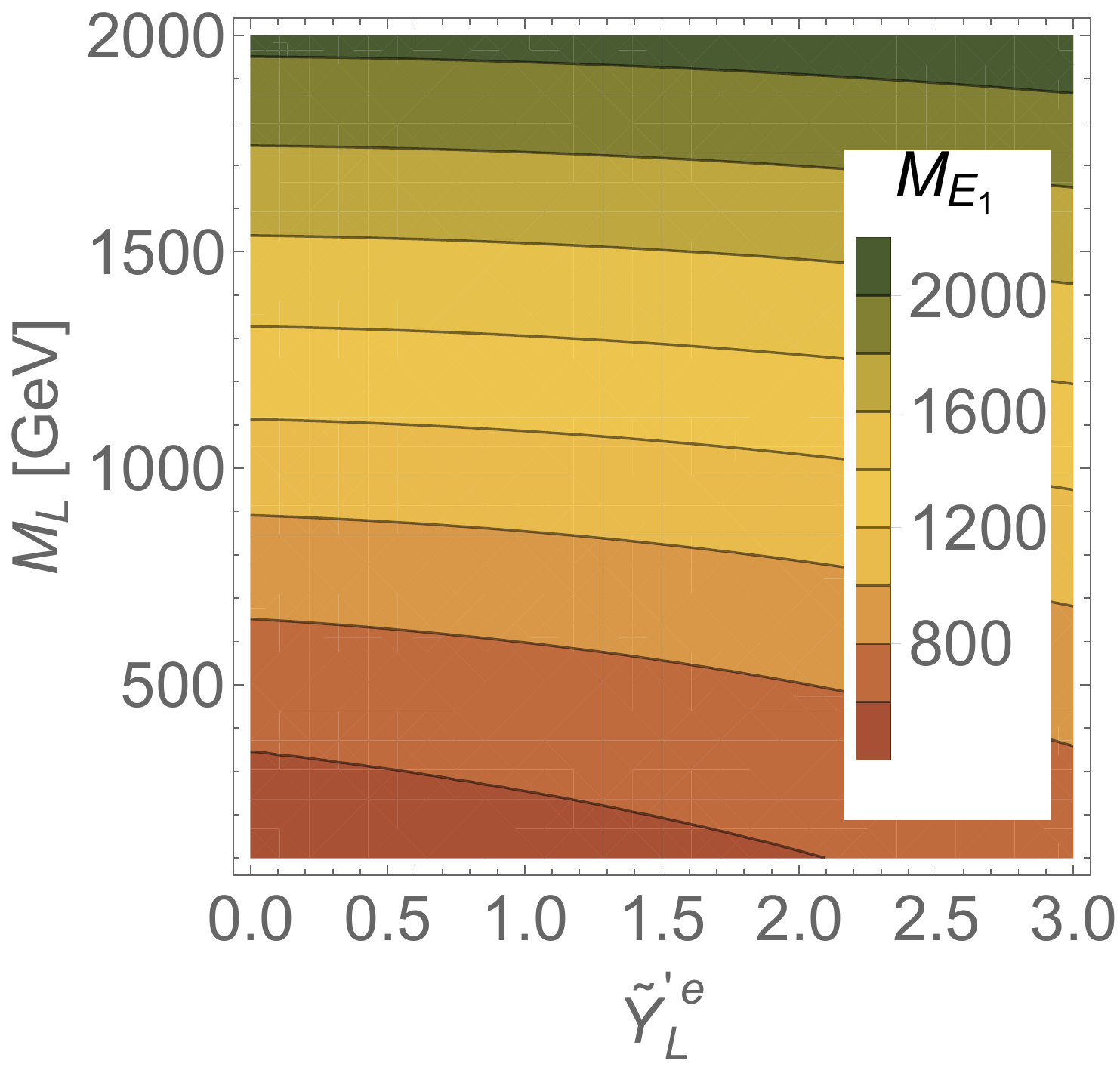} \\
    \includegraphics[width=3in,height=2.2in]{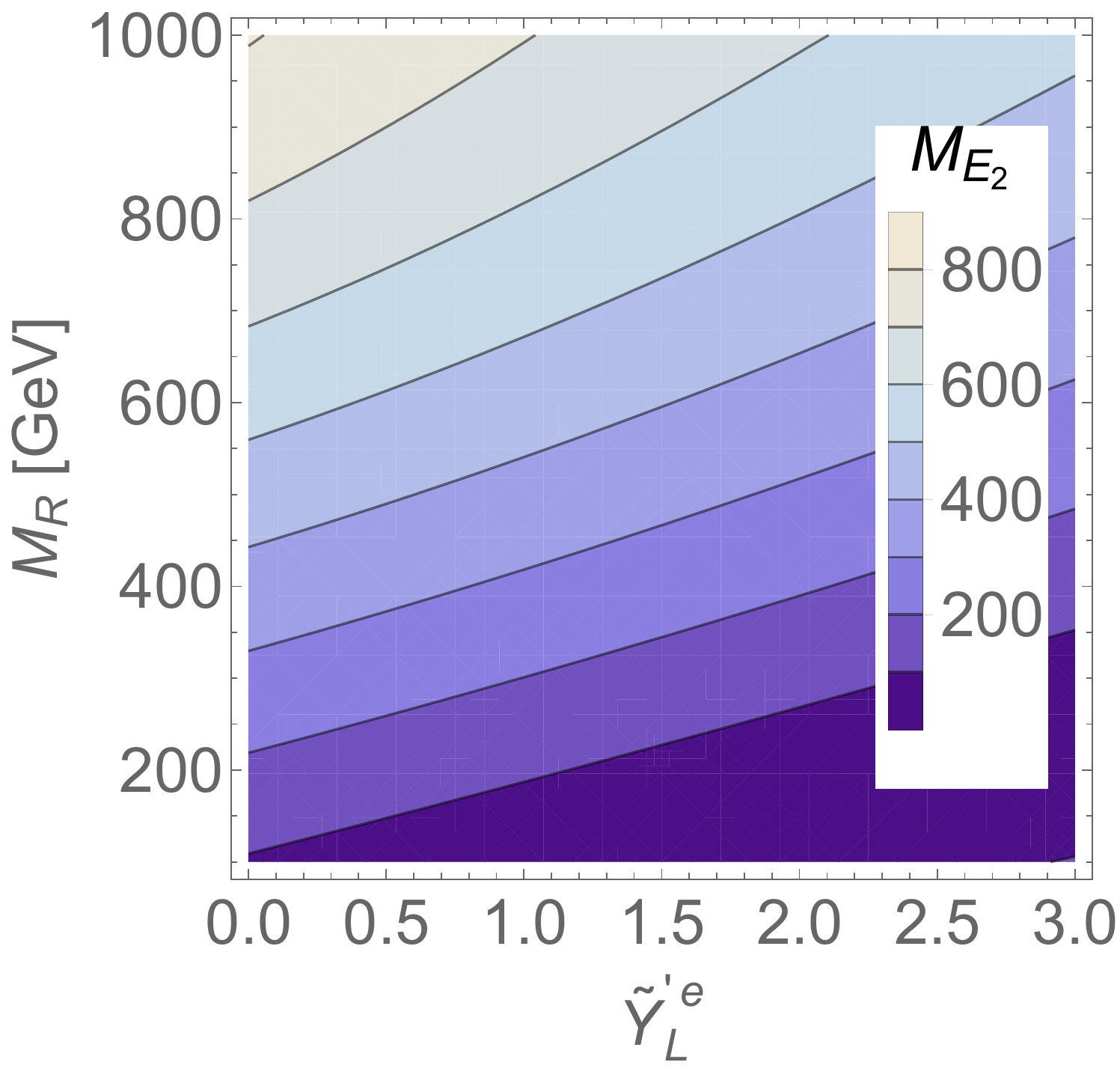} &
        \includegraphics[width=3in,height=2.2in]{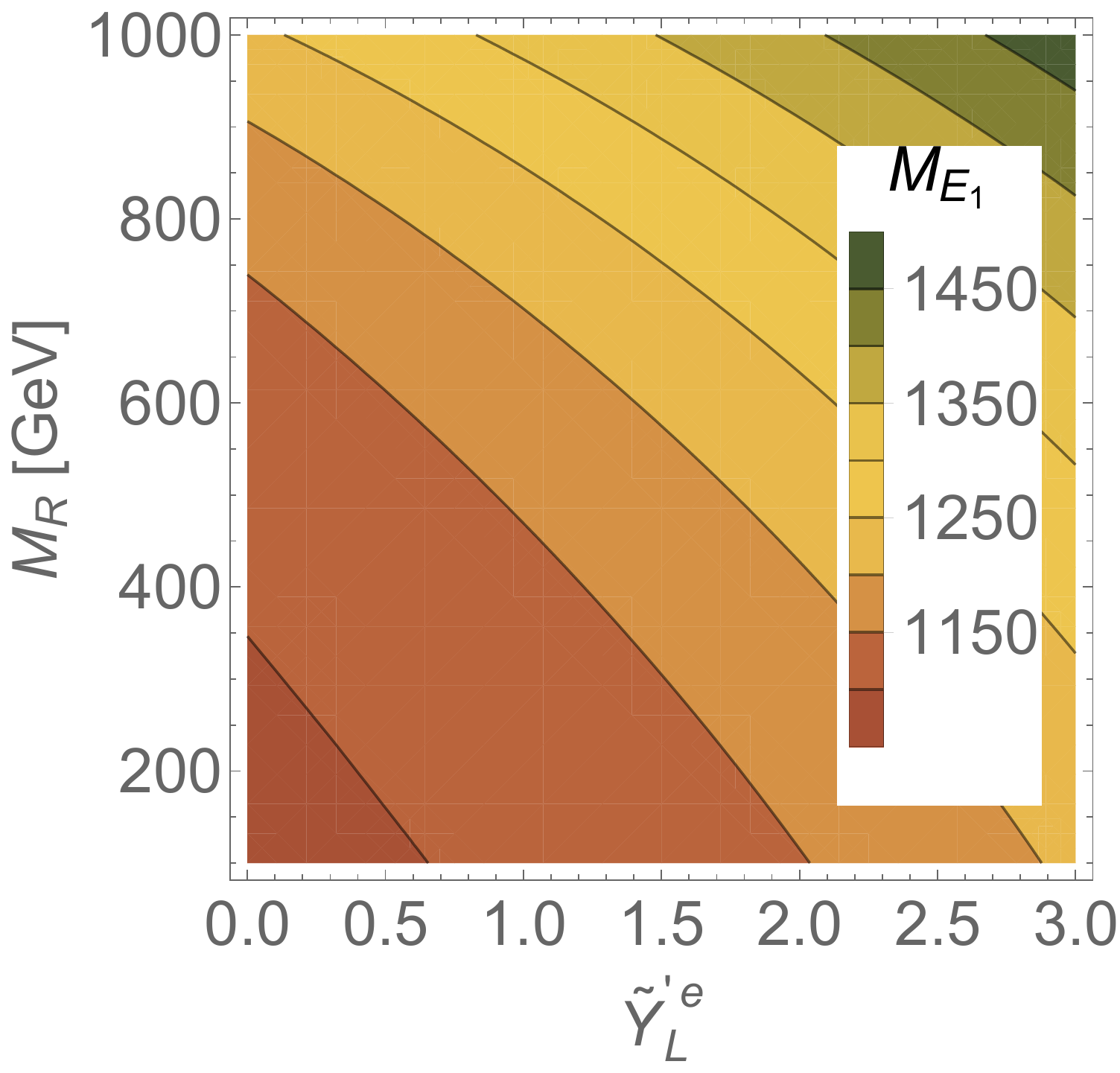} 
        \end{tabular}
     \caption{ \it  
  Left panel: Contour plots showing the dependence of the  lightest vectorlike charged lepton masses ($M_{E_2}$) on
$M_L, \tilde{Y}_L^{\prime\,e}$ for $M_R=275$ GeV (top) and on $M_R, \tilde{Y}_L^{\prime\,e}$ for $M_L=1$ TeV (bottom).  Right panel: Same contour graphs, but for the heavier vectorlike  charged lepton mass ($M_{E_1}$). As before, $Y^{\prime\,e}_L=0.1, Y^{\prime\,e}_R=2.5$ and the panels on the right indicate the color-coded mass values for the contours.}
\label{fig:masses_E1}
\end{figure}

In  Figs. \ref{fig:masses_E2} and \ref{fig:masses_E1} we show the mass dependence of the charged vectorlike leptons (both $E_2$ and $E_1$)  on the various parameters of the model. Specifically, we show the contour plots of $M_{E_2}$ and $M_{E_1}$ in the
$(M_L, M_R)$ plane and in the planes correlating $M_L$ and $M_R$ individually to the Yukawa 
couplings $\tilde{Y}_L^{\prime \, e}$ and $\tilde{Y}_R^{\prime \, e}$. We observe that for the 
same set of parameter ranges, the lighter charged VL lepton mass ($M_{E_2}$) can at most reach 1 TeV
while the mass of the heavier state ($M_{E_1}$) cannot be lower than 1~TeV. In this regard, we should mention that
while scanning over the mass ranges, we impose the direct search limit on $M_{E_2}>101.9$ GeV
given by LEP~\cite{Olive:2016xmw}.
We also note that, as the plots indicate, the lightest vectorlike lepton $E_2$ is mostly right-handed, while the heavier one $E_1$ is mostly left-handed. We note that $M_{E_2}$ is the most sensitive to the parameters $\tilde{Y}^{\prime\,e}_R$ and $M_R$, while $M_{E_1}$ is the most sensitive to the parameters $\tilde{Y}^{\prime\,e}_L$ and $M_L$. 

\begin{figure}[]
\centering
\begin{tabular}{cc}
     \includegraphics[width=2.5in,height=2.2in]{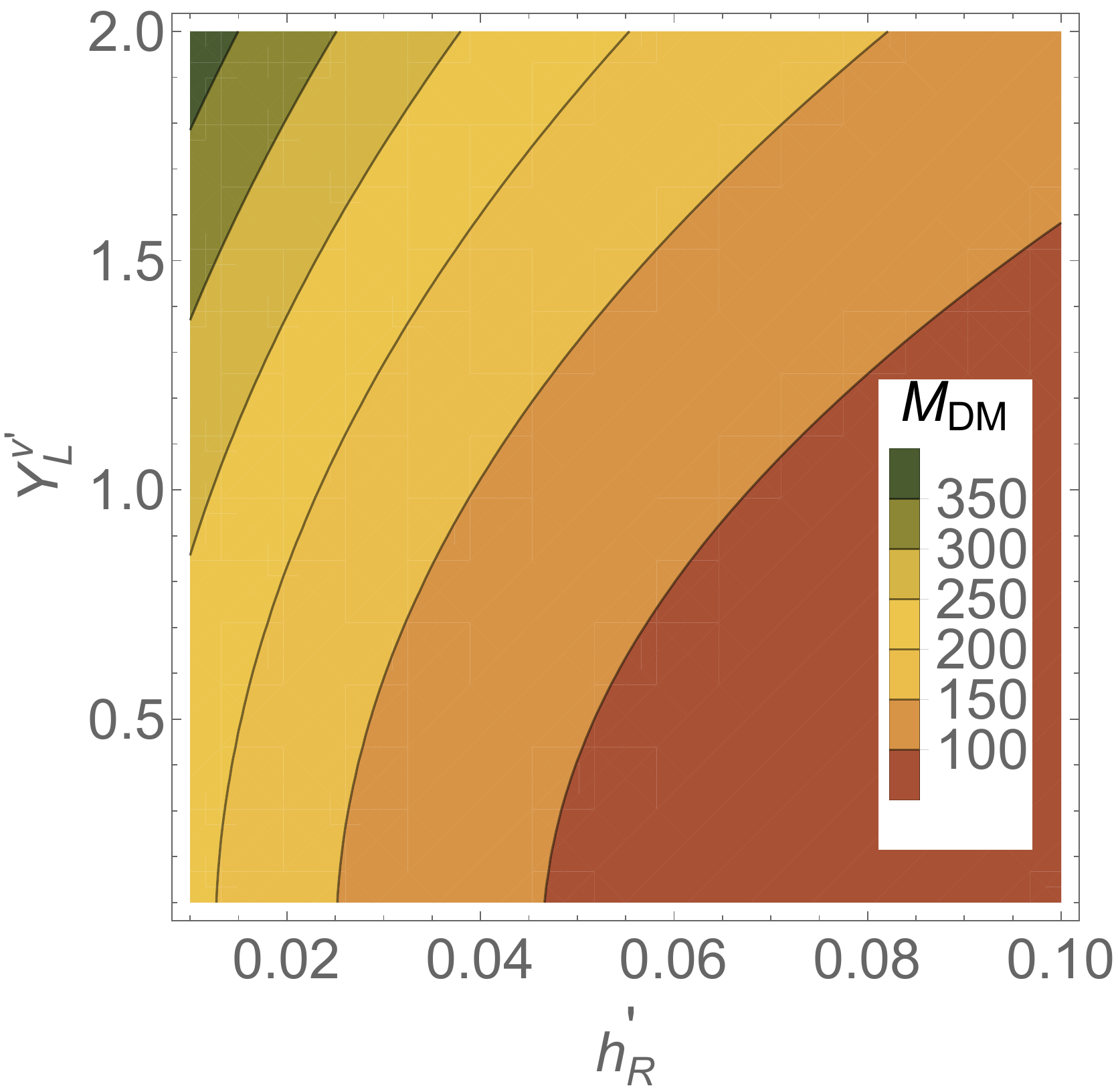}
&  \includegraphics[width=2.5in,height=2.2in]{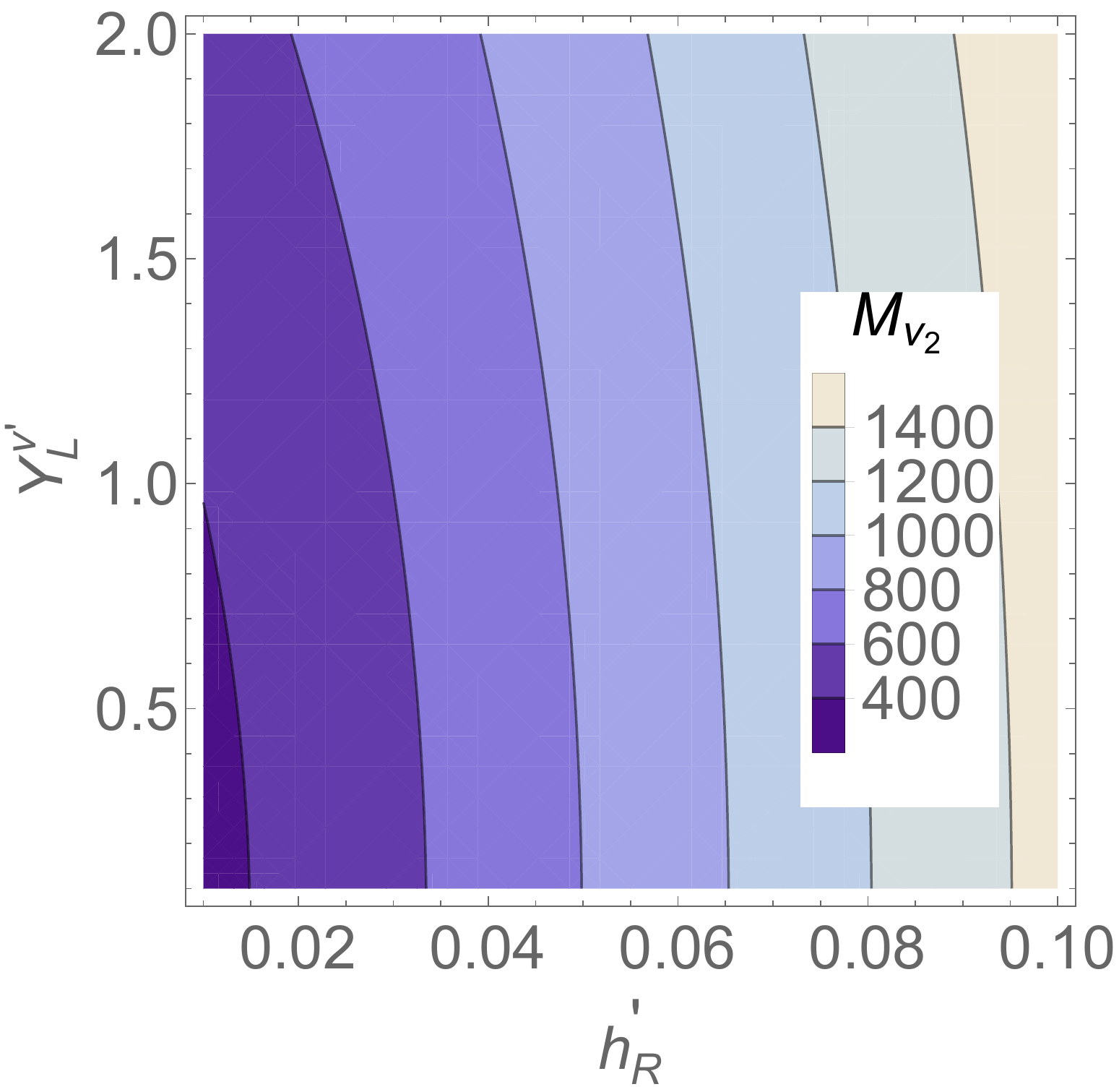} \\ 
\includegraphics[width=2.5in,height=2.2in]{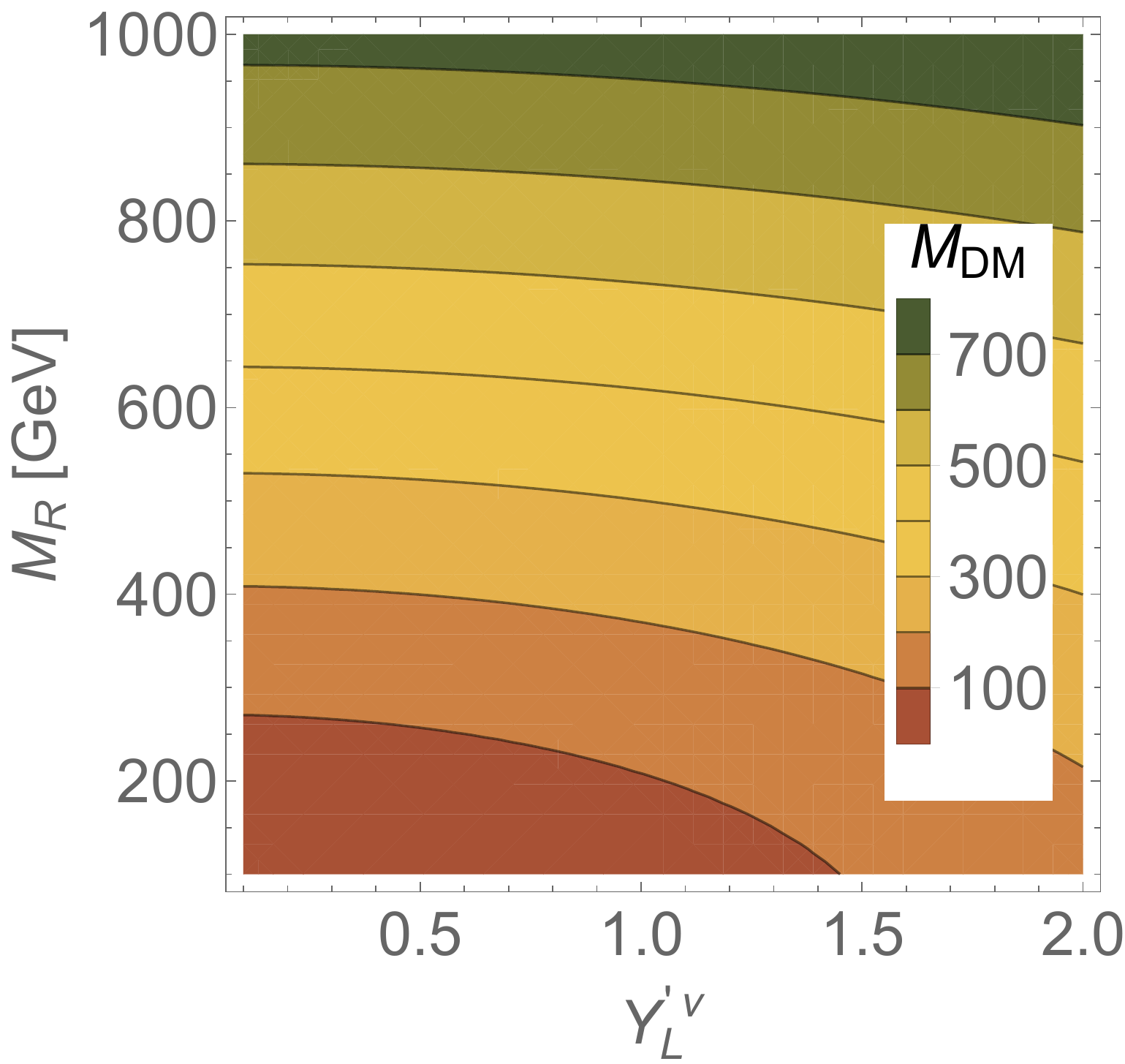} &
 \includegraphics[width=2.5in,height=2.2in]{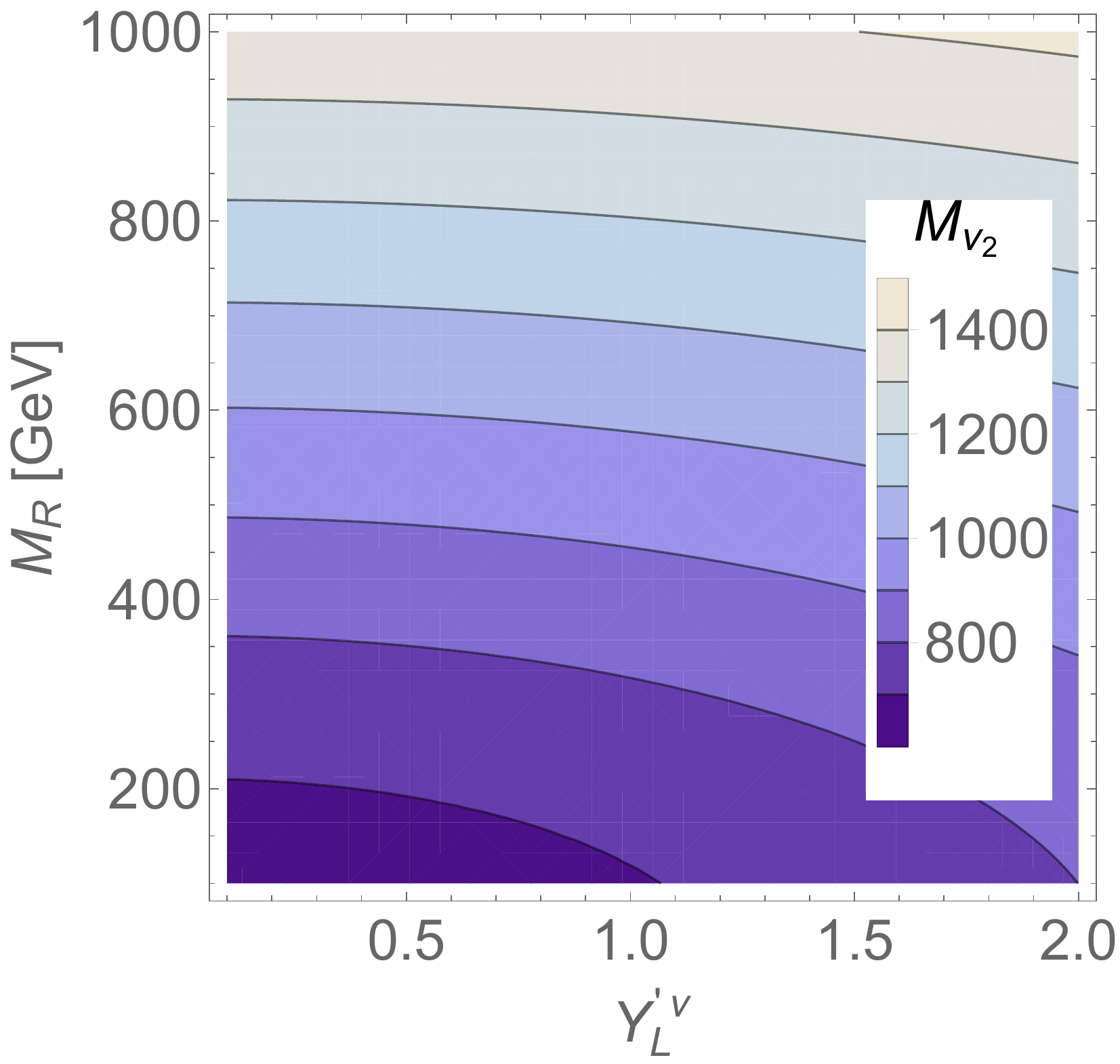} \\
\includegraphics[width=2.5in,height=2.2in]{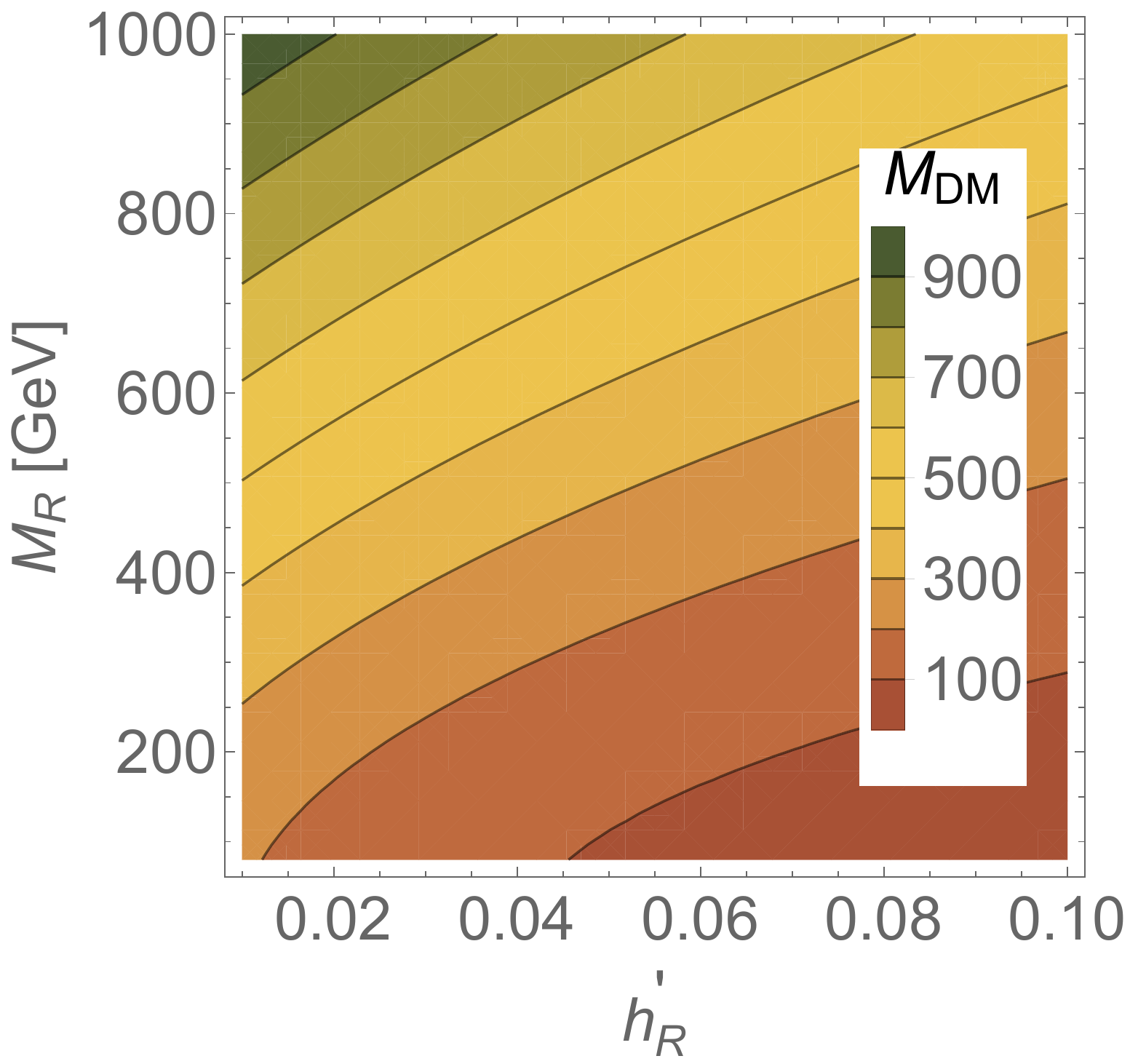} &
\includegraphics[width=2.5in,height=2.2in]{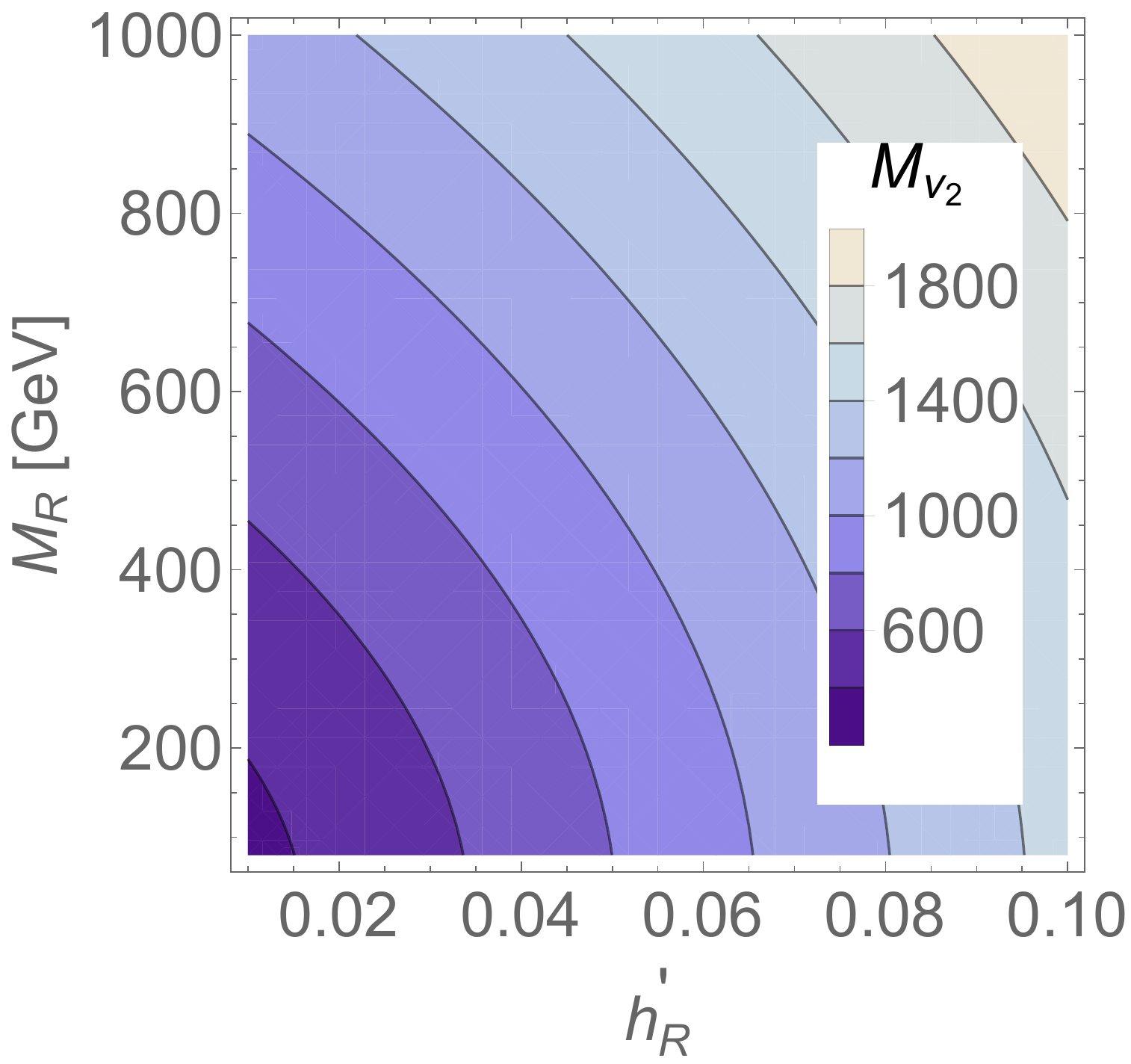}                                                                        
\end{tabular}
     \caption{\it  Contour graphs showing the dependence of the vectorlike neutrino mass, for the lightest state $\nu_1$ (left panel) and  for the heavier state $\nu_2$ (right panel) as a function of $Y^{\prime\,\nu}_L$ and $h^\prime_R$, 
with $M_R$=275 GeV (top),  as a function of $M_R$ and  $Y^{\prime\,\nu}_L$ for $h^\prime_R=0.045$ (middle), 
and as a function of $M_R$ and $h^\prime_R$ for $Y^{\prime\,\nu}_L=1.5$ (bottom). 
The panels on the right indicate the color-coded mass values for the contours.}
\label{fig:masses_Mnu1_Mnu2}
\end{figure}

In Fig. \ref{fig:masses_Mnu1_Mnu2}, we plot the dependences on the parameter space for the vectorlike neutrino masses
($M_{\nu_1}$ and $M_{\nu_2}$). Here $\nu_1$ is the lightest neutrino state and the DM candidate (the brown-colored graphs on the left) and the second vectorlike lightest neutrino is $\nu_2$ (the blue-colored contours on the right). 
The variation is shown only with the relevant parameters, namely, the bare mass terms $M_L$ and $M_R$ 
which tune the masses of these neutral VL lepton candidates and with the Yukawa couplings
$Y_L^{\prime \, \nu}$ and $h_R^\prime$ which control the DM annihilation cross section
and set its relic density. In consequence to the parameter scanning, we now proceed to our
analysis on DM sector and collider signatures.  

\section{Dark Matter}
\label{sec:dm}

For the vectorlike neutrino to be a viable candidate for dark matter, it must satisfy conditions of providing the right level of relic abundance from thermal dark matter production in the early Universe. In addition, as the lack of any dark matter signals in either direct or indirect dark matter detection experiments confront our theoretical expectations,  these must satisfy increasingly severe constraints from experiments. For the dark matter analysis,  we extend the left-right model in Ref.~\cite{Roitgrund:2014zka} to include vectorlike leptons using {\tt FeynRules} \cite{Alloul:2013bka} and  extract the resulting file in {\tt CalcHEP} \cite{Belyaev:2012qa} to implement the model into  {\tt micrOMEGAs} \cite{Belanger:2014vza}. We use {\tt micrOMEGAs} to calculate the relic density ($\Omega_{DM} h^2$), the spin-independent cross section ($\sigma^{SI}$), the annihilation cross section ($\langle \sigma v \rangle$), and the neutrino and muon fluxes, which are the  most constrained observables for our model.  We analyze these in turn below.

\subsubsection{Relic Density}
\label{subsubsec:rd}

First, we analyze the consequences of having the lightest vectorlike neutrino as our dark matter candidate. Using the results in the previous sections, we explore the parameter space of the model which yields the correct relic density of dark matter, determined very precisely as the amount of nonbaryonic dark matter in the energy-matter of the Universe to be $\Omega_{DM}h^2=0.1199\pm 0.0027$ 
\cite{Ade:2013zuv}, with $\Omega_{DM}$ being the energy density of the dark matter with respect to the critical energy density of the universe, and $h$ being the reduced Hubble parameter. 

\begin{figure}[htbp]
\begin{center}
    \includegraphics[width=3.5in,height=2.5in]{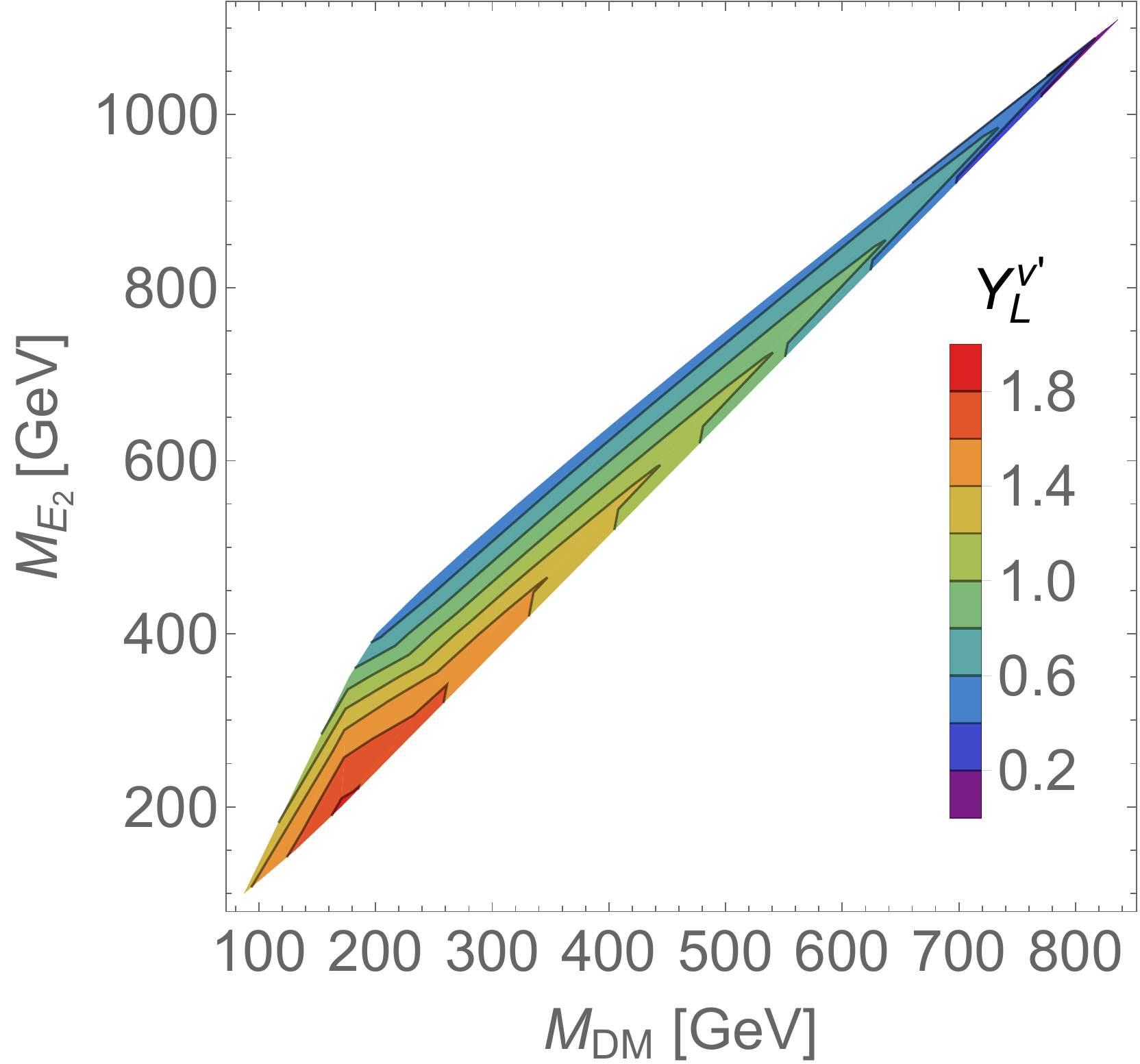}
\end{center}
     \caption{\it  Contour plots for the allowed relic density as a function of the vectorlike lepton mass $M_{E_2}$ and the vectorlike neutrino (the dark matter candidate) mass (in GeV) for  $M_L=1$ TeV. We impose the restriction $0.1144 \leq \Omega_{DM} h^2 \leq 0.1252$. The color code for the Yukawa coupling $Y_L^{\nu \prime}$ is indicated on the right.  All the other parameters are fixed at our BP1 values in Table \ref{table:bp}.}
\label{fig:relic_ML=1}
\end{figure}

 In Fig.~\ref{fig:relic_ML=1}, we show the $2\sigma$ allowed range of relic density: $0.1144 \leq \Omega_{DM} h^2 \leq 0.1252$, as constrained by WMAP \cite{Komatsu:2010fb} and Planck \cite{Ade:2013zuv}, in the
$(M_{\rm DM}-M_{E_2^\pm})$ plane by varying $Y_L^{\nu \prime}$ and fixing $h^\prime_R=0.045$. It should be mentioned
here that $M_{E_2^\pm}$ is directly related to $M_R$ while $M_{\rm DM}$ depends on $M_R\,, h_R^\prime\,, {\rm and}\, Y_L^{\nu \prime}$.
Consequently, the DM relic density depends mainly upon  model parameters $M_R\,, h_R^\prime\,, {\rm and}\, Y_L^{\nu \prime}$, while the annihilation cross section is most sensitive to $Y^{\prime\,\nu}_L$, the DM coupling to the SM
Higgs doublet. 
For low $M_{DM}$,  the dominant contribution to the DM annihilation cross section comes from the $s$-channel diagram 
where the DM pair self-annihilate through the neutral SM Higgs mediation.
 As $M_{DM}$ increases, $t$-channel contributions (via the DM itself) to $Z_L h$ annihilation modes become dominant.

 In Fig.~\ref{fig:relic_ML=1} the contours indicate the parameter region that respects the relic constraints for the range of $Y^{\prime\,\nu}_L$,  shown in the color-coded column on the right. It is to be noted here that  the mass splittings $(M_{E_2^\pm}-M_{DM})$ can be small only for a small range of allowed parameter space near $Y_L^{\nu \prime} \gsim 1.7$. For the rest of
the parameter space, the allowed mass splitting is quite large. 
Coannihilation of the DM candidate with 
other states does not occur in this scenario, since the
other heavy states ($\nu_2, \, \nu_3,\, \nu_4$) are much heavier than the
DM candidate. On the other hand, co-annihilation with the vectorlike charged leptons $E_2^\pm$ would happen only 
 if the mass splitting can be as low as 3-4 GeV. However, the inclusion of both the $2\sigma$ 
upper and lower bounds on relic density constraints evades the possibility of having a small
mass difference between $E_2^\pm$ and DM (3-4 GeV), as this can only yield an under-abundant DM relic.


\subsubsection{Direct Detection}
\label{subsubsec:dd}
Direct detection experiments look for signals emerging from dark matter scattering off normal matter (neutrons or protons). As the dark matter only interacts weakly, such events are very rare, and  direct detection experiments require very  accurate background rejection. However, these are important, as the expected signals test the nature of the dark matter.

\begin{figure}[htbp!]
\center
\begin{center}
$\begin{array}{cc}
\includegraphics[height=2.5in,width=3.2in]{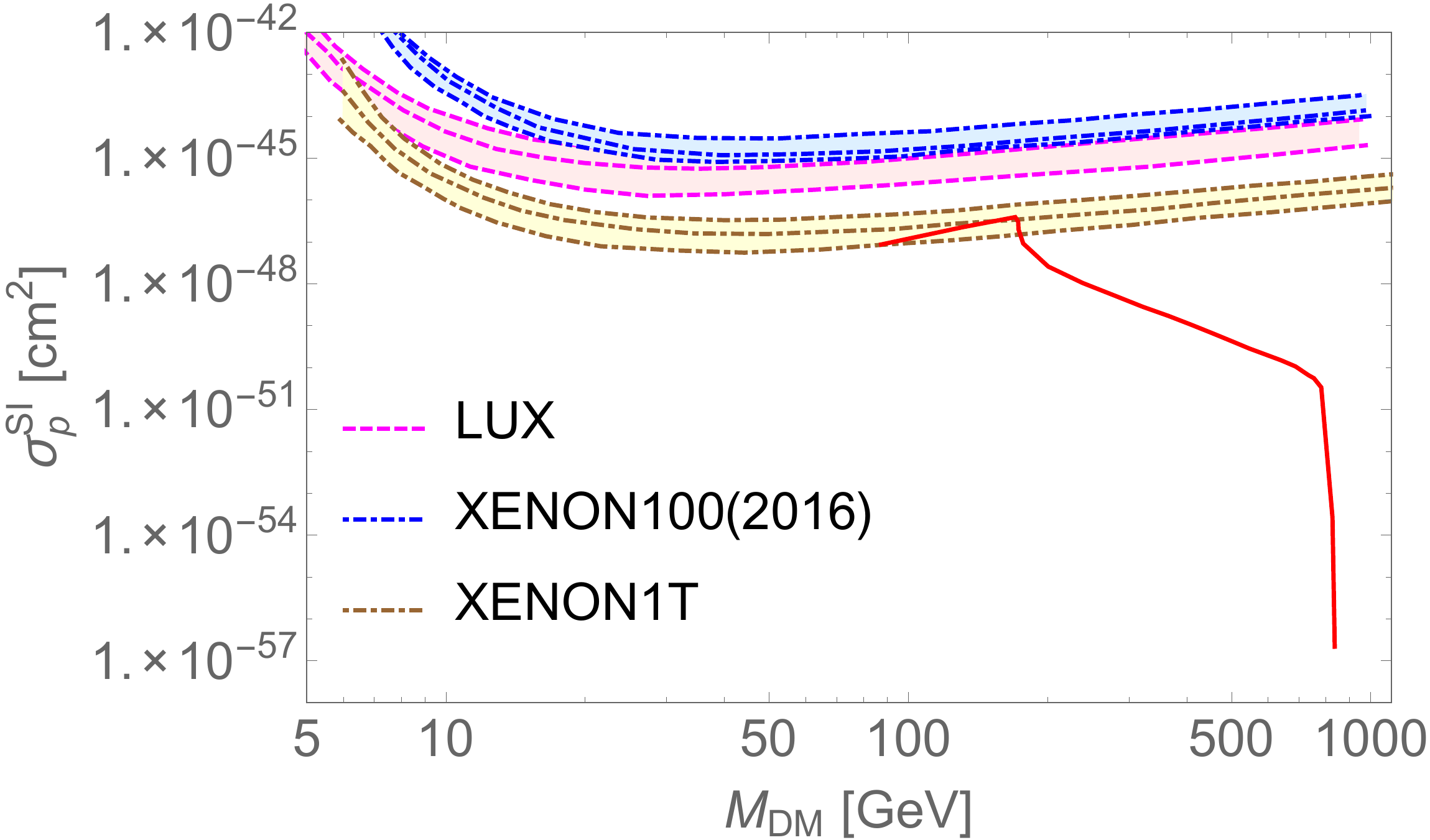}
& \includegraphics[height=2.50in,width=3.2in]{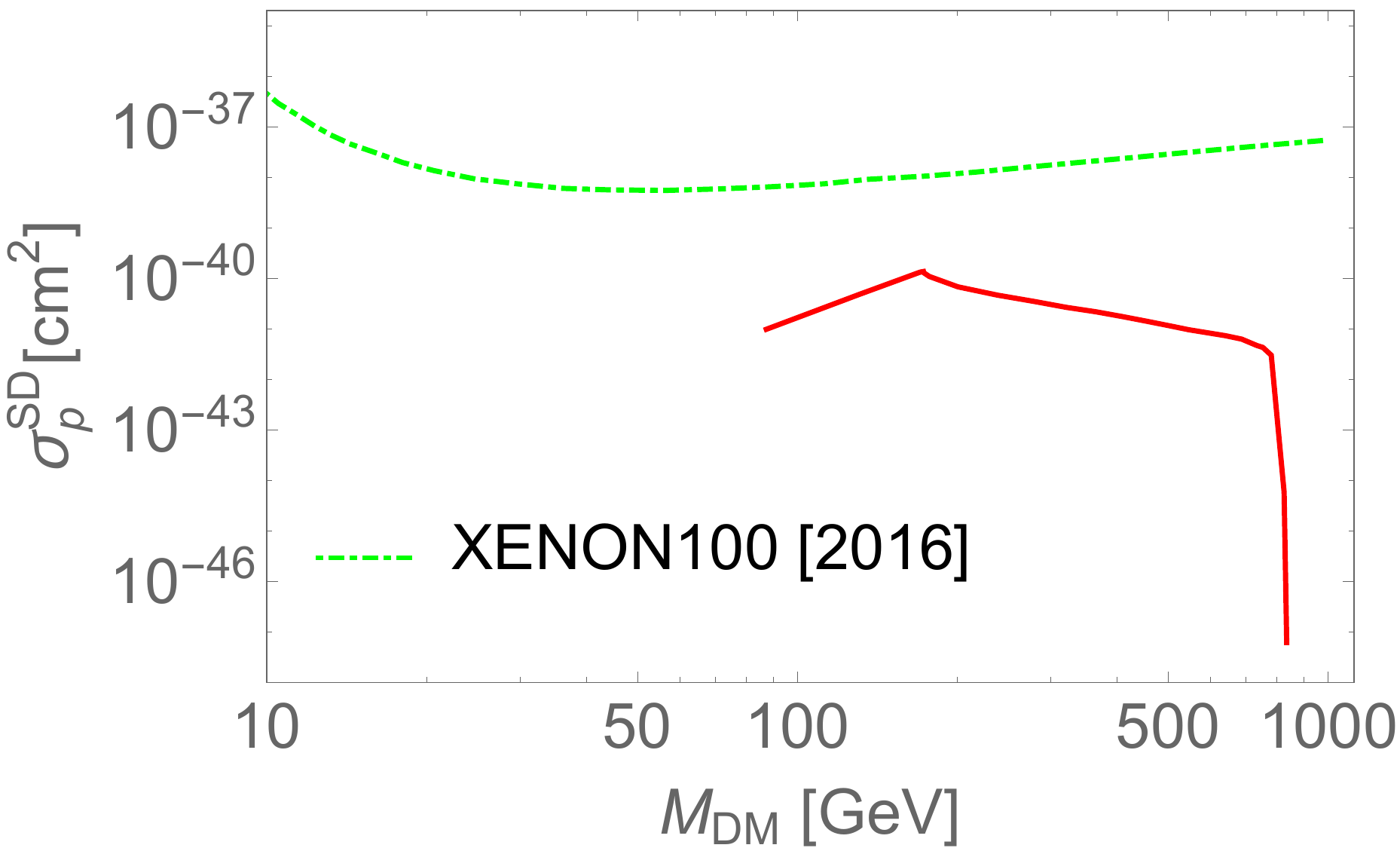}
\end{array}$
\end{center}
\caption{\it Left: Spin-independent cross-section of the proton as a function of the dark matter mass in the left-right symmetric model with vectorlike leptons (red curve), and the experimental upper limits from XENON100  \cite{Lavina:2013zxa} (dashed blue curve),  from  LUX \cite{Akerib:2013tjd} (dashed pink curve), and from XENON1T \cite{Aprile:2015uzo}, all with $2\sigma$ expected sensitivity (dashed yellow curve). 
Right: Spin-dependent cross-section as a function of the dark matter mass (red curve) and the experimental limit from XENON100  \cite{Lavina:2013zxa}. 
We include only points in the parameter space where relic density constraints are satisfied.  All the other parameters are fixed at our BP1 values in Table \ref{table:bp}.}  
\label{fig:SI}
\end{figure}

The interaction of dark matter with detector nuclear matter can be spin-dependent or spin-independent.
The spin-dependent scattering can only happen with odd-numbered nucleon in the nucleus
of the detector material, while in spin-independent (scalar) scattering, the coherent scattering of all
the nucleons in the nucleus with the DM are added in phase. Consequently, 
in direct detection experiments, the experimental sensitivity to spin-independent (SI) scattering is much 
larger than the sensitivity to spin-dependent scattering, which experiences an enhancement in 
scattering from large target nuclei. In our case, 
the $Z$, $Z_R$ boson mediators influence the former, while Higgs boson 
exchanges usually dominate the latter. The most stringent bounds on the spin-independent $\sigma_{SI}$ cross section 
in terms of the dark matter mass come from the XENON100 \cite{Lavina:2013zxa} and  LUX 
\cite{Akerib:2013tjd} experiments, which have seen no dark matter interaction events yet. We explore the spin-independent cross section and compare this against the constraints  from XENON100 (dashed blue curve), the LUX experiment (dashed pink curve) 
and the projected XENON1T (dashed yellow curve) in Fig.~\ref{fig:SI}, where, on the left, we plot the spin-independent dark matter cross section from 
direct searches as a function of the dark matter mass. As seen in the figure, the cross section predicted by our model (continuous red curve) mostly lies below the experimental bounds for $M_{DM}$ values between 87.4 - 836.5 GeV (the exact region where we get the correct relic density), except in the region 70 - 150 GeV, where our theoretical expectations lie within  the  $2\sigma$ expected sensitivity from XENON1T  \cite{Aprile:2015uzo}.
 Note that in Fig.~\ref{fig:SI}, the scattering cross section drops suddenly for masses around $M_{DM}\simeq 600$ GeV. 
The reason lies in the fact that we have plotted {\it only} those points that satisfy the relic constraints. Scanning over the parameters, we found that it was difficult to satisfy relic constraints for a variety of Yukawa couplings and whenever possible, we looked for the largest scattering cross-section. 
 Increased precision may rule out lower dark matter regions of the parameter space. On the right, we plot the spin-dependent cross section, and the recent experimental limit from XENON100 \cite{Lavina:2013zxa}: the constraints imposed are much milder, and our cross section is smaller than the bound imposed by the data by 1-2 orders of magnitude.

\subsubsection{Indirect Detection}
\label{subsubsec:annihilation}
Indirect-detection experiments look for signals arising from pair annihilation of dark matter particles into SM particles. 
There are large number of final states that can be looked at, including $\mu^+,\, \bar{d},\, \bar{p},\, \gamma$-line and $\gamma$-continuum spectra. Since our dark matter candidate is primarily right-handed, 
the  Higgs bosons (especially $\Delta^0_R$), the $Z$ and the $Z_R$ can all act as mediators and enhance the dark matter-pair annihilation cross-section into fermion pairs. The coupling between the dark matter particle and SM mediators 
must produce an acceptable annihilation rate and, besides satisfying direct detection constraints, must be 
 sufficient to produce the correct relic density. 
 
\begin{figure}[htbp!]
\centering
\includegraphics[height=2.5in,width=3.5in]{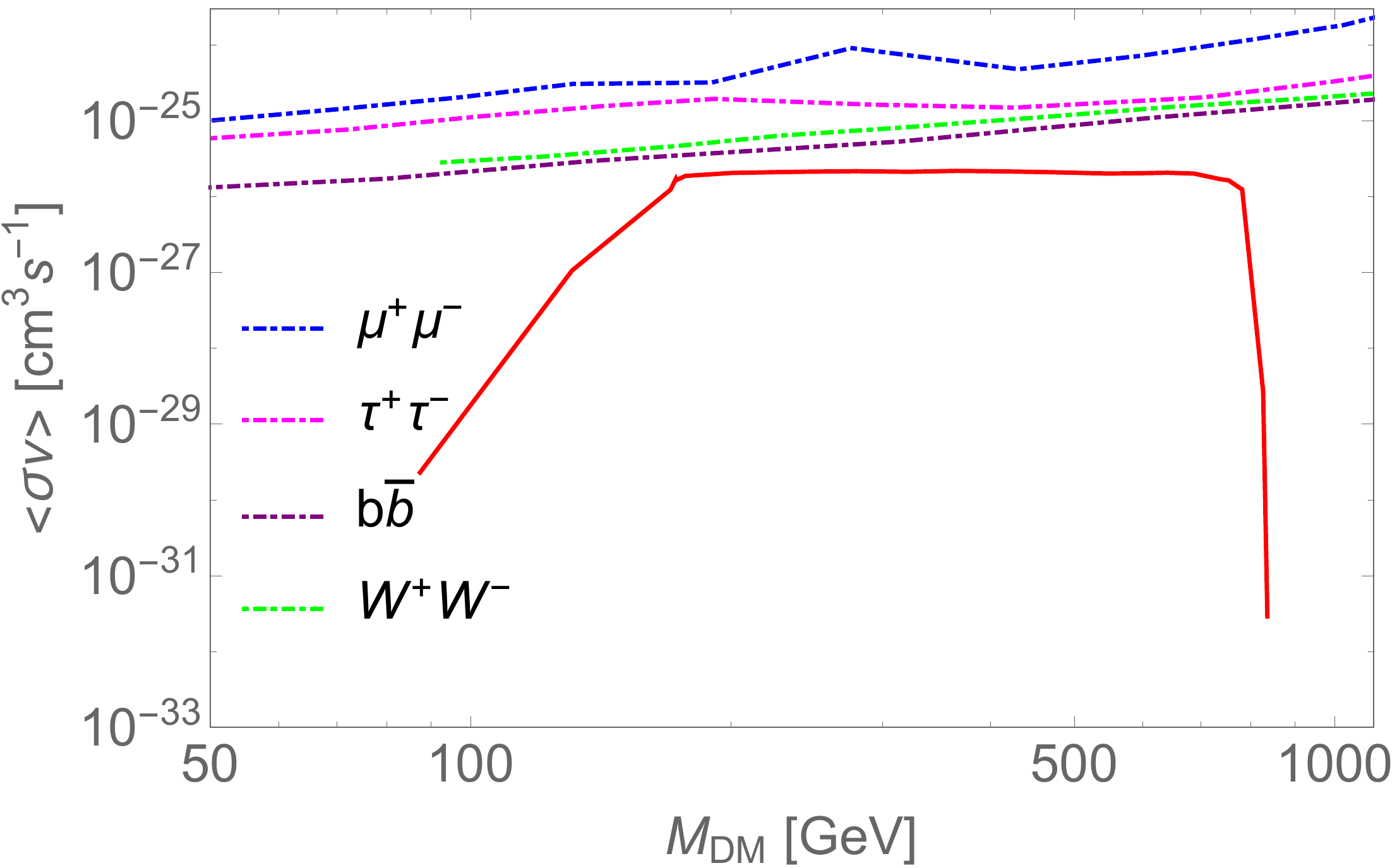}
\caption{\it  Annihilation cross-section of dark matter into SM particles as a function of the  dark matter mass (red curve). We compare this with the combined indirect detection limits from Fermi-LAT and MAGIC Collaboration on gamma rays arising from annihilations in dwarf spheroidal galaxies \cite{Ahnen:2016qkx}. The dashed curves represent annihilation into $\mu^+ \mu^-$ (blue), $\tau^+ \tau^-$ (pink), $ b\bar{b}$ (purple) and $W^+W^-$ (green). We include only points in the parameter space where relic density constraints are satisfied.  All the other parameters are fixed at our BP1 values in Table \ref{table:bp}.}
\label{fig:ACS}
\end{figure}

The most stringent constraints on dark matter annihilation cross sections have been derived from the Fermi gamma-ray space telescope (Fermi-LAT) \cite{Daylan:2014rsa},  used to search for dark matter annihilation products from dwarf spheroidal galaxies and the Galactic Center, which probe annihilation cross sections into photons. To obtain the correct value for the dark matter density the annihilation cross section should be  $\langle \sigma v\rangle \sim 3 \times 10^{-26} {\rm cm^3/s}$. 
In Fig.~\ref{fig:ACS} we show the annihilation cross section of dark matter as a function of the dark matter mass $M_{DM}$ and compare it with the constraints on the dark matter annihilation cross section for the  most restrictive channels, $\mu^+\mu^-$,  $b\bar{b}$, and especially the $W^+W^-$ channels,  at 95\% C.L., found from examining continuum gamma-ray spectra from the dwarf spheroidal galaxy
Segue-I \cite{Daylan:2014rsa,Ahnen:2016qkx}.  
 The red line shows a sudden drop at high mass values which is due to our choice of discrete points
that satisfy the relic density constraints, similar to our previous case for direct detection
cross section. For light $M_{\rm DM}$ ($\lsim 100-115$ GeV), the dominant annihilation mode is into $b {\bar b}$ (90\%) (through the SM like Higgs), while for larger $M_{\rm DM}$ (from 120 GeV up to 800 GeV) 
 $Z_L h$ becomes the dominant annihilation mode (again, about 90\%) because of DM annihilation through the $t$-channel.
We note that a substantial region of the parameter space 
survives the limits from indirect detection, 
although in the region $M_{DM}\in (175-300)$ GeV, our theoretical prediction is close to the experimental limits.

\begin{figure}[htbp!]
\centering
\begin{tabular}{cc}
\includegraphics[height=2.5in,width=3in]{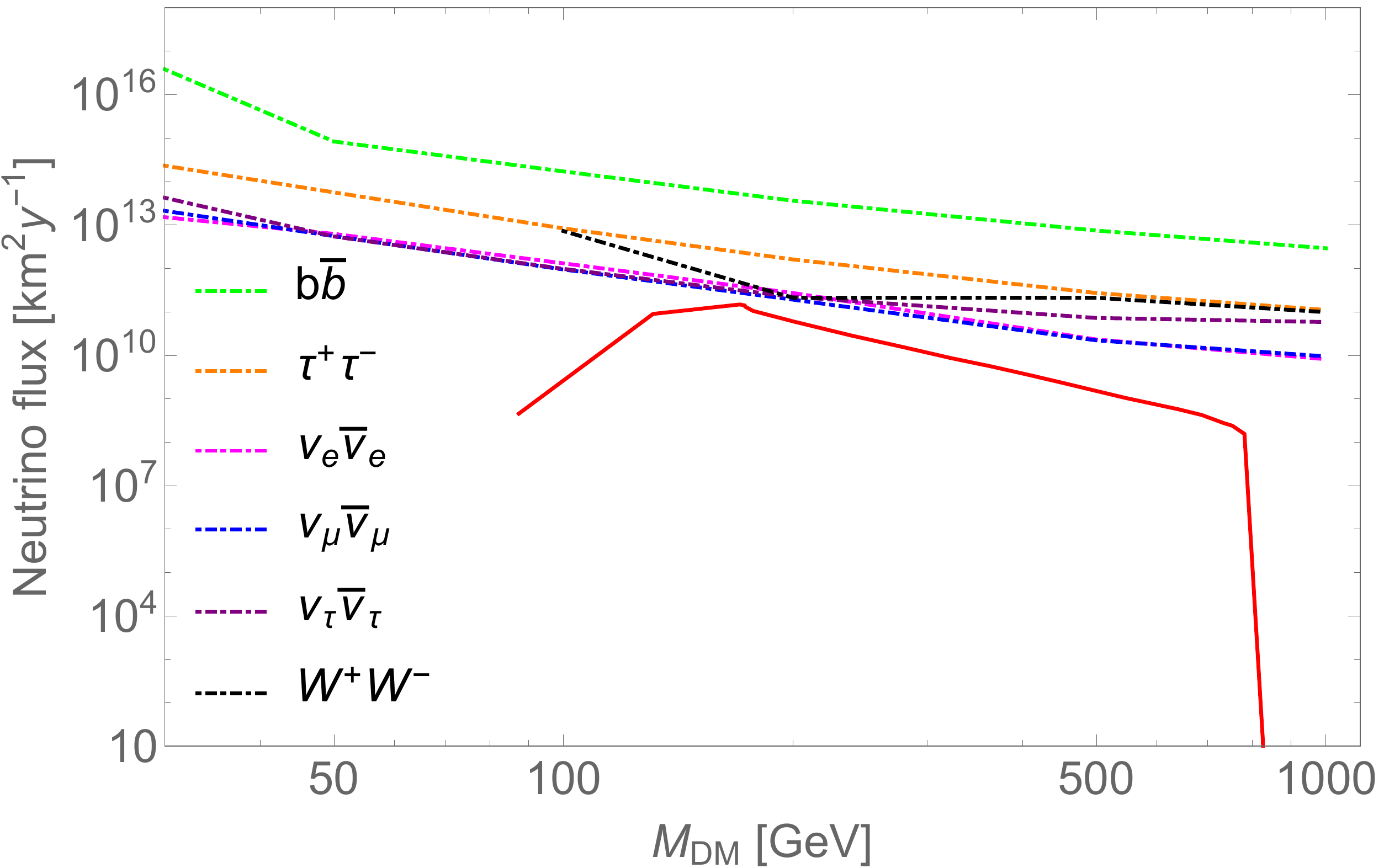}&
\includegraphics[height=2.5in,width=3in]{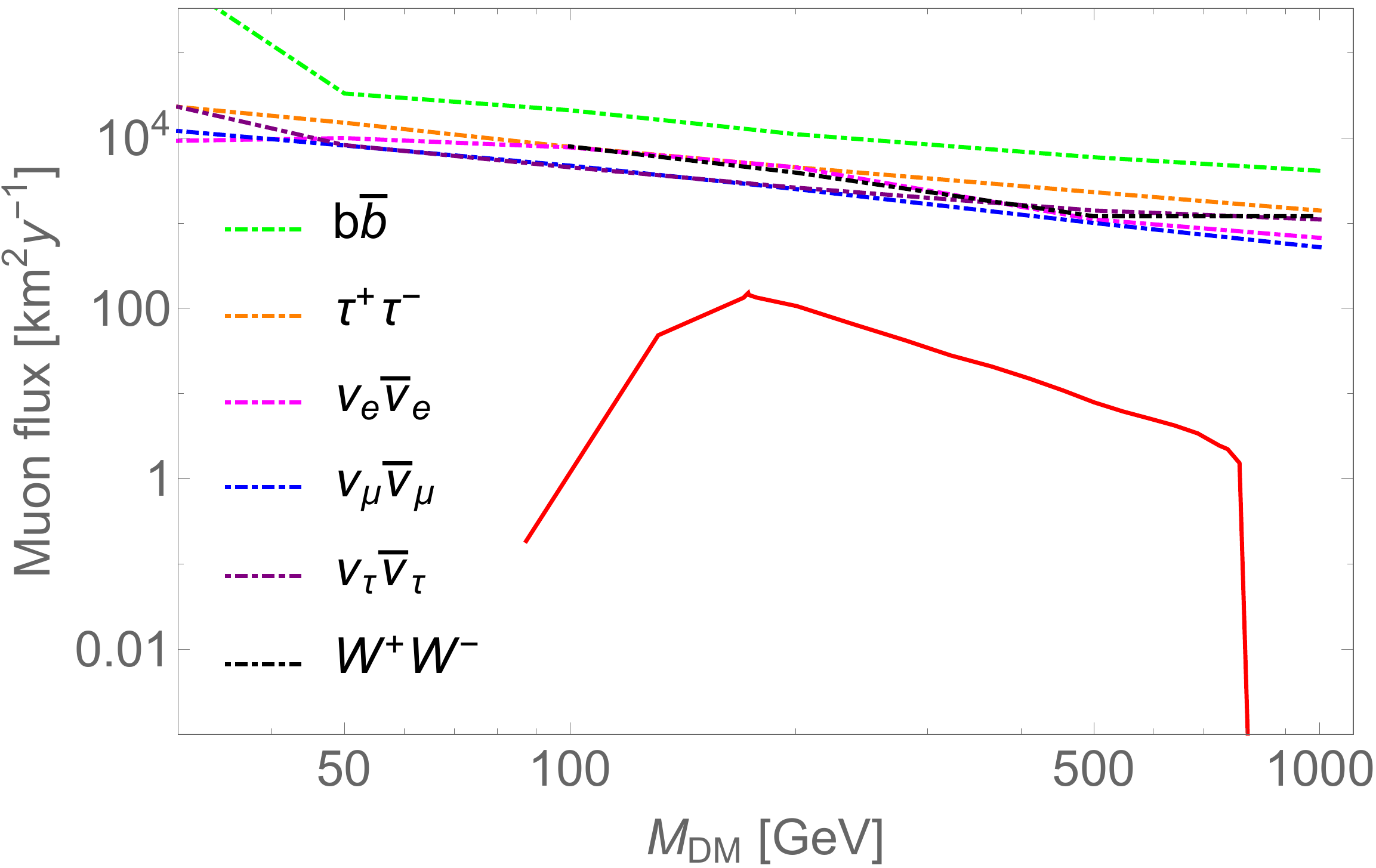}
\end{tabular}
\caption{\it  Neutrino (left) and muon (right) fluxes as a function of dark matter mass (red curve). The dashed curves represent 90 \% C.L. upper limits from Baikal \cite{Avrorin:2014swy}. We include only points in the parameter space where relic density constraints are satisfied.  All the other parameters are fixed at our BP1 values in Table \ref{table:bp}.}
\label{fig:neutrino_flux}
\end{figure}

The dark matter can annihilate into cosmic rays, over much different
annihilation channels, for processes which are model dependent. The emission of most particles can be modeled 
by using leptonic scenarios equally as well as hadronic scenarios. This ambiguity does not exist for high-energy neutrinos, as they
can be created efficiently only in hadronic interactions via the decay of charged pions. The
detection of a high-energy astrophysical neutrino source would then be a signal of accelerated hadrons. 
Each annihilation channel provides a unique neutrino
energy spectrum and since the probability of
neutrino detection depends sensitively on its energy, different
neutrino signals can be expected from different
annihilation channels. Also, since neutrinos interact only weakly with matter, they are insensitive
to radiation fields and are accessible to cosmological distance scales. However, the same effect yields low 
cross-sections, and the backgrounds from existing atmospheric neutrinos are significant. 
For our vectorlike neutrino dark matter candidate, annihilation in the Galaxy into ordinary neutrinos $\nu \bar{\nu}$ may be of significance. During propagation, 
neutrinos oscillate between flavors,  but after traveling across cosmological
distances, the coherence between different flavor
states are lost and, as they reach Earth, neutrinos become mass eigenstates. These experiments also include limits on the muon flux, which incorporates limits for $b \bar{b},\, \tau^+\tau^-$ and $W^+W^-$ channels, for the purpose of comparing with other neutrino telescope experiments. In Fig. \ref{fig:neutrino_flux} we plot the flux as a function of the dark matter mass (neutrino flux in the left panel, muon flux in the right panel) and compare it with the experimental limits from Baikal \cite{Avrorin:2014swy}.

\subsection{Direct DM Searches at the LHC}
\label{subsec:DM_at_LHC}
DM searches can be performed at the LHC where in general, the DM particles would be invisible, and reveal their presence only as missing transverse energy. The direct searches at the LHC involve looking at the associated particles which
come from ISR (initial state radiation) or from their associated production with the DM candidate.
The LHC DM searches concentrate mainly on the $mono-X,(X = jet, \gamma, Z, W)$ signals 
where the DM particle is produced either in association with one or more QCD jets, 
or with a vector boson $V=\gamma, Z, W$.
The strongest constraint placed by the recent ATLAS searches on mono-jet signal excludes a signal cross section above 
19 fb at a 95\% C.L~\cite{Aaboud:2016tnv}.  
In our model, the cross sections for the vectorlike neutrino production with jets are small (${\cal O}(0.1~\rm fb)$);
therefore, they safely satisfy the current experimental limits and can only be detected with
higher detector sensitivity. However, here we present a more viable
detection channel for the vectorlike neutrinos, where they are produced from the secondary decay 
of charged vectorlike leptons.
\section{Collider Searches}
\label{sec:collider}

In this section, we will analyze our findings in light of the collider searches
for the new exotic vectorlike leptons~\cite{Dermisek:2014qca,Dermisek:2014cia}. As already explained in the previous section,
the imposition of an extra parity symmetry provides a viable cold Dark Matter candidate.
More explicitly, the lightest neutral vectorlike lepton which is the 
lightest among the physical mass eigenstates defined in Eq.~(\ref{eq:vneutrino-eigenvalues})
acts as the good DM candidate.  Also, the notable feature of the DM particle is that
it is dominantly right-handed and thus can easily yield correct relic density 
within the 2$\sigma$ range of Planck's latest relic density value $\Omega_{DM}h^2=0.1199\pm 0.0027$, as we showed in Sec. \ref{sec:dm}.
We have shown that there exists ample parameter space which satisfies all the DM constraints, 
including relic density and direct detection data. However, in addition to the DM vectorlike neutrino,
the model consists of extra charged vectorlike (VL) leptons, which can decay into final states including DM.  An obvious question would be 
whether in the allowed parameter region, 
there could be any signature of these charged leptons in the existing
or upcoming collider experiments. 

\subsection{Benchmark Points}
\label{subsec:bench}

At this point, we would like to review our choice of benchmark points and 
explain their plausibility. The  
vectorlike lepton sector of this model depends upon 14 parameters,  
among which 12 are the Yukawa couplings
connecting the non-standard charged and neutral leptons ($Y_L$, $Y_R$, $h_L$ and $h_R$s) to the Higgs bosons and, the rest are the 
two bare mass terms ($M_L$, and $M_R$) for vectorlike leptons. 
We are interested in the region of parameter space where the DM constraints are satisfied 
and the VL charged leptons are kinematically accessible both to the LHC with $\sqrt{s} = 14 $ TeV 
and at the proposed $e^+e^-$ international liner collider with $\sqrt{s} = 1$ TeV.
In view of these, we first fix
most of the Yukawa couplings and we choose 
different masses for the VL leptons by varying the bare mass parameters, $M_L$ and $M_R$.  
It can be noted that the dominant right-handed nature of the DM $(\nu_1)$ and 
the lighter charged VL lepton ($E_2^\pm$) demand a relatively smaller $M_R$ than $M_L$, for both  
to acquire masses of the order of few hundred GeV. With this procedure, for a 
particular choice of DM mass, we get the correct relic density and direct detection cross section 
by mainly tuning the Yukawa parameter $h_R^\prime$. In Table~\ref{table:fixedparam}, we display
the values for the Yukawa coupling parameters of our choice, while in Table~\ref{table:bp}, we give the
masses of the lightest charged and neutral leptons and the corresponding relic density,
and direct detection cross section,  both spin dependent and spin-independent, for the
three benchmark points of our interest. 
\begin{table}[htbp!]
	\centering
	\begin{tabular}{|c|c|c|c|c|c|c|c|c|c|c|c|c|}
		\hline 
		Yukawa Parameters & ${Y_L^{\prime}}^e$ &$ \tilde{Y}_L^{\prime \, e}$ & $ Y_R^{\prime \, e}$ & $\tilde{Y}_R^{\prime \, e}$ & $Y_L^{\prime \, \nu}$ & $Y_R^{\prime \, \nu}$ & $\tilde{Y}_L^{\prime \, \nu}$ &  $\tilde{Y}_R^{\prime \, \nu}$ &  $h_L^\prime$ & $h_L^{\prime \prime}$ & $h_R^{\prime \prime}$ & $\kappa_1$ [GeV] \\ \hline
		 Value  & 0.1 & 0.0 & 0.0 & 2.5 & 1.5 & 0.04   & 0.1 & 0.1 & 0.8 & 0.7 & 0.0 & 246  
		       \\ \hline
	\end{tabular}
	\caption{\it Fixed parameters for all benchmark points.}
	\label{table:fixedparam}
\end{table}
\begin{table}[htbp!]
	\centering
	\begin{tabular}{|p{2cm}|c|c|c|c|c|c|c|c|}
		\hline
		Benchmark Points & $h_R^\prime$ & $M_{L}$ & $M_{R}$ & $M_{\rm DM}$ & $M_{E_2^\pm}$ & $\Omega_{\rm DM} h^2$ & $\sigma_{\rm SD}$  & $\sigma_{\rm SI}$  \\ 
		 &  & $(\rm GeV)$ & $(\rm GeV)$ & $(\rm GeV)$ & $(\rm GeV)$ &  & (pb) &  (pb) \\ \hline
		BP1 & 0.045 & 1000 & 275 & 173 & 275 & 0.116 & $1.3\times10^{-4}$ & $2.55\times10^{-11}$ \\ \hline
		BP2 & 0.033 & 2000 & 350 & 258 & 350 & 0.112 & $4.3\times10^{-5}$ & $4.3\times10^{-12}$ \\ \hline
		BP3 & 0.032 & 2500 & 400 & 299 & 400 & 0.117 & $3.1\times10^{-5}$ & $2.42\times10^{-12}$ \\ \hline
	\end{tabular}
	\caption{\it Benchmark points for vectorlike leptons, including masses with corresponding
		DM relic density and direct detection cross section.}
	\label{table:bp}
\end{table}

\subsubsection{Higgs Signal Strength}\label{subsubsec:higgs_strength}
In our model, the lightest CP-even scalar state resembles the 125 GeV Standard Model-like Higgs boson discovered at the LHC. Therefore,
it is important to check the signal strengths for the production and decays of this Higgs state in this model relative to the current experimental data. Since the tree-level couplings of the lightest CP-even Higgs boson with all the Standard Model particles remain unchanged, we do not expect any deviations in the tree level decay channels of this Higgs boson from that of the Standard Model one. 
The gluon fusion production is not affected by leptons, but the loop induced decay modes of Higgs into the diphoton channel will get extra contributions 
from singly and doubly charged scalars and also from the exotic vectorlike charged lepton loops. The new 
VL charged leptons are expected to contribute destructively (with respect to the dominant $WW$ loop contribution), as the fermion loop comes with a  negative sign, while the charged scalar loops may enhance or suppress 
the decay depending on the sign of the coupling of the charged scalars to the 
lightest CP-even Higgs boson. For brevity, we do not give the general expression for
the Higgs to diphoton decay width, which can, however, be found in literature~\cite{Shifman:1979eb,Gunion:1989we,Djouadi:2005gj,Djouadi:2005gi}.
 The test is done only for the charged lepton mass values chosen in the above-mentioned benchmark
points while the singly and doubly charged scalar masses are chosen as 243 and 305 GeV, respectively,
throughout the analysis. The implications of these charged scalars in 
our collider study will be mentioned later. 
The Higgs couplings to these nonstandard charged scalars and the VL leptons are fixed for
all benchmark points since the Yukawa parameters are kept fixed. Moreover, there is no additional 
contribution to Higgs production through gluon fusion, hence only the ratio of the partial decay width
Higgs to diphoton channel between the model prediction and that of the SM value represents our
signal strength.  According to the latest result from LHC Run II at 13 TeV, the Higgs to diphoton signal strength is   
$\displaystyle \frac{\mu_{exp}}{\mu_{SM}}= \frac {\left [ \sigma (pp \to h) BR(h\to \gamma \gamma) \right]_{exp}}{\left [ \sigma (pp \to h) BR(h\to \gamma \gamma) \right]_{SM}}=0.85^{+0.22}_{-0.20}$ \cite{ATLAS:2016nke}.
For our benchmark points, the signal strengths are, for BP1, BP2 and BP3, respectively,  0.59, 0.73 and 0.80,
 which are all within the $2\sigma$ range of experimental data.

\subsection{Searches at the LHC}
\label{subsec:LHC}
In this section we consider the pair production of the lightest charged 
vectorlike leptons, namely $E_2^\pm $, at the LHC.
	\begin{eqnarray}
	&~ p~ p& \rightarrow ~E_{2}^+~{E_{2}^-},  \label{eq:s1} 
	\end{eqnarray}	
Instead of scanning over the multi-dimensional parameter space ,
we choose three benchmark points which are allowed by the constraints
coming from the dark matter relic density and direct detection cross section. 
%


In the chosen parameter region, these charged leptons 
can only decay to a SM $W$ boson in conjunction with the DM particle $(\nu_1)$ 
with 100\% branching fraction: $~E_2^\pm \rightarrow ~W^\pm {{\nu_1}}$, thus
leading to $W^+W^- + \mET$ final state, where  $\mET$ arises due to 
the presence of heavy neutral $\nu_1$ particle, which is  the cold 
dark matter candidate. Depending
upon the decay mode of the Standard Model $W$ boson, there are three possible final states :
\begin{eqnarray}	
&(i)&	2\ell^\pm + 0j + \mET  \,, \nonumber ({\rm both~W~decay~leptonically}) \\
&(ii)&	1\ell^\pm + 2j + \mET \,, \nonumber ({\rm one~W~decays~ leptonically,~the~other~one~decays~hadronically})\\
&(iii)&	0\ell^\pm + 4j + \mET  \,, ({\rm both~W~decay~hadronically}).
\label{final_states}
\end{eqnarray}
where, $\ell = e, \mu $, and $j$ corresponds to light quark jets.  
At this point, it should be mentioned that the final 
states closely resemble the pair production of SM $W$.
However,  in this case, 
one may expect to see some deviation in the shape 
of the $\mET$ distribution compared to that of the SM $W$-pair signal. This change
may be attributed to the fact that for the signal, the missing transverse energy
comes from the massive neutral particle whereas in the SM background almost
massless neutrinos are the decay products. Hence, we expect that 
the SM processes which contribute to the background for the 
SM $W$-pair signal will also play the same role in our signal process.
Although the cross sections are
 higher for the final states $(ii)$ and $(iii)$, listed
in Eq.~(\ref{final_states}), these  final states are difficult to measure
at the LHC because of the large background 
contributions mainly arising from $t\bar t$, single top, 
$W^\pm + \rm jets$ and other di-boson productions, all of which are 
difficult to suppress. Therefore, the di-leptons and missing transverse energy $(2 \ell^\pm + \mET)$ final
state is the only possible channel to probe this vectorlike heavy lepton signal at the
LHC. 
As already mentioned, our signal process  
mimics the exact SM $W$ pair production process, but with a different $\mET$ 
spectrum. Hence, it is expected that the application of same event selection method
for SM $W$ pair production to our signal process will suppress the
other dominant SM backgrounds to the final state $(i)$. Therefore, to have 
an overview of signal significance, throughout our analysis, we only consider
the SM $W^+W^-$ process as the dominant background.

For our analysis, we supplement  the model in \cite{Roitgrund:2014zka} with VL leptons, using FeynRules~\cite{Alloul:2013bka}
which gives the UFO model files required in {\tt Madgraph5}~\cite{Alwall:2014hca}
to generate the signal events at the LO
parton level. The SM background events
are also generated using {\tt Madgraph5}. 
The unweighted parton level events are passed 
through the {\tt Pythia}(v6.4)~\cite{Sjostrand:2006za} to simulate
showering and hadronization effects, including fragmentation. The
detector simulation is done using the {\tt
Delphes}(v3)~\cite{deFavereau:2013fsa}. Finally, we perform the cut
analyses using {\tt MadAnalysis5}~\cite{Conte:2012fm}.
At this point, we would like to mention that at the detector
level, the criteria for the isolation of 
electrons and muons at the final state are performed using the method 
described in Ref.~\cite{ATLAS:2016rin}, where electrons are isolated 
with the {\tt Tight} criterion defined in \cite{ATLAS:2016iqc} and muons are isolated 
using the {\tt Medium} criterion defined in \cite{Aad:2016jkr}. 
Jets are reconstructed using the anti-$k_t$ clustering algorithm 
with radius parameter $\Delta R = 0.4$ with minimum 
$p_T=20$ GeV and jets originating from the fragmentation of
$b$-hadrons ($b$-jets), if any, are tagged with 85\% tagging efficiency, and 
10\% and 1\% mis-tagging efficiency for $c$-quark and light-quark jets respectively.
The leading order (LO) production cross-sections are calculated using the 
{\tt NNPDF3.0} parton distributions. 
Before discussing the cut analyses, we show the histograms for the signal
and background after imposing the basic cuts described previously. 

\begin{figure}[htbp!]
\centering
\begin{tabular}{cc}
\includegraphics[height=60mm,width=8cm]{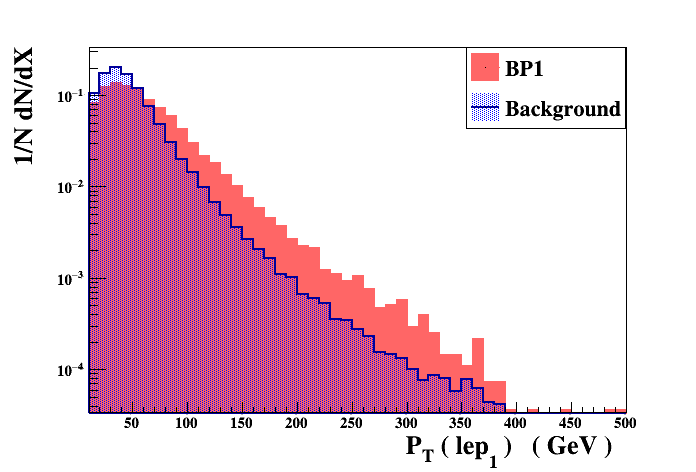} &
\includegraphics[height=60mm,width=8cm]{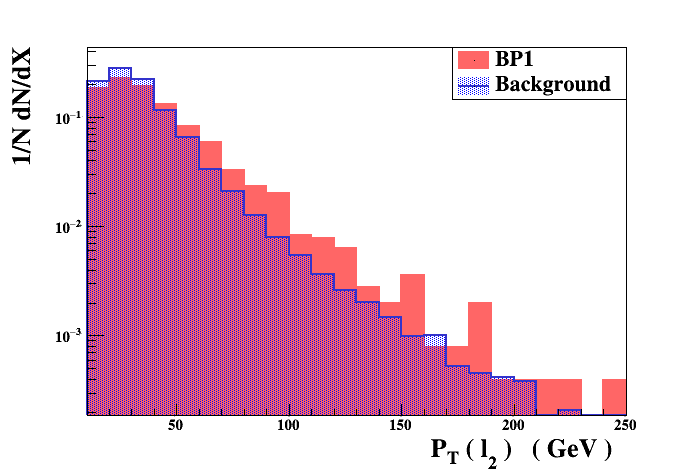} \\
\end{tabular}
\caption{\it Distribution for the transverse momentum of the hardest (left panel) and second hardest (right panel) lepton for  
benchmark point BP1.  $N$ is the total number of events before any cut. Distributions for benchmark points BP2 and BP3 are very similar, so we do not plot them as well.}
\label{ptlep}
\end{figure}
\begin{figure}[htbp!]
\centering
\begin{tabular}{cc}
\includegraphics[height=60mm,width=8cm]{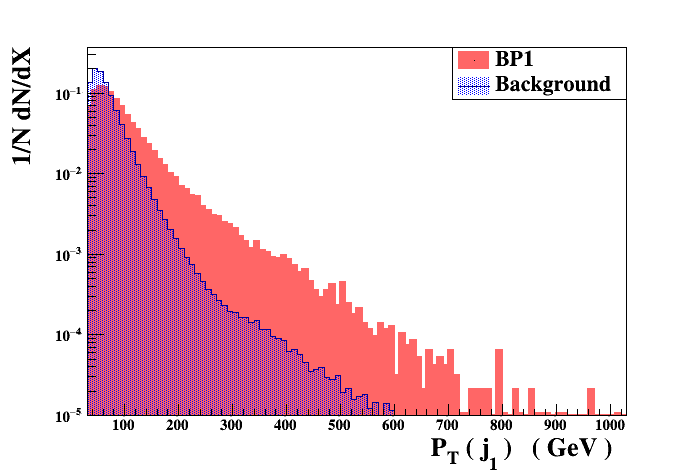} &
\includegraphics[height=60mm,width=8cm]{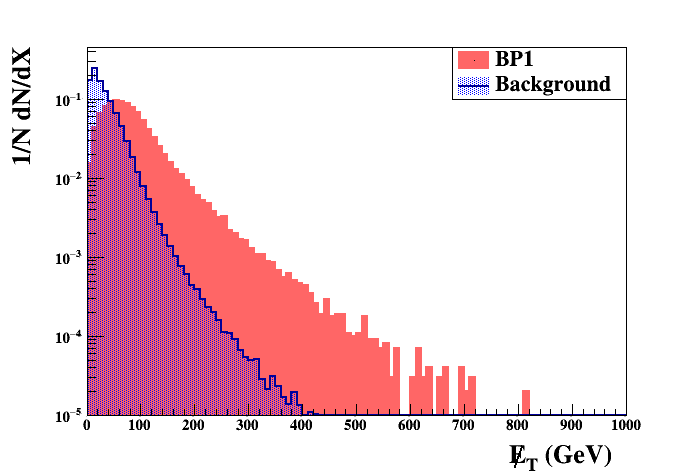} \\
\end{tabular}
\caption{\it Transverse momentum of the hardest jet and missing transverse energy distribution
 for  benchmark point BP1.  $N$ is the total number of events before any cut. Distributions for benchmark points BP2 and BP3 are very similar, so we do not plot them as well.}
 \label{met}
\end{figure}

In Fig.~(\ref{ptlep}), we show the distribution of transverse 
momentum $(p_T)$ for the hardest (left panel) and second hardest (right panel) charged leptons
in the event for the three benchmarks in Table \ref{table:bp}.   
As expected, one may note that the signal and background distribution
follow almost the same shape, including the 
visible Jacobian peak at half the $W$ mass $(m_W/2)$, except for a little smearing effect 
mainly in the signal distribution for the highest $p_T$ lepton, because in the signal 
case, the decaying $W$ boson gets an extra inherent $p_T$ from its parent 
vectorlike charged lepton. Thus, it is important to note that 
we are not allowed to impose larger $p_T$ cuts on the charged leptons 
than that applied in the SM $W$ pair production case~\cite{ATLAS:2016rin}.
The deviation in the distributions of different benchmark points
is self-explanatory from their different cross sections.
In Fig.~(\ref{met}), we depict the distribution of the $p_T$ of
hardest jet and also the missing transverse energy $\mET$~ for 
all the three benchmark points. Analogous to the $p_T$ distribution 
of the hardest lepton, the hardest jet also shows a tail 
at the high $p_T$ end. But since we have already chosen two lepton final states, 
the extra jets are only coming from the initial state radiation (ISR). 
The common kinematical feature of ISR jets~\cite{Krohn:2011zp} is a crucial
dependence on the mass scale that is being probed at the collider experiment
and usually the transverse momentum ($p_T$) of the ISR jets 
is higher with heavier BSM particles at the final state. Therefore, 
an upper cut on the $p_T$ of any extra jets
would yield a negative contribution to the signal significance
for our final state, since the application of jet
veto would be more stringent for the signal than the background.
Note that the jet transverse momentum distribution, shown in Fig.~\ref{met},
is obtained before applying any cuts and with inclusive decay modes of the $W$ boson,
and therefore the effect of ISR jets cannot
be seen from here.
On the other hand, the loss of significance can be gained
from the large $\mET~$ in the signal events. As already discussed,
for the background events, the missing transverse energy arises only from
the neutrinos or the mis-measurement of jets and photons, while,
for the signal events, the lightest neutral VL lepton $(\nu_1)$ gives 
the dominant contribution to $\mET$, which is the stable DM candidate.
The large mass range of this $\nu_1$, as shown in Table~\ref{table:bp}, 
as well as its inherently large $p_T$ due to the mass difference between it and  
the decaying $E_2^\pm$ significantly enhances the $\mET$ for the
signal events. Therefore, one can expect that demanding missing transverse energy $\mET > 60$ GeV
may help in suppressing the background and simultaneously improving the 
signal significance.

We now discuss the effect of selection cuts imposed over the 
basic cuts. It should be mentioned here that the analysis
is done for a LHC run at 14 TeV and the expected reach of integrated
luminosity is 3000 $\rm fb^{-1}$. Therefore, we seek to examine 
the maximum reach of signal significance at 3000 $ \rm fb^{-1}$,
with the significance defined by
\begin{eqnarray}
S &=& \frac{N_S}{\sqrt{N_S + N_B}} \,,
\label{significance}
\end{eqnarray}
where $N_S$ and $N_B$ represent the number of signal and background
events respectively. 

Akin to the selection cuts imposed in Ref. \cite{ATLAS:2016rin}, 
we list the selection cuts that are imposed in our case in Table~\ref{tab:lhc_cuts}. 

\begin{table}[htbp!]
	\centering
	\begin{tabular}{|c|c|} \hline \hline
		Cut name & Selection criteria \\ \hline
		C1 & Number of jets with $p_T(j) > 30$ GeV and $|\eta| < 4.5$ = 0  \\ \hline
		C2 & At least two leptons with $p_T(\ell) > 25$ GeV \\ \hline
		C3 & Number of additional leptons with  $p_T(\ell) > 10$ GeV = 0\\ \hline
		C4 & $\mET > 60$ GeV \\ \hline
		C5 & Number of b-tagged jets with $p_T(b) > 20 $ GeV = 0 \\ \hline
	\end{tabular}
	\caption{\it Selection Cut requirements.}
	\label{tab:lhc_cuts}
\end{table}

We then pass our simulated signal and background events through the cut selection and
check the corresponding significance reach at the highest possible integrated luminosity
that can be attained at the LHC. We sum up this in Table~\ref{table:cut_flow}. As can be
seen from this table, the maximum significance $\sim 3 \sigma $ is attained for the BP1, which 
has the largest production cross-section for the vectorlike lepton pairs. 
For the other two benchmarks, the signal significance is rather poor. From this 
analysis it is very clear that it would be extremely difficult to probe the vectorlike leptons 
scenario at the 14 TeV LHC run even with the highest possible luminosity attainable
at that energy. Thus we are motivated to look for the same signal process for the same benchmark points 
at the upcoming International $e^+ e^-$ Linear Collider (ILC) experiment.  

\begin{table}[htbp!]
	\centering
	\begin{tabular}{c|p{3cm}|c|c|c|c|c|p{2cm}|}
		\cline{2-8}
	 & {Production cross section ($\rm fb$)} & \multicolumn{5}{|p{4cm}|}{Effective cross sections in $fb$ after Cuts} & Significance reached at ${\cal L}_{\rm int} = 3000 \rm fb^{-1}$\\ \cline{3-7}
	 &  & C1 & C2 & C3 & C4 & C5 & \\ \hline 
	\multicolumn{1}{|c|}{SM Background} & 70940 & 6608 & 558.9 & 558.9 & 178.2 & 177.2 & -- \\ \hline 
	\multicolumn{1}{|c|}{BP1} & 138.4 & 9.43 & 1.33 & 1.33 & 0.69 & 0.69 & 2.83 \\ \hline 
	\multicolumn{1}{|c|}{BP2} & 53 & 3.69 & 0.47 & 0.47 & 0.22 & 0.22 & 0.90 \\ \hline 
	\multicolumn{1}{|c|}{BP3} & 31 & 1.94 & 0.29 & 0.29 & 0.15 & 0.15 & 0.01 \\ \hline 
   \end{tabular}
     \caption{\it Effective cross section obtained after each cut for both background and signal and the 
   	respective significance reach at 3000 $fb^{-1}$ integrated luminosity at 14 TeV LHC.}
 \label{table:cut_flow}
\end{table}

To conclude this section, we have shown that pair production of VL leptons is not very promising at the LHC, and would require very high luminosity to disentangle the signal from the background.

\subsection{Searches at the ILC}
\label{subsec:ILC}

In view of the fact that the plausible signal for probing the lightest charged VL leptons
at the LHC seems difficult to observe and may require much higher luminosity than can be reached,
we look for the possibility of probing the same signal at the upcoming International Linear Collider (ILC).
The ILC is favored for its clean signal and less background noise, which at LHC originates mainly from
the QCD processes. At the LHC, higher order perturbative QCD corrections as well as nonperturbative QCD effects 
give rise to large systematic uncertainties in theoretical calculations and hence 
precision measurements do not seem to be feasible. On the contrary, the initial state 
particles ($e^-~ \& ~e^+$) at the ILC are pointlike elementary particles and only interact
through electroweak interactions with only a few-percent-level modification in radiative corrections.
This is why the ILC results provide better precision and thus help the theoretical understanding 
of the Standard Model signal and background processes, which may also shed some light on the 
presence of subtle new physics interactions. In addition to this, the ILC will also 
be furnished with polarized electron and positron beams
so that the processes can be completely characterized based on each initial and final polarization
state.
For the signal process, we consider exclusive leptonic final states and hence a better significance
than at the LHC is naturally expected. In the following section, we perform the analysis for the pair production
of the lightest charged VL leptons, namely, $E_2^\pm$, at the ILC. Equivalent to Eq.~(\ref{eq:s1}),
the process of interest in this case is
	\begin{eqnarray}
	&~ e^+ ~ e^- & \rightarrow ~E_{2}^+~{E_{2}^-}\,.  \label{eq:s2} 
	\end{eqnarray}	

\begin{figure}[ht!]
\centering
\includegraphics[height=7cm,width=10cm]{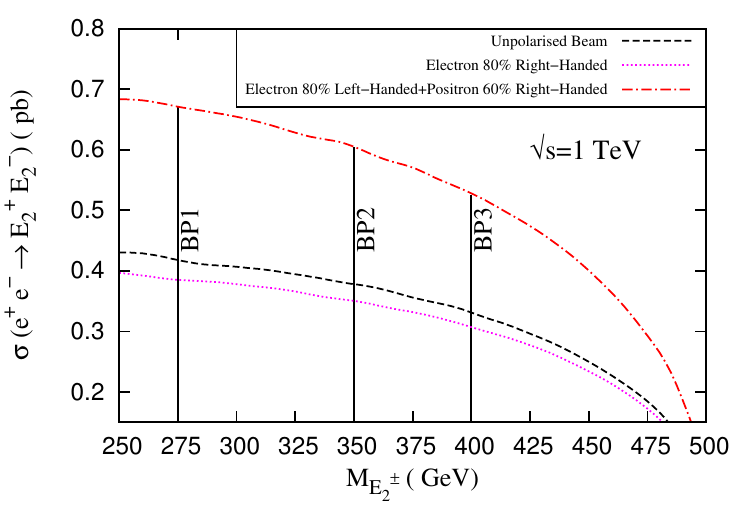}
\caption{\it Signal cross-section for different polarization states of incoming electron and positron beams
at the ILC for various VL charged lepton masses.}
\label{ilc_xsec}
\end{figure}

To compare with our previous result on the searches at the LHC, here we also consider only the di-lepton final state, as mentioned 
in $(i)$ of Eq.~(\ref{final_states})\footnote{A similar study on VL charged lepton search at the ILC is done in Ref.~\cite{Ari:2013wda}.
}. 
In addition to this, we also perform the analysis of VL leptons production using some particular choice of polarization for the incoming 
electron and positron beams. Explicitly, we consider three distinct combinations~\cite{Vauth:2016pgg}, 
\begin{enumerate}[$(a)$]
\item Both the electron and positron beams  are unpolarized.
\item The electron beam is 80\% left polarized, and the positron beam is 60\% right polarized.
\item The electron beam is 80\% right polarized, and the positron beam is unpolarized.
\end{enumerate}
For an extensive review on the physics case for the polarized beam at the ILC, we refer to Ref.~\cite{Baer:2013cma}.
In Fig.~\ref{ilc_xsec}, we show the production cross section for our process, given in Eq.~(\ref{eq:s2}),
at the ILC center of mass energy 1 TeV. We indicate our chosen benchmark points on
the graphs. It is to be noted that the polarization states of the initial electron and/or positron
beams change the signal cross section significantly. The highest production cross section can be 
reached for the combination $(b)$, defined previously. The collider analysis is done using {\tt Madgraph5}.
In compliance with LHC searches, here too the dominant SM background for di-lepton final states is
SM W-boson pair production. $WWZ$ and $ZZ$ will also contribute to the
background, {\it albeit} with small cross-sections. 

Before imposing selection cuts, we check the distributions of various 
kinematical variables at the parton level. To do so, some basic cuts 
are enforced first, such as 
\begin{itemize}
\item The minimum transverse momentum of the charged lepton at the final state should be greater than 10 GeV 
(${p_T}_{\ell} > 10$~GeV), and
\item The pseudo-rapidity of each charged lepton must be within 2.5 ($|\eta_{\ell}| < 2.5$). 
\end{itemize}

\begin{figure}[htbp!]
\centering
\begin{tabular}{cc}
\includegraphics[height=60mm,width=8cm]{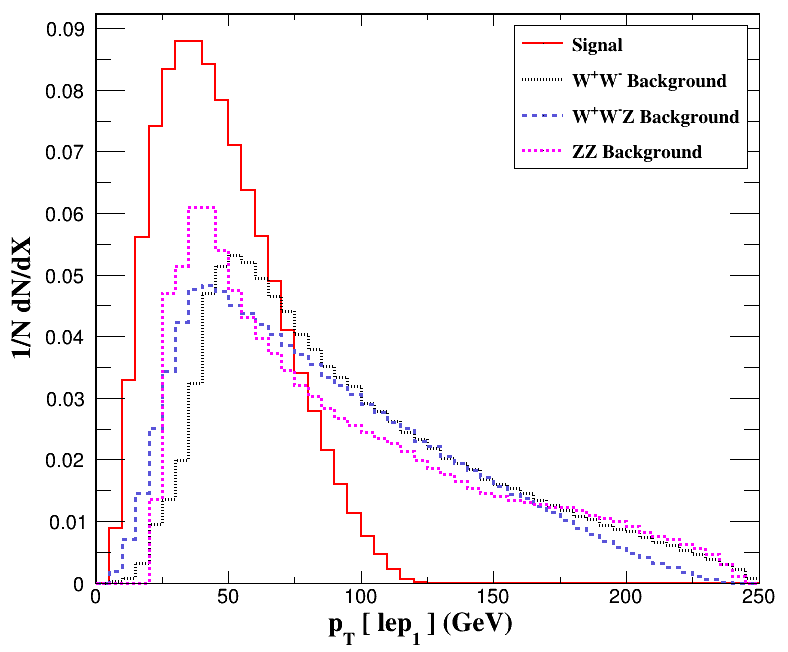} &
\includegraphics[height=60mm,width=8cm]{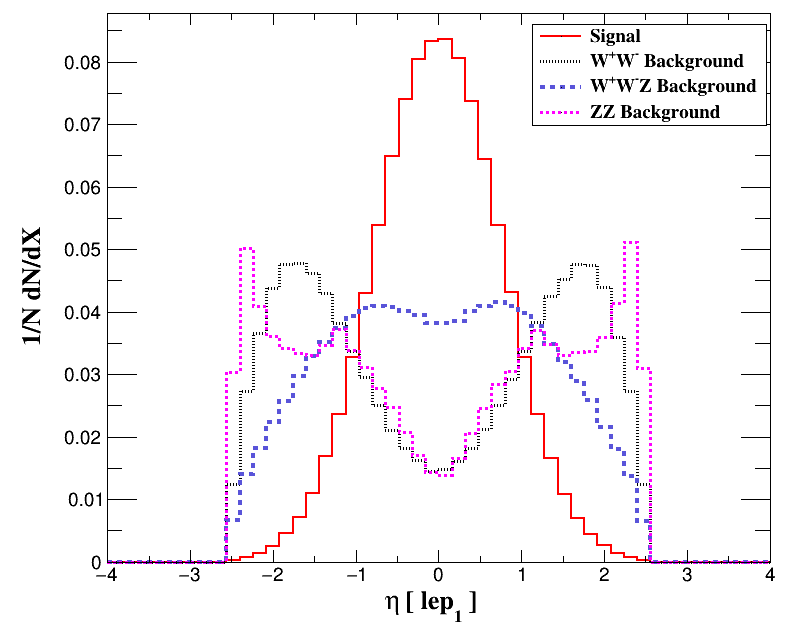} \\
\includegraphics[height=60mm,width=8cm]{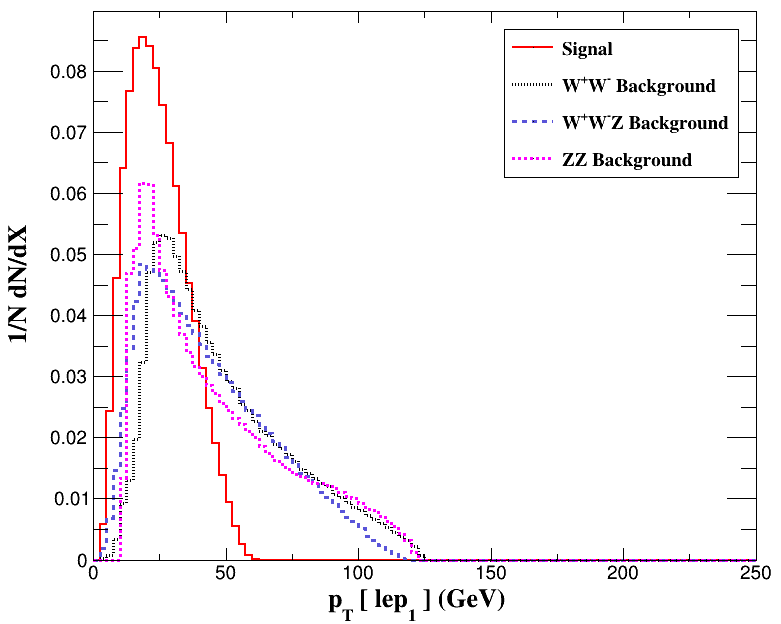} &
\includegraphics[height=60mm,width=8cm]{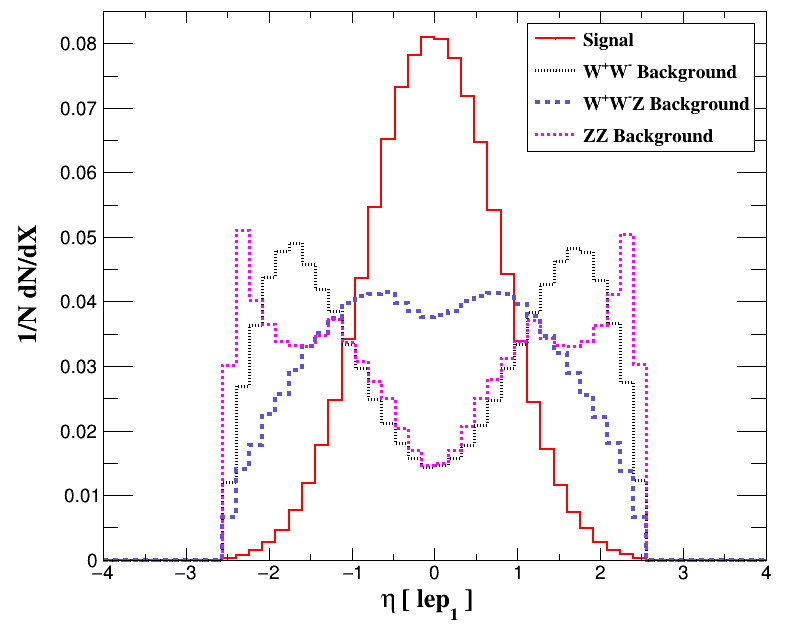} \\
\includegraphics[height=60mm,width=8cm]{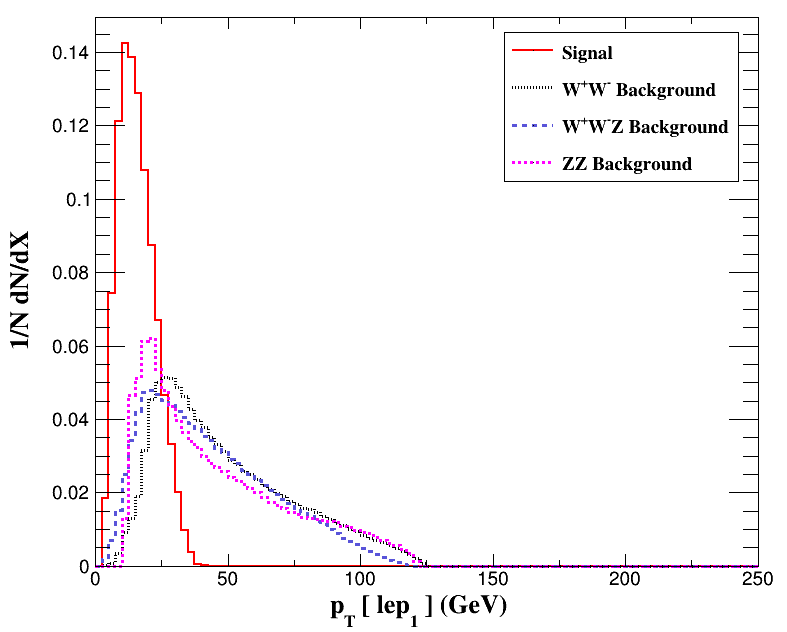}  &
\includegraphics[height=60mm,width=8cm]{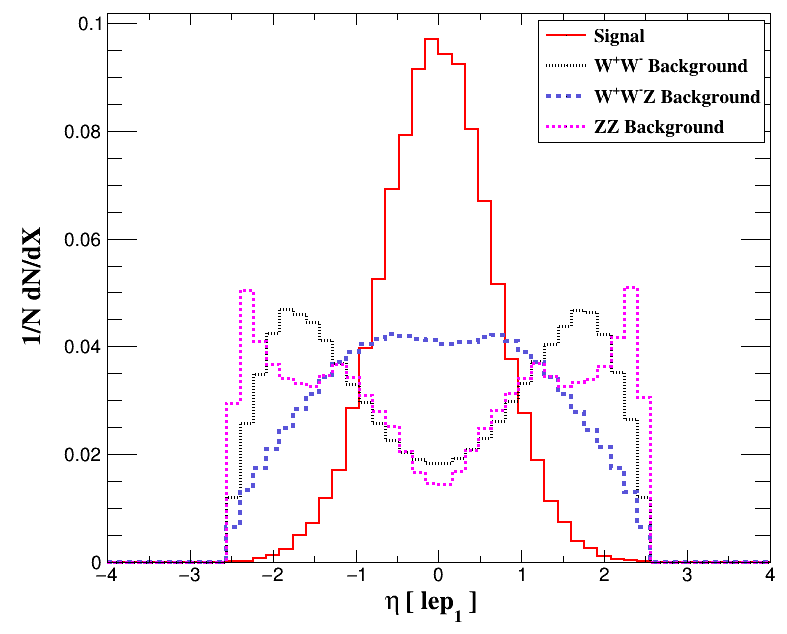} 
\end{tabular}
\caption{\it Transverse momentum and pseudo-rapidity distribution of the hardest lepton 
 for BP1 and for the combinations $(a)$ unpolarized (top), $(b)$ both-polarized (middle) and $(c)$ only electron-polarized(bottom).}
 \label{ilc_pt1}
\end{figure}

\begin{figure}[htbp!]
\centering
\begin{tabular}{cc}
\includegraphics[height=60mm,width=8cm]{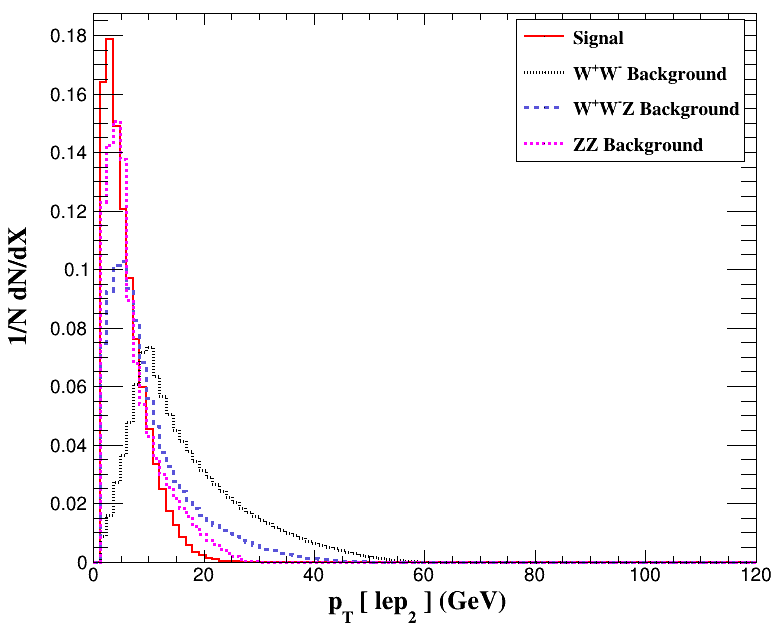} & 
\includegraphics[height=60mm,width=8cm]{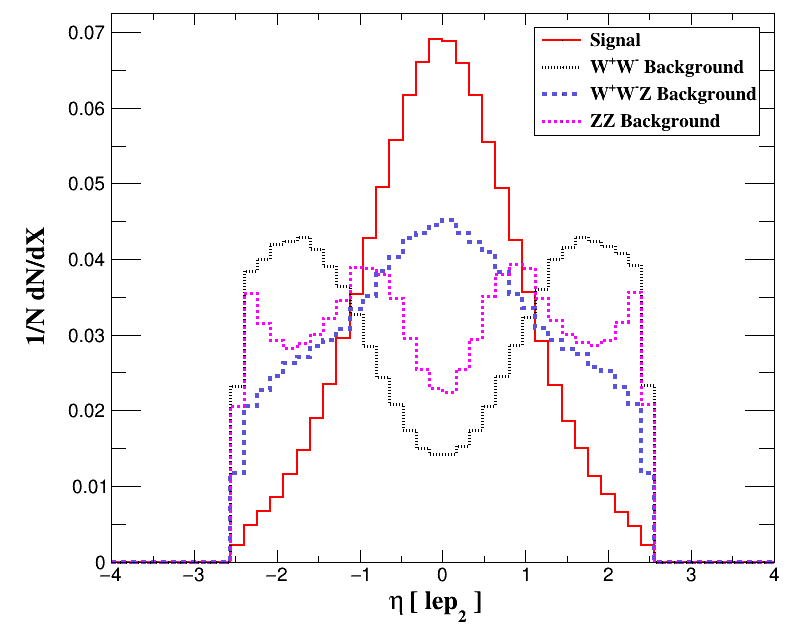} \\
\includegraphics[height=60mm,width=8cm]{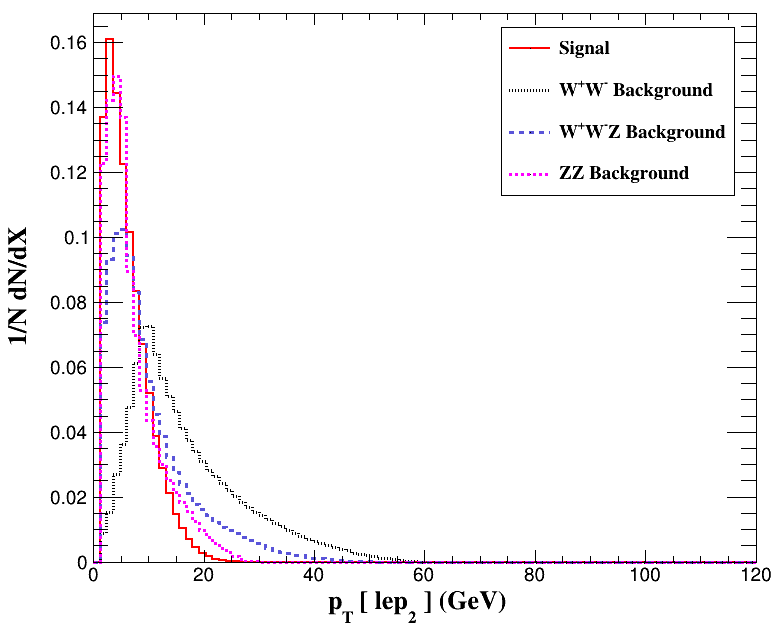} &
\includegraphics[height=60mm,width=8cm]{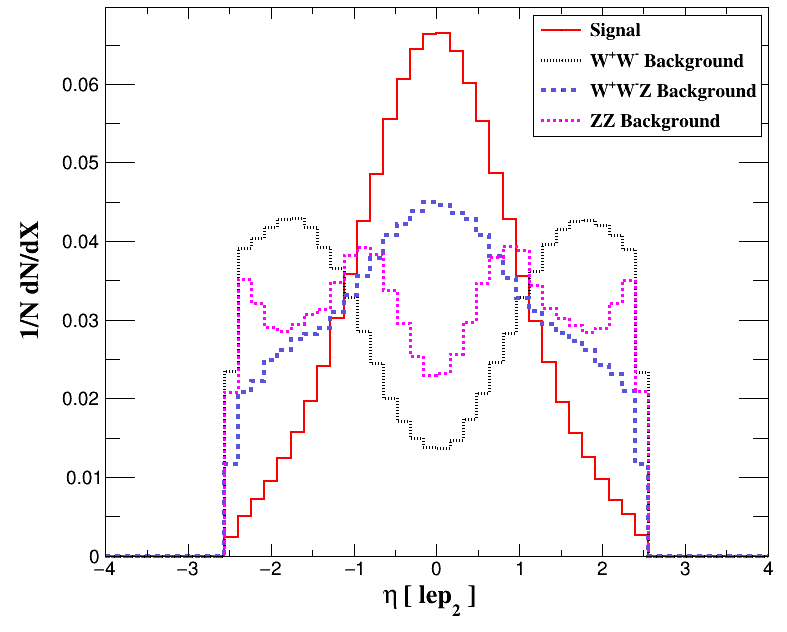} \\
\includegraphics[height=60mm,width=8cm]{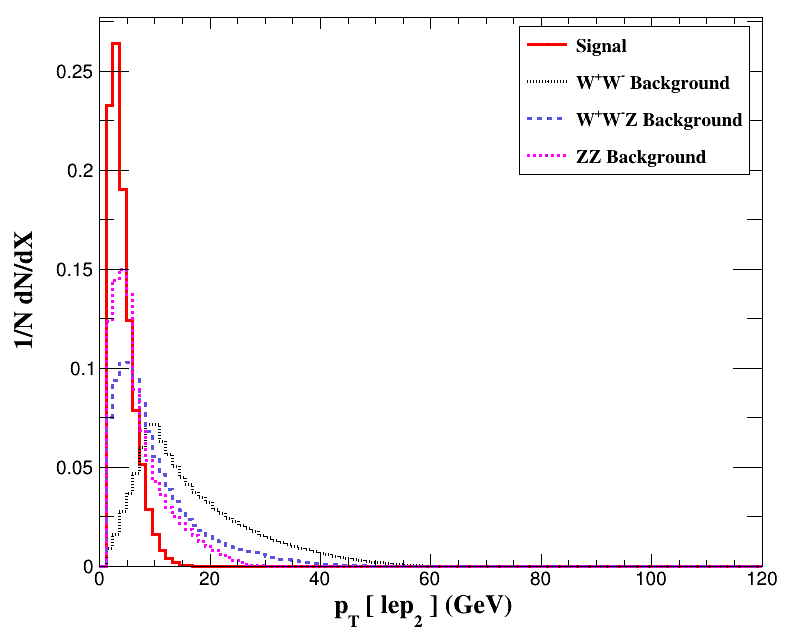}  &
\includegraphics[height=60mm,width=8cm]{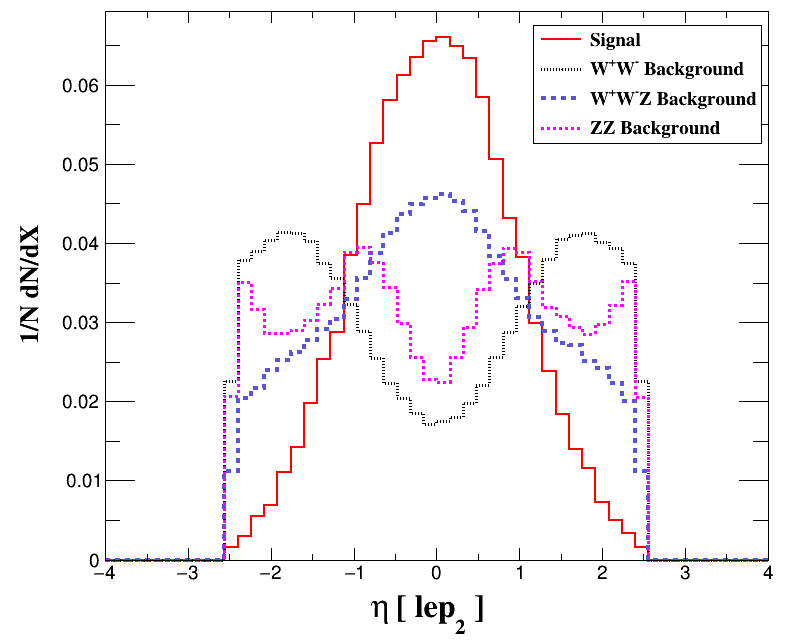} 
\end{tabular}
\caption{\it Transverse momentum and pseudo-rapidity distribution of the second hardest lepton 
 for BP1 and for combination $(a)$ unpolarized (top), $(b)$ both-polarized (middle) and $(c)$ only electron-polarized (bottom).}
 \label{ilc_pt2}
\end{figure}

In Fig.~\ref{ilc_pt1}, the transverse momentum $(p_T)$ and pseudo-rapidity $(\eta)$
distribution for the hardest lepton in the final state is shown for all three polarization combinations.
The same is shown in Fig.~\ref{ilc_pt2}, but for the second hardest lepton. We mention here that
as an example plot, we only show the distributions for our first benchmark point (BP1), which is the most promising for 
the experimental detection of our chosen process.

\begin{figure}[htbp!]
\centering
\begin{tabular}{ccc}
\includegraphics[height=50mm,width=5.2cm]{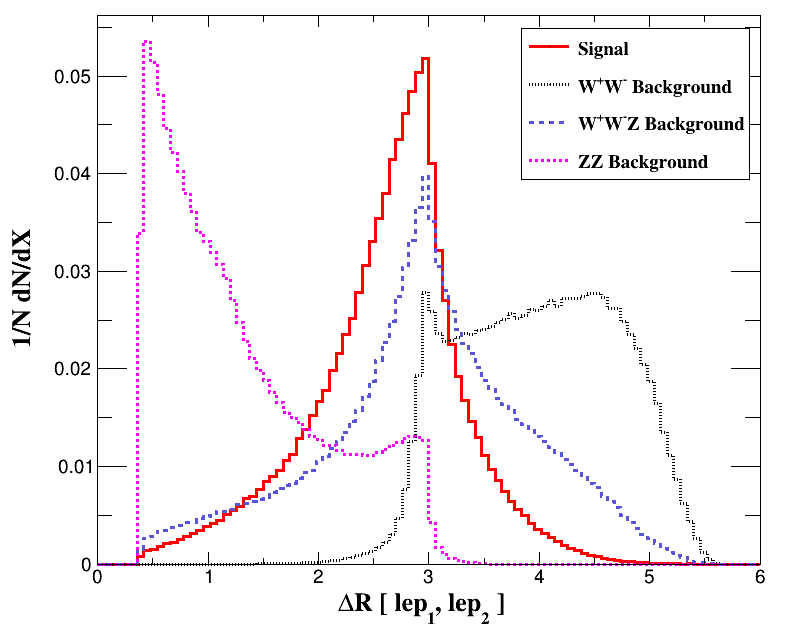} &
\includegraphics[height=50mm,width=5.2cm]{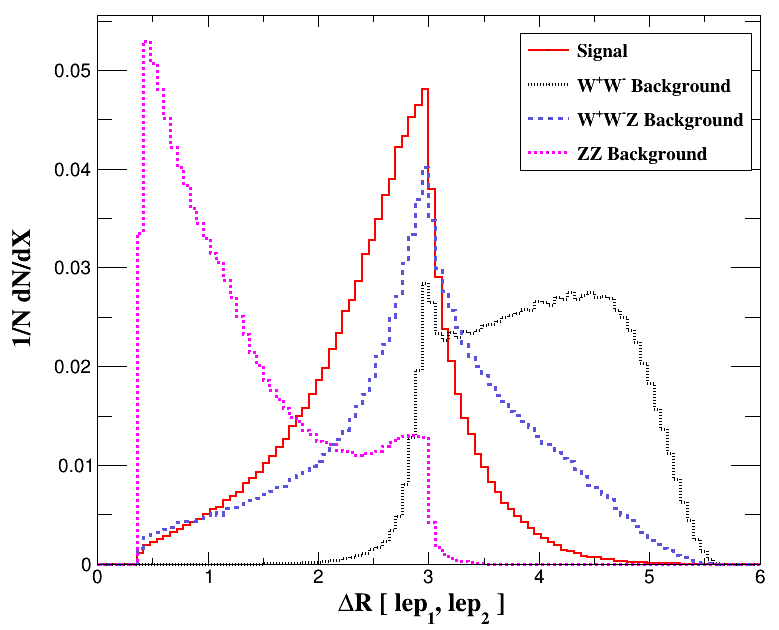} &
\includegraphics[height=50mm,width=5.2cm]{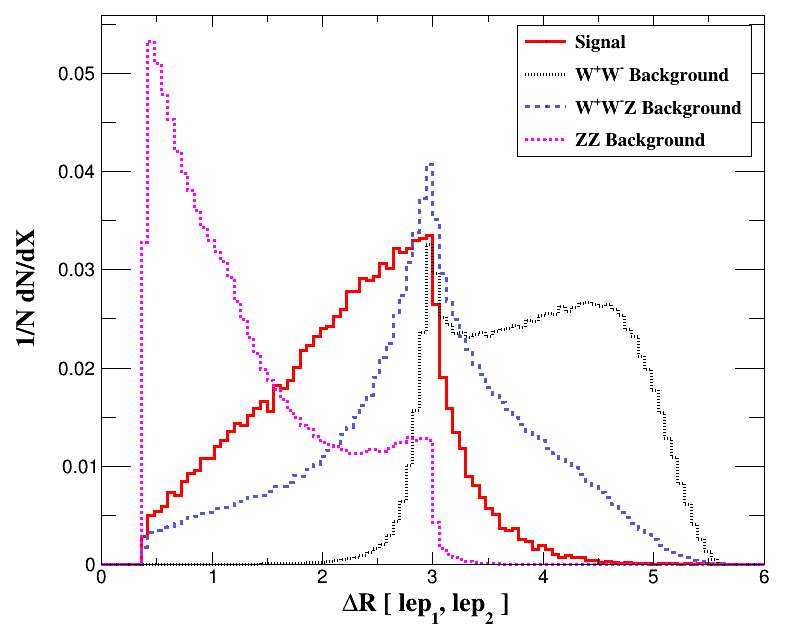}  
\end{tabular}
\caption{\it $\Delta R$ distribution of the two leptons 
 for BP1 and for combination $(a)$ unpolarized (left), $(b)$ both-polarized (middle) and $(c)$ only electron-polarized (right) respectively.}
\label{ilc_deltar}
\end{figure}

The point to be noted here is that for all the cases, the lepton $p_T$ spectrum for the 
SM background is relatively harder than that of the signal distribution. This feature can be 
understood from the fact that for a SM background, leptons originate from the direct production $W^\pm$ bosons,
whereas in the signal process they come from the cascade decay of heavy leptons $E_2^\pm$. This also 
explains why the pseudo-rapidity distribution of the leptons is mostly central for the 
signal, unlike the $WW$ and $ZZ$ cases where leptons show peaking behaviour at large pseudo-rapidities. However, 
due to the 3 body kinematics of the $WWZ$ process, the pseudo-rapidity distribution of leptons coming from this 
process are evenly distributed over the full rapidity range (-2 to +2). Keeping this in mind, one can also similarly interpret the distribution of
$\Delta R$ between the two leptons, which we show in Fig.~\ref{ilc_deltar}, where $\Delta R$
is defined as the measure of angular separation between the two charged leptons by means of their
difference in pseudo-rapidity $(\eta)$ and azimuthal angle $(\phi)$ : $\Delta R = \sqrt{(\Delta\eta)^2 + (\Delta\phi)^2}$. 
Both of the charged leptons are generated from one single $Z$ boson for $ZZ$ background
and therefore the separation is less than the signal and $WW$ background, where the leptons 
come from two $W$'s with large angular separation.

At this point, it is worth mentioning that the di-lepton final state may also arise from pair 
production of the singly charged scalar states. The charged Higgs masses only depend
on the scalar quartic coupling of Eq.~(\ref{eq:pot_htm}) and are 
independent of the parameter space that we chose for our benchmark points
in VL lepton searches. 
But the production cross section times the branching ratio to di-lepton final states
is much lower than what we get from $E_2$ pair production, and so it contributes negligibly 
to final significance\footnote{The scalar sector of this model can, however, be detected
from other interesting final states but this is beyond the scope of this work and  will be addressed  in a future project.}. In this respect, it should also be 
mentioned that the same recipe also applies to our studies for the LHC searches but 
similar to this case, the effective cross section for the pair production of singly charged
scalars decaying to di-lepton final states at the LHC is too small to 
add anything to the final signal significance\footnote{A study on the heavier state of the scalar sector of the model
at the High Energetic Future Hadron Collider can be found in Ref.~\cite{Dev:2016dja}.}. 

Investigating the distributions of the above-mentioned kinematic variables, we choose our final selection cuts.
First of all, from Fig.~\ref{ilc_deltar} one can see a significant 
deviation in the spectrum of $\Delta R (\ell_1, \ell_2) $ between our signal 
and the SM background, mainly from $ZZ$ and $WWZ$ processes. We thus select events with di-leptons, where the angular 
separation between the two leptons must satisfy our first selection cut, $2 < \Delta R (\ell_{1}$,$\ell_{2})$ $< 3.5$.
Next, following the distributions in Fig.~\ref{ilc_pt1},
we see that a choice of the hardest lepton $p_T$ between 25 GeV and 160 GeV along with a
pseudo-rapidity within the central region $(\mid \eta(\ell_1)\mid < 1.5)$ may enhance the signal significance by
considerably reducing the main $WW$ background. However, one should know that these
additional cuts are chosen by only analyzing the unpolarized scenario and are kept the same for the other two 
polarization combinations. The modification of selection cuts according to specific
polarization cases will further increase our final signal significance. The last selection cut
on the transverse momentum and pseudo-rapidity of the second hardest lepton follows the same
logic as the selection cut on the hardest lepton.
All the selection cuts are displayed in Table~\ref{ilc_cuts}.
\begin{table}[htbp!]
	\centering
	\begin{tabular}{|c|c|} \hline \hline
		Cut name & Selection criteria \\ \hline
		C1 & $2 < \Delta R (\ell_{1}$,$\ell_{2})$ $< 3.5$ \\ \hline
	        C2 & At least one lepton with $ 25 < p_T(\ell_{1}) < 160$ GeV and $|\eta(\ell_{1})| < 1.5$  \\ \hline
		C3 & At least two leptons with $p_T(\ell_{2}) > 20$ GeV  $|\eta(\ell_{2})| < 1.5$  \\ \hline
	
	\end{tabular}
\caption{The cuts implemented for ILC searches.}
\label{ilc_cuts}
\end{table}
\def\nl{\cline{1-5}&&&&&\\[-2.5ex]}
\begin{table}[htbp!]
	\centering
	\begin{tabular}{|c|p{2cm}|c|c|c|p{1.0cm}|}
		\hline
		 \multicolumn{2}{|c|}{}& \multicolumn{3}{|p{3.5cm}|}{Effective Cross-section (fb) after the cut} & ${\cal L}_{5\sigma}$ ($\rm fb^{-1}$) \\ \hline
		SM-background & Production Cross-sec. (fb) 
		& C1 & C2 & C3 &  \\ \nl
		$W^{+}$ $W^{-}$ & 56.5 & 12.62 & 1.36  & 1.16 &  \\ \nl
	         $W^{+}$ $W^{-}$Z & 0.44 & 0.21 & 0.057 & 0.037 &  \\ \nl
	         Z Z & 2.13 & 0.46 & 0 & 0 &  \\ \nl
		Total background & \multicolumn{3}{c|}{ } & 1.197 &   \\ \hline \hline
		BP1& 17 & 12.31 & 10.4 & 8.81 &  3.22 \\ \hline 
	       BP2 & 14.5 & 9.44 & 9.02 & 7.45 & 3.89 \\ \hline 
	        BP3 & 12.8 & 7.68 & 7.33 & 5.85 & 5.16 \\ \hline 
	        \end{tabular}
	 \caption{\it Effective cross sections after each cut for both background and signal, and the 
   	 integrated luminosity required for $5\sigma$ significance  (${\cal L}_{5\sigma}$) at {\rm 1 TeV} at the ILC
   	 for an unpolarized incoming beam.}
\label{cut_flow_unpol}
\end{table}

The cut-flow and the required integrated luminosity for a $5\sigma$ discovery reach 
are given in Tables~\ref{cut_flow_unpol}, \ref{cut_flow_pol} and \ref{cut_flow_empol},
respectively, for polarization combinations $(a)\,,~(b)\,,$ and $(c)$, i.e., for completely 
unpolarized initial states, states where both the electron $(80\% {\rm left} )$ and positron $(60\% {\rm right})$ beams are polarized and states with only a
right-polarized electron $(80\%)$ beam. The significance is calculated using the same relation given in
Eq.~(\ref{significance}). 
\begin{table}[htbp!]
	\centering
	\begin{tabular}{|c|p{2cm}|c|c|c|p{1.0cm}|}
		\hline
		\multicolumn{2}{|c|}{}& \multicolumn{3}{|p{3.5cm}|}{Effective Cross-section(fb) after the cut} & ${\cal L}_{5\sigma}$ ($\rm fb^{-1}$)\\ \hline
		SM-background & Production Cross-sec. (fb) 
		& C1 & C2 & C3 &  \\ \nl
	 $W^{+}$ $W^{-}$ & 162 & 35.75 & 3.93  & 3.35 &  \\ \nl
	 $W^{+}$ $W^{-}$Z & 1.2 & 0.64 & 0.17 & 0.11 &  \\ \nl
	 Z Z & 4.4 & 1.0 & 0 & 0 &  \\ \nl	
	 	Total background & \multicolumn{3}{c|}{ } & 3.46 &   \\ \hline \hline	
	BP1 & 27.7 & 20.58 & 17.13  & 14.76  & 2.09   \\ \hline 
	BP2 & 23.5 & 16.1 & 15.52 & 13.02  & 2.44  \\ \hline 
	BP3 & 20.64 & 12.89 & 12.39 & 10 & 3.37 \\ \hline 
   \end{tabular}
   \caption{\it Effective cross sections after each cut for both background and signal, and the 
      	 integrated luminosity required for $5\sigma$ significance  (${\cal L}_{5\sigma}$) at {\rm 1 TeV} at the ILC for a both-polarized incoming beam.}
   \label{cut_flow_pol}
\end{table}

Let us understand the aftermath of selection cuts for all three cases.
As expected, the $\Delta R$ selection cut (C1) seems quite competent in suppressing 
mainly the SM background. While almost 75\% signal events pass the cut, only 22\% of 
$W^+W^-$ background events remain unaffected. The signal significance is better 
when the other two cuts (C2) and (C3) are applied on top of (C1). 
Quantitatively, almost 90\% of background events fail to overcome  the
cut 2 (C2) selection barrier, while around 85\% of signal events survive.
The last cut (C3), however, may not play a convincing role 
in enhancing the signal significance, but the requirement of a second lepton is mandatory to 
avoid other unwanted SM backgrounds. 
\begin{table}[htbp!]
	\centering
	\begin{tabular}{|c|p{2cm}|c|c|c|p{1cm}|}
		\hline
		\multicolumn{2}{|c|}{}& \multicolumn{3}{|p{3.5cm}|}{Effective Cross-section(fb) after the cut} & ${\cal L}_{5\sigma}$ ($\rm fb^{-1}$)\\ \hline
		SM-background & Production Cross-sec. (fb) 
		& C1 & C2 & C3 & 
		\\ \nl
		$W^{+}$ $W^{-}$ & 11.7 & 2.89 & 0.31  & 0.27 &  \\ \nl
	 $W^{+}$ $W^{-}$Z & 0.09 & 0 & 0 & 0 &  \\ \nl
	 Z Z & 1.4 & 0.32 & 0 & 0 &  \\ \nl
		Total background & \multicolumn{3}{c|}{ } & 0.27 &   \\ \hline \hline
	BP1 & 15.62 & 10.99 & 9.45 & 7.86 &  3.29 \\ \hline 
	BP2 & 13.4 & 8.3  & 7.83 & 6.37 & 4.08  \\ \hline 
	BP3 & 11.9 & 6.85 & 6.5 & 5.11 & 5.16 \\ \hline 
	 \end{tabular}
   \caption{\it Effective cross sections after each cut for both background and signal and the 
      	 integrated luminosity required for $5\sigma$ significance (${\cal L}_{5\sigma}$) at {\rm 1 TeV} at the ILC for an
      	 incoming beam with only the
   	 electron beam polarized.}
   \label{cut_flow_empol}
\end{table}

Reviewing the cut-flow tables, we observe that the required integrated luminosity 
for a 5$\sigma$ discovery reach is quite low, about  2-5 $\rm fb^{-1}$, which  
can easily be reached even in the first run of the ILC. The best possible 
channel turns out to be the benchmark point BP1, with both the initial electron and
positron beams polarized as 80\% left and 60\% right helicity states, 
and where the required integrated luminosity for discovery reach is as low as
2.09 $\rm fb^{-1}$. All  the rest of the benchmarks and beam polarization states are also promising and can easily be tested 
at the upcoming ILC searches.

 Finally, we would like to point out that we have intentionally chosen those benchmark points where the
mass difference between $E_2^\pm$ and the DM (or $\nu_1$) is more than 80 GeV so that the secondary $W$ bosons
produced from $E_2^\pm$ only decay on shell. However, there are few available parameter points that can
survive the 2$\sigma$ Planck relic density constraints where the mass difference is less than 80 GeV.
For such points $E_2$ only decays to three-body final states. 
We choose two example benchmark values consistent with  Fig.~\ref{fig:relic_ML=1}, as shown in Table~\ref{offshell_w} for
which the relic density lies within the $2\sigma$ Planck limit. In Table~\ref{offshell_w}, we give the 
masses of both the DM candidate and the vectorlike charged lepton $E_2^\pm$ along with the respective production
cross section for $E_2^\pm$ pair production at 1 TeV at the ILC.  Note that such points can only 
be obtained for large Yukawa couplings $Y_L^{\prime \nu} \gsim 1.7$, while fixing the other couplings at the values 
given in Table~\ref{table:fixedparam} and  $h_R^\prime = 0.045$.
We analyze both sample points for unpolarized beams at 1~TeV at the ILC with the 
same cuts mentioned previously. The cut analysis indicates that a 5$\sigma$ signal significance 
 requires only 8.65 and 3.06 $\rm fb^{-1}$ integrated luminosity, respectively, for the two points, at 1~TeV at the ILC.
Therefore, it is evident that our ILC search prospect is promising regardless of the mass difference between
the two states.

\begin{table}[htbp!]
	\centering
	\begin{tabular}{|c|c|c|c||c|p{1cm}||c|p{1cm}||c|p{1cm}||c|p{1cm}||c|p{1cm}|}
	\hline
	 $M_{\rm DM}$ & $M_{E_2^\pm}$ & $Y_L^{\prime \nu}$ &  $\Omega_{\rm DM} h^2$ & $(a)~ \sigma_{\rm prod}$ & ${\cal L}_{5\sigma}$ & $(b)~ \sigma_{\rm prod}$ & ${\cal L}_{5\sigma}$ & $(c)~\sigma_{\rm prod}$ & ${\cal L}_{5\sigma}$  \\ 
	 	 $(\rm GeV)$ & $(\rm GeV)$ &  &  & (fb) &  $(\rm fb^{-1})$ & (fb) &  $(\rm fb^{-1})$& (fb) &  $(\rm fb^{-1})$\\ \hline
		160 & 184 & 1.8 & 0.119 & 8.4 & 8.65 & 7.73 & 9.4 & 13.44 & 5.4\\ \hline
        173 & 238 & 1.7 & 0.118 & 15.72 & 3.06 & 14.46 & 3.3& 25.15 & 1.9  \\ \hline
	 \end{tabular}
   \caption{\it Two distinct parameter points where $(M_{E_2^\pm} - M_{\rm DM}) < 80$ GeV and the 
     corresponding values for relic density with the $E_2^\pm$ pair production cross-section 
     and the integrated luminosity required for $5\sigma$ significance  (${\cal L}_{5\sigma}$) at {\rm 1 TeV} at the ILC for
     the following cases:
     (a) Both the electron and positron beams are unpolarized  (columns 5 and 6).
     (b) The electron beam is 80\% right polarized and the positron beam is unpolarized (columns 7 and 8).
     (c) The electron beam is 80\% left polarized and the positron beam is 60\% right polarized  (columns 9 and 10).}
   \label{offshell_w}
\end{table}

Overall, we see that the VL charged leptons, if they are light $\sim 500~\rm GeV$ or so have a clearer signature at the ILC than at the
LHC, where they are extremely difficult to probe, even with the highest possible reach of integrated luminosity.
\section{Conclusion}
\label{sec:conclusion}
We present a complete and thorough investigation of the effects of introducing vectorlike leptons into left-right symmetric models. Our aim is to adjoin one missing piece, a dark matter candidate, into the model. In keeping with the symmetries of the model, two vectorlike doublets are introduced, one left-handed and one right-handed, together with their mirrors. A discrete parity symmetry forbids mixing of vectorlike particles with ordinary leptons: this is introduced for simplicity, as mixing can occur, but, given constraints from flavor-changing decays such as $\mu \to e \gamma$, only with the third family, and even there the mixing is constrained to be small so as not to spoil low energy phenomenology results. 

However, in the absence of mixing with SM leptons, the vectorlike leptons mix among themselves, and the lightest state (electrically neutral, and mostly right-handed) is stable and can serve as a dark matter candidate. We show that, for a large range of the parameter space, this vectorlike neutrino obeys constraints from the relic density abundance. In direct detection, the limits on spin-dependent cross sections do not restrict the parameter space, while the spin-independent cross section falls below the LUX and XENON100 limits, whereas XENON1T puts pressure on the lighter (70-150 GeV) region of dark matter mass, which lies within its 2$\sigma$ sensitivity curves, rather than below. For indirect detection, we analyze the annihilation cross section into SM particles and show that it is safely below the Fermi-LAT limits, and the muon and neutrino fluxes coming from cosmic rays, also agree with experimental bounds from neutrino telescopes. 

Finally, we investigate the distinctive signals of this scenario at colliders. At the LHC, the pair production of the lightest vectorlike charged leptons, each decaying further into a $W$ boson and dark matter yielding $W^+W^- + \mET$, is analyzed and compared to the background coming from SM $W$ pair production. We devise three benchmarks obeying all dark matter and Higgs signal constraints, and show that, with judicious background cuts and at HL (high luminosity) LHC, one benchmark could reach $\sim 3 \sigma$ signal to background significance. 
At this point, we advocate the idea of testing our model at an upcoming electron-positron collider experiment (ILC), in particular for  
our search channel. The ILC experiment is generally preferred over the hadron colliders because of
its clean environment and ability to provide high precision measurements. Here, we mainly opt for 
the ILC because of its two main special characteristics. First, the final state with VL charged leptons can be easily
probed with much less SM background interference, and second, the ILC provides us with its 
distinct feature of polarized incoming electron and positron beams, which makes the search channel
easier to investigate. In particular, we consider three distinct combinations of beam polarizations,
 named  $(a), ~(b),$  and $(c)$, respectively, for the cases where both incoming beams are unpolarized, 
where the electron beam is 80\% right-polarized and the positron beam is 60\% right-polarized, 
and where only the electron beam is 80\% right-polarized with a
completely unpolarized positron beam. Since, our signal resembles mostly  SM $W$ pair production,
the enhancement (suppression) in the production cross-section with an incoming beam polarization 
follows the same rule as SM $WW$ background does. 
Similar as for the LHC scenario, we illustrate our search strategies using the
same benchmark points as before.  
The dominant background here is also the SM $W^+W^-$ pair production, with small contributions from $W^+W^-Z$ and $ZZ$. 
The most convenient kinematic variables to distinguish the signal from background are the pseudo-rapidity 
of the two charged leptons in the final state and specifically the angular separation ($\Delta R$) between them. 
Strategic cuts on these variable lead to large signal significance for the pair production of vectorlike leptons.
Moreover, the choice of the polarized beam combination $(b)$, where the electron beam is 80\% left polarized and 
positron beam 60\% right polarized, renders the best possible signal significance. A $5\sigma$ discovery reach,
 in this case, can easily be attained even with integrated luminosity as small as 2 $\rm fb^{-1}$ for the highest production cross section benchmark BP1. The other polarization combinations are
also impressive and can be tested even at the very first run of the ILC.

 Our analysis strategy demonstrates the viability of the model 
 prediction both for DM detection and for collider signatures at the upcoming ILC. Our left-right model with dark matter is thus quite predictable and easily testable, perhaps at the HL-LHC, and certainly at the ILC albeit, given the CM energies available at the linear collider, for relatively light vectorlike leptons, with masses $M \le 500$~GeV. 

\section{Acknowledgments}
MF  thanks NSERC for partial financial support under Grant
No. SAP105354. NG would like to thank the Council of Scientific and Industrial Research (CSIR), Government of India for financial support.
\section{Appendix: eigenvalues for the vectorlike neutrino mass matrix}

We list the eigenvectors of the vectorlike neutrino mass matrix given in Eq. \ref{eq:neutrinomasses}. 
\begin{eqnarray}
| \nu_1 \rangle&\simeq&\frac{ M_{\nu_1} m_{\nu}^\prime}{\sqrt{M^2_{\nu_1}(M_R^2+M^2_{\nu_1})+m^{\prime\,2}_\nu (M_L^2+M^2_{\nu_1})}}  |\nu_L^{\prime} \rangle
 +\frac{M^2_{\nu_1}}{\sqrt{M^2_{\nu_1}(M_R^2+M^2_{\nu_1})+m^{\prime\,2}_\nu (M_L^2+M^2_{\nu_1})}}  |\nu_R^{\prime\, c} \rangle \nonumber \\
 &+&\frac{M_L m_\nu^\prime}{\sqrt{M^2_{\nu_1}(M_R^2+M^2_{\nu_1})+m^{\prime\,2}_\nu (M_L^2+M^2_{\nu_1})}}  |\nu_R^{\prime \prime\, c}\rangle
+\frac{M_R M_{\nu_1}}{\sqrt{M^2_{\nu_1}(M_R^2+M^2_{\nu_1})+m^{\prime\,2}_\nu (M_L^2+M^2_{\nu_1})}}  |\nu_L^{\prime \prime} \rangle
\nonumber\\
| \nu_2 \rangle&\simeq&\frac{M_{\nu_2}m_{\nu}^\prime}{\sqrt{M^2_{\nu_2}(M_R^2+M^2_{\nu_2})+m^{\prime\,2}_\nu (M_L^2+M^2_{\nu_2})}}  |\nu_L^{\prime} \rangle
 +\frac{M^2_{\nu_2}}{\sqrt{M^2_{\nu_2}(M_R^2+M^2_{\nu_2})+m^{\prime\,2}_\nu (M_L^2+M^2_{\nu_2})}}  |\nu_R^{\prime\, c} \rangle \nonumber \\
& +&\frac{M_L m_\nu^\prime}{\sqrt{M^2_{\nu_2}(M_R^2+M^2_{\nu_2})+m^{\prime\,2}_\nu (M_L^2+M^2_{\nu_2})}}  |\nu_R^{\prime \prime\, c}\rangle 
+\frac{M_R M_{\nu_2}}{\sqrt{M^2_{\nu_2}(M_R^2+M^2_{\nu_2})+m^{\prime\,2}_\nu (M_L^2+M^2_{\nu_2})}}  |\nu_L^{\prime \prime} \rangle \nonumber\\
|\nu_3 \rangle &=&\frac{1}{\sqrt{2}} \left( |\nu_L^\prime \rangle +| \nu_R^{\prime \prime \, c} \rangle \right ) \nonumber \\
|\nu_4 \rangle &=&\frac{1}{\sqrt{2}} \left( |\nu_L^\prime \rangle - | \nu_R^{\prime \prime \, c} \rangle \right ) 
\end{eqnarray}


\bibliographystyle{JHEP.bst}
\bibliography{ref.bib}

\end{document}